\documentclass[english, 12pt]{article}
\usepackage[latin9]{inputenc}
\usepackage{geometry}
\usepackage{float}
\usepackage{amsthm} 
\usepackage{amsmath}
\usepackage{amssymb}
\usepackage{mathtools}
\usepackage{graphicx}
\usepackage{caption} 
\usepackage{subcaption}
\usepackage{setspace}
\usepackage{esint}
\usepackage[authoryear]{natbib}
\usepackage{hyperref}
\usepackage{comment}
\usepackage[usenames,dvipsnames]{color}
\usepackage{ulem}
\usepackage{array}
\usepackage{textcomp}
\usepackage{enumitem}

\usepackage{multirow}
\makeatletter
\counterwithout{figure}{section}
\numberwithin{equation}{section}
\numberwithin{figure}{section}

\newcommand{\nc}{\newcommand}

\nc\marketp{marketplace}
\nc\marketpd{market} 

\usepackage[capitalize]{cleveref}

	\crefname{figure}{Figure}{Figures}
	\Crefname{figure}{Figure}{Figures}
	
\counterwithout{figure}{section}
	
\theoremstyle{plain}
\newtheorem{theorem}{\protect\theoremname}
	\crefname{theorem}{Theorem}{Theorems}
	\Crefname{theorem}{Theorem}{Theorems}

 \theoremstyle{definition}
  
	\crefname{defn}{Definition}{Definition}
	\Crefname{defn}{Definition}{Definition}

   \theoremstyle{definition}
  
	\crefname{obs}{Observation}{Observation}
	\Crefname{obs}{Observation}{Observation}
	
	   \theoremstyle{definition}
  \newtheorem{remark}{\protect\remarkname}
	\crefname{remark}{Remark}{Remark}
	\Crefname{remark}{Remark}{Remark}

 \theoremstyle{definition}

  \theoremstyle{definition}
  
  \crefname{assumption}{Assumption}{Assumptions}
	\Crefname{assumption}{Assumption}{Assumptions}
	
  \theoremstyle{plain}
  \newtheorem{cor}{\protect\corollaryname}
  
  \theoremstyle{plain}
  \newtheorem{prop}{\protect\propositionname}
  \crefname{prop}{Proposition}{Propositions}
	\Crefname{prop}{Proposition}{Propositions}
	
  \theoremstyle{plain}
  \newtheorem{lemma}{\protect\lemmaname}
  \crefname{lemma}{Lemma}{Lemmata}
	\Crefname{lemma}{Lemma}{Lemmata}
	
  \theoremstyle{plain}

  \theoremstyle{remark}
  
  \crefname{rem}{Remark}{Remarks}
	\Crefname{rem}{Remark}{Remarks}
	
  \theoremstyle{plain}
  
	\crefname{claim}{Claim}{Claims}
	\Crefname{claim}{Claim}{Claims}
  
  \theoremstyle{plain}

  \newtheorem*{assumption*}{\protect\assumptionname}

\makeatother

\usepackage{babel}
  \providecommand{\claimname}{Claim}
  \providecommand{\conjecturename}{Conjecture}
  \providecommand{\corollaryname}{Corollary}
  \providecommand{\definitionname}{Definition}
  \providecommand{\observationname}{Observation}
  \providecommand{\assumptionname}{Assumption}
  \providecommand{\examplname}{Example}
  \providecommand{\factname}{Fact}
  \providecommand{\propositionname}{Proposition}
  \providecommand{\lemmaname}{Lemma}
  \providecommand{\remarkname}{Remark}
	\providecommand{\theoremname}{Theorem}

\newcommand{\pdfcolor}{blue}
\hypersetup{
    colorlinks=true,
    linkcolor=\pdfcolor,
    citecolor=\pdfcolor,
    urlcolor=blue}
\usepackage[usenames,dvipsnames]{color}

\global\long\def\piX{\pi_X}
\global\long\def\cl{\text{cl}}

\newcommand{\kr}[1]{\left\lVert{#1}\right\rVert_{KR}}

\nc{\pd}[1]{\textcolor{Blue}{PD comment: #1}}
\nc{\ak}[1]{\textcolor{Red}{AK comment: #1}}

\newcommand{\ul}{\underline}
\newcommand{\ol}{\overline}
\newcommand{\df}{\mathrm{d}}

\newcommand{\bdis}{\begin{displaymath}}
\newcommand{\edis}{\end{displaymath}}
\newcommand{\beq}{\begin{equation}}
\newcommand{\eeq}{\end{equation}}
\newcommand{\bea}{\begin{eqnarray*}}
\newcommand{\eea}{\end{eqnarray*}}
\newcommand{\bean}{\begin{eqnarray}}
\newcommand{\eean}{\end{eqnarray}}

\newcommand{\Tau}{\mathcal{T}}
\newcommand{\R}{\mathbb{R}}

\newcommand{\E}{\mathbb{E}}

\DeclareMathOperator{\argmax}{arg\,max}
\DeclareMathOperator{\argmin}{arg\,min}
\DeclareMathOperator{\supp}{supp}
\DeclareMathOperator{\ext}{ext}

\DeclareMathOperator{\rbd}{rbd}
\DeclareMathOperator{\Gr}{Gr}
\newcommand{\1}{\mbox{\bf 1}}

\DeclareMathOperator{\relint}{relint}
\DeclareMathOperator{\bd}{bd}
\DeclareMathOperator{\interior}{int}
\DeclareMathOperator{\epi}{epi}
\DeclareMathOperator{\Lip}{Lip}

\newcommand{\ccV}{\widehat{V}}
\newcommand{\ceV}{\overline{V}}

   \defcitealias{onuchic}{Onuchic \textsuperscript{\textregistered} Ray}

\begin{document}
\onehalfspace

\title{The Persuasion Duality\thanks{We thank Isa Chavez, Elliot Lipnowski, Benny Moldovanu, Jacopo Perego, Xiaoyun Qiu, David Rahman, Doron Ravid, Chris Shannon, Jakub Steiner, Marciano Siniscalchi, Bruno Strulovici, and Alexander Wolitzky for helpful comments and suggestions. Dworczak thanks the University of Oxford and the University of Zurich for hospitality and support during the academic year 2021-2022, when some of the work on this paper was undertaken; he also gratefully acknowledges the support of the Alfred P. Sloan Fellowship. Kolotilin gratefully acknowledges support from the Australian Research Council Discovery Early Career Research Award DE160100964.} }
\author{Piotr Dworczak\thanks{Department of Economics, Northwestern University.}\; and Anton Kolotilin\thanks{School of Economics, UNSW Business School.}}
 
\date{}
\maketitle 
\begin{center}
\today
\end{center} 

\bigskip

\begin{abstract} 

We present a unified duality approach to Bayesian persuasion. The optimal dual variable, interpreted as a price function on the state space, is shown to be a supergradient of the concave closure of the objective function at the prior belief. Strong duality holds when the objective function is Lipschitz continuous. 

When the objective depends on the posterior belief  through a set of moments, the price function induces prices for posterior moments that solve the corresponding dual problem. Thus, our general approach unifies known results for one-dimensional moment persuasion, while yielding new results for the multi-dimensional case. In particular, we provide a necessary and sufficient condition for the optimality of convex-partitional signals, derive structural properties of  solutions, and characterize the optimal persuasion scheme in the case when the state is two-dimensional and the objective is quadratic.


\end{abstract}

\thispagestyle{empty}

\setcounter{page}{1}

\section{Introduction}

\cite{kamenica2011} show that the optimal signal in a Bayesian persuasion problem concavifies the objective function in the space of posterior beliefs over the state (see \citealp{BM2019} and \citealp{kamenica2019} for excellent overviews of the burgeoning literature on Bayesian persuasion).  
Although conceptually attractive, concavification is not always a tractable approach. Thus, several recent papers (see \citealp{Kolotilin2017}, \citealp{dworczak2019},  \citealp{DK},  \citealp{GP2}, and \citealp{KCW}) used duality theory to characterize the optimal signal.

In this paper, we present a unified duality approach to the Bayesian persuasion problem. Our approach builds on and extends the geometric duality of \cite{Gale1967}. The primal and the dual problems correspond to finding, respectively, the concave closure and the concave envelope of the objective function. We show that the optimal dual variable is a supergradient of the concave closure of the objective function at the prior belief (Section \ref{sec:duality}). Moreover, the dual variable can be represented as a price function on the state space. Because concave functions on finite-dimensional spaces have a supergradient at any interior point, strong duality always holds when the state space is finite. It may fail, however, when the state space is infinite; we prove that strong duality holds if the objective function is Lipschitz. %

If the objective function depends only on a finite set of moments of the posterior distribution (the ``moment persuasion" case analyzed in Section~\ref{sec:moment}),  prices for states induce prices for moments. The resulting price function is convex, lies above the graph of the objective function, and exhibits all other properties of the optimal dual variable known from the analysis of one-dimensional moment persuasion. Thus, our results generalize and unify the duality results established by \cite{Kolotilin2017}, \cite{dworczak2019}, \cite{DK}, and \cite{KCW} for the one-dimensional case. When the state space is multi-dimensional or the objective function depends on more than one moment, our generalized duality approach yields new results and insights. If the objective function is differentiable, the price function can be constructed explicitly as the upper envelope of hyperplanes that are tangent to the objective function at the conjectured support of  moments. Using this construction, we derive a necessary and sufficient condition for the optimality of a convex-partitional signal (an extension of the one-dimensional notion of a monotone-partitional signal), and establish a multi-dimensional analog of the bi-pooling result due to \cite{ABSY} and \cite{KMS}.

We use these tools to characterize the optimal signal in the classical model of \cite{rayo2010optimal} in which the state is two-dimensional and the objective function is a quadratic form (Section~\ref{sec:application}). We show that the ``bait and switch" pooling strategy of \citeauthor{rayo2010optimal} results from a trade-off between the conflicting goals of disclosing as much information as possible about a sum of two variables, while disclosing  as little information as possible about their difference. Under regularity conditions, duality permits us to represent the optimal signal as a convex partition of the two-dimensional state space into negative-sloped line segments. That is, the optimal signal discloses a weighted sum of the two dimensions,  with a weight that may depend on the induced posterior moment. We  further characterize cases in which the weight is constant, such as when the optimal signal is a sum of the two dimensions.

A contemporaneous paper \cite{malamud} also made progress on analyzing multi-dimensional moment persuasion, relying on different tools.\footnote{Using the theory of real analytic functions, \citeauthor{malamud} establish a remarkably powerful result that, under a regularity condition on the prior and the objective function, there exists an optimal deterministic signal. This result forms the foundation of their analysis. Relying on metric geometry and the theory of the Hausdorff dimension, they show that optimal signals correspond to low-dimensional manifolds.} While some of our results in Section \ref{s:cp} are related to theirs, we believe the two approaches to be complementary: for example, \citeauthor{malamud} allow the state space to be non-compact, while we cover cases when optimal signals are non-deterministic. The precise relationship to this and other papers is discussed in more detail throughout the paper in the context of specific results.

We briefly note that---despite our focus on Bayesian persuasion as the leading application---the methods we develop can be applied in any problem in which a linear objective is maximized over distributions of posteriors subject to a Bayes-plausibility constraint. Such optimization programs arise in various models with multiple interacting Receivers and in the analysis of rational-inattention and information-acquisition problems. We further discuss alternative applications and directions for future research in Section \ref{s:concl}.

\section{Model}\label{sec:model}

Let $(\Omega,\rho)$ be a compact metric space, where $\rho$ is a metric on $\Omega$. We will also refer to $\Omega$ as a measurable space, in which case the $\sigma-$algebra should be understood as the Borel $\sigma-$algebra induced by the metric $\rho$. The set of Lipschitz functions on $\Omega$, denoted by $\Lip(\Omega)$, is the set of functions $p:\,\Omega\to \mathbb{R}$ such that 
\[
\lVert p \rVert_L:=\sup\left\{\frac{|p(\omega)-p(\omega')|}{\rho(\omega,\omega')}:\, \omega,\omega'\in \Omega,\, \omega\neq\omega'\right\}<\infty.
\]
A function $p\in \Lip(\Omega)$ is $L$-Lipschitz if $\lVert p\rVert_L\leq L$. Let $\Lip_1(\Omega)$ denote the set of 1-Lipschitz functions on $\Omega$.

 Let $M(\Omega)$ be the set of finite signed Borel measures on $\Omega$, and $\Delta(\Omega)$ be the subset of probability measures. On the linear space $M(\Omega)$, we define the Kantorovich-Rubinstein norm: for each $\mu\in M(\Omega)$,
\[
\kr \mu := |\mu(\Omega)|+ \sup\left\{\int_\Omega p(\omega)\df\mu(\omega):\, p\in \Lip_1(\Omega),\, p(\omega_0)=0\right\},
\]
where $\omega_0$ is an arbitrary fixed element of $\Omega$. Since $(\Omega,\rho)$ is a compact metric space, Theorem 6.9 and Remark 6.19 in \cite{villani2009} yield that $\kr{\cdot}$ metrizes the weak$^\star$ topology on $\Delta(\Omega)$ and that $(\Delta(\Omega),\kr{\cdot})$ is a compact metric space.
Let $\Delta(\Delta(\Omega))$ be the set of Borel probability measures on $\Delta(\Omega)$, endowed with the Kantorovich-Rubinstein distance. Then,  $\Delta(\Delta(\Omega))$ is also a compact metric space.

We now formally define the persuasion problem, as in \cite{kamenica2011}. The state space is $\Omega$, and there is a prior belief $\mu_0\in \Delta (\Omega)$. An objective function $V:\Delta(\Omega)\rightarrow \R$ is bounded and upper semi-continuous. We will be imposing increasingly stronger assumptions on $V$ to derive increasingly stronger results throughout the paper.

 The persuasion problem  is to find a {\it distribution of posterior beliefs} $\tau\in \Delta(\Delta(\Omega))$ to\footnote{Formally, $\int_{\Delta (\Omega)} \mu \df \tau (\mu)=\mu_0$ is understood as $\int_{\Delta (\Omega)} \mu (B) \df \tau (\mu)=\mu_0(B)$ for all measurable $B\subset \Omega$. The same comment applies whenever we integrate functions with values in the space of measures. An alternative approach is to use the Bochner integral instead of the familiar Lebesgue integral.}
\begin{equation}\label{primal}
\begin{gathered}
\text{maximize  $\int_{\Delta (\Omega)} V(\mu) \df \tau (\mu)\quad$}\\
\text{subject to $\int_{\Delta (\Omega)} \mu \df \tau (\mu) = \mu_0$}. \tag{P}
\end{gathered}
\end{equation}
We will denote by  $\Tau(\mu_0)$ the set of feasible distributions of posteriors, that is,
\[
\Tau(\mu_0) = \left\{\tau\in \Delta(\Delta(\Omega)):\, \int_{\Delta (\Omega)} \mu \df \tau (\mu) = \mu_0 \right\}.
\]
We define the \textit{concave closure} of $V$ at $\mu_0$ to be the value of the persuasion problem:
\[
\ccV(\mu_0):=\sup_{\tau\in \Tau(\mu_0)} \int_{\Delta (\Omega)} V(\mu) \df \tau (\mu).
\]
That is, $\ccV(\mu_0)$ is the supremum of $z\in \mathbb{R}$ over all $(z,\mu_0)$ that can be expressed as a convex combination of $(V(\mu),\mu)$ with $\mu\in \Delta(\Omega)$.\footnote{\cite{kamenica2011} define the concave closure of $V$ as the smallest concave function that lies above $V$. Instead, we defined it as the value of the persuasion problem. In the general case of compact metric $\Omega$, the equivalence of these definitions follows from Proposition 3 in the Online Appendix of \cite{kamenica2011}.}

\begin{figure}[h]
\centering
\includegraphics[scale=0.4]{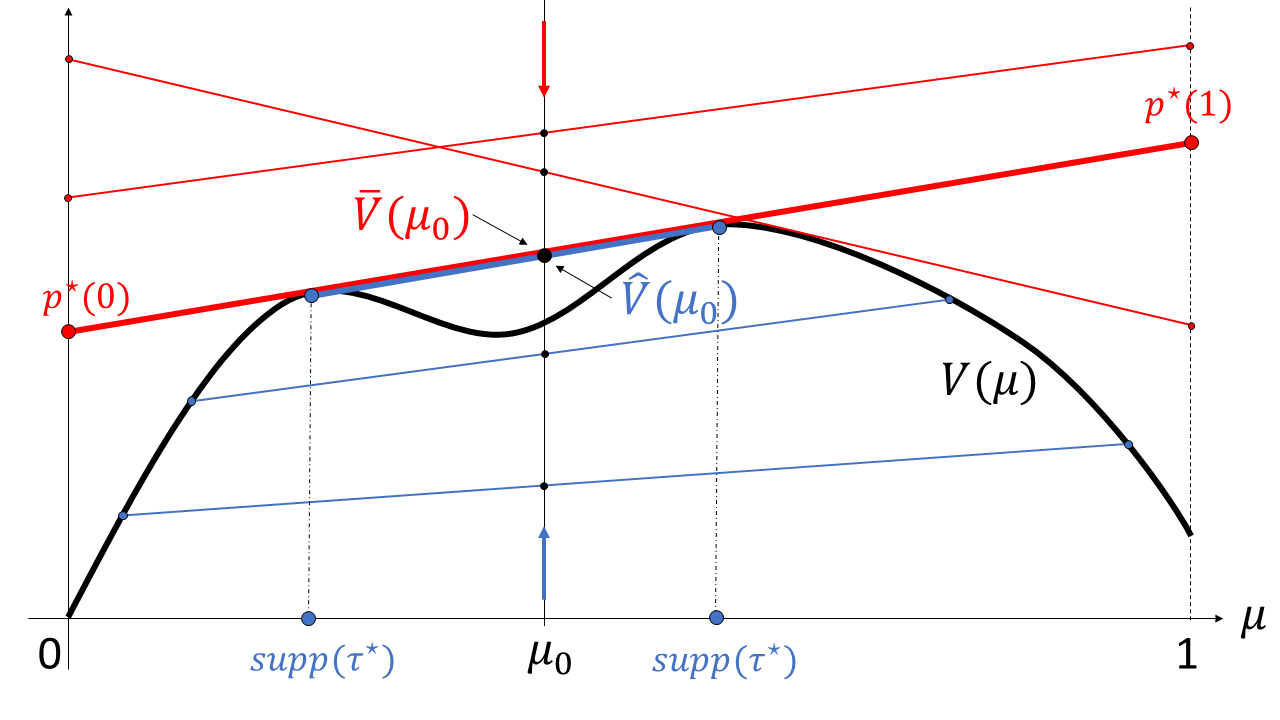}
\caption{The concave closure and the concave envelope at $\mu_0$ in the binary-state case. The concave closure is the inner construction of the convex hull of the graph of $V$: It involves maximizing the value at $\mu_0$ over all convex combinations of points on the graph of $V$ (exemplified by blue lines in the figure). The concave envelope is the outer construction of the convex hull of the graph of $V$: It involves minimizing the value at $\mu_0$ over all affine functions lying above the graph of $V$ (exemplified by the red lines in the figure).} 
\label{fig1}
\end{figure}

The dual problem is to find a {\it price function} $p\in \Lip(\Omega)$ to
\begin{equation}\label{dual}
\begin{gathered}
\text{minimize} \int_{\Omega} p (\omega) \df \mu_0 (\omega)\quad\\
\text{subject to } V(\mu)\leq\int_\Omega p(\omega)\df \mu(\omega)  \text{ for all } \mu \in \Delta(\Omega).\tag{D}
\end{gathered}
\end{equation}
We will denote by  $\mathcal P(V)$ the set of feasible price functions, that is,\footnote{In an earlier draft \cite{wp}, we considered a dual problem with a continuous $p$ (but not necessarily Lipschitz). While that approach allowed for strong duality to hold under slightly more permissive assumptions, we could not find any economic applications exploiting that additional generality. The current formulation, inspired by a comment from Doron Ravid, leads to a more elegant exposition.}
\[
\mathcal P(V)=\left\{p\in \Lip(\Omega):\, V(\mu)\leq\int_\Omega p(\omega)\df \mu(\omega)  \text{ for all $\mu \in \Delta(\Omega)$}\right\}.
\]
We define the \textit{concave envelope} of $V$ at $\mu_0$ to be the value of the dual problem:
\[
\ceV(\mu_0):=\inf_{p \in \mathcal P(V)} \int_{\Omega} p(\omega) \df \mu_0(\omega).
\]
By Definition 7.4 in  \cite{aliprantis2006}, the concave envelope of $V$ at $\mu_0$ is the infimum of values taken at $\mu_0$ by all continuous affine functions on $M(\Omega)$ that bound $V$ from above on $\Delta(\Omega)$. Our definition is equivalent: By Theorem 0 in \cite{hanin},\footnote{\cite{hanin} credits the result to  \cite{KR}. The version of the result that we use is formulated in Exercise 8.10.143 in \cite{bogachev2007}; see also Theorem 7.3 in \cite{edwards}.} the space dual to $(M(\Omega),\,\kr{\cdot})$ is the space $\Lip(\Omega)$, modulo the constant functions. Hence, any continuous linear function on $(M(\Omega),\,\kr{\cdot})$ can be represented as $\int_{\Omega} p(\omega) \df \mu(\omega)$ for some $p\in \Lip(\Omega)$.\footnote{The distinction between affine and linear functions is immaterial here since a continuous affine function $\int_\Omega p(\omega)\df \mu(\omega) +c$  coincides with the continuous linear function $\int_\Omega (p(\omega)+c)\df \mu(\omega)$ on $\Delta(\Omega)$.}  The construction of the concave closure and the concave envelope are illustrated in Figure \ref{fig1} (in the binary-state case). 

We interpret the persuasion problem as a linear production problem of \cite{gale1989}. The states are economic resources, and the probability measure $\mu_0$ is a producer's endowment of resources. The set $\Delta (\Omega)$ is the set of linear production processes available to the producer. A process $\mu \in \Delta (\Omega)$ operated at unit level consumes the measure $\mu$ of resources and generates income $V(\mu)$. A production plan $\tau$ describes the level at which each process $\mu$ is operated. The primal problem is for the producer to find a production plan that exhausts the endowment $\mu_0$ and maximizes the total income.

To interpret the dual problem, imagine that there is a wholesaler who wants to buy out the producer. The wholesaler sets a unit price $p(\omega)$ for each resource $\omega$. The producer's (opportunity) cost of operating a process $\mu$ at unit level is thus $\int_\Omega p(\omega)\df \mu (\omega)$. A price function $p$ is feasible for the wholesaler if the income generated by each process of the producer is not greater than the cost of operating the process, which makes the producer willing to sell all the resources. The dual problem is for the wholesaler to find feasible prices that minimize the total cost of buying up all the resources.\footnote{In the persuasion context, similar interpretations of the dual variable as a price function appear in \cite{dworczak2019}, \cite{GP2}, and \cite{KCW}.}

\section{Duality}\label{sec:duality}

In this section, we establish weak and strong duality for the persuasion problem:
\begin{itemize}
\item \textbf{Weak duality} states that $\ccV(\mu_0)\leq \ceV(\mu_0)$, that is, the concave closure is bounded above by the concave envelope.
\item \textbf{No duality gap} requires the equality  $\ccV(\mu_0)= \ceV(\mu_0)$, that is, the concave closure and the concave envelope coincide.
\item \textbf{Primal and dual attainment} additionally require existence of solutions to the primal and the dual problems, respectively. We use the term \textbf{strong duality} when both primal and dual attainment (and hence also no duality gap) hold.\footnote{The exact use of these terms varies across authors. For example, \cite{villani2009} uses the term strong duality to refer to primal attainment and no duality gap. Our convention is consistent with the economics literature where strong duality typically includes existence of solutions to the dual problem (see \citealp{daskalakis} and \citealp{kleiner} for recent examples). } 
\end{itemize}

For the case of a binary state, these properties are illustrated in Figure \ref{fig1}. Weak duality follows from the fact that any red line (any affine function lying above the graph of $V$) achieves a higher value at $\mu_0$ than any blue line (any convex combination of points on the graph of $V$). No duality gap states that the infimum over values that red lines can take at $\mu_0$ is equal to the supremum over values that the blue lines can take at $\mu_0$. Finally, strong duality requires that these extrema are attained (as depicted by the bold red and blue lines in the figure). 
 
Weak duality serves as a verification tool. If we can find a feasible $\tau\in \Tau(\mu_0)$ and a feasible $p\in \mathcal P(V)$ such that $\int_{\Delta(\Omega)}V(\mu)\df \tau(\mu)=\int_\Omega p(\omega) \df \mu_0(\omega)$, then $\tau$ is optimal. Within our interpretation, weak duality states that the total income generated by the producer cannot exceed the total cost of the resources under feasible prices, which make the producer willing to sell the resources. Thus, if there exists a plan for the producer and feasible prices for the wholesaler that equalize the total income with the total cost, then this plan must be optimal for the producer, and the prices must be optimal for the wholesaler.  However, weak duality does not guarantee that such solutions can be found.

No duality gap ensures that the bound imposed by weak duality is tight. Thus, a  feasible $\tau\in \Tau(\mu_0)$ is optimal if and only if it achieves the value of the concave envelope $\ceV(\mu_0)$. The absence of a  duality gap still does not guarantee that the optimality of $\tau$ can be verified by finding a feasible price function $p$. 

Finally, primal and dual attainment ensure that the solutions to both the primal and the dual problems exist, and hence optimality of the primal solution can be demonstrated by exhibiting a dual solution.  Within our interpretation, strong duality states that there exists a feasible plan for the producer and feasible prices for the wholesaler such that the cost of each operated process is equal to the income it generates. In the remainder of this section, we establish weak duality, no duality gap, primal attainment, and---under additional conditions---dual attainment.

\begin{theorem}[Weak Duality]\label{weak}
$\ccV(\mu_0)\leq \ceV(\mu_0)$. 
\end{theorem}

\begin{proof}
The proof is relegated to Appendix \ref{app:weak_duality}.
\end{proof}

As the (standard) proof reveals, weak duality does not even require the weak assumptions on $V$ that we imposed (it is only needed that the primal and the dual problems are well defined). Under our assumptions, weak duality is subsumed by the following stronger claim.

\begin{theorem}[No duality gap and primal attainment]\label{t:duality}
There is no duality gap, 
\begin{equation}\label{eq_C}
\ccV(\mu_0)=\ceV(\mu_0), \tag{O}
\end{equation}
and the value of the concave closure $\ccV(\mu_0)$  is attained by some feasible $\tau\in\Tau(\mu_0)$. 
\end{theorem}

\begin{proof}
The proof is relegated to Appendix \ref{app:weak}.
\end{proof}

The primal  problem \eqref{primal} corresponds to maximizing an upper semi-continuous function $V$ over the compact set of feasible distributions $\Tau(\mu_0)$, so existence of a solution follows from the Weierstrass Theorem. No duality gap is a consequence of hyperplane separation. However, instead of explicitly relying on a version of the hyperplane separation theorem, we show that the second concave conjugate (double Legendre transform) of the concave closure equals the concave envelope. The Fenchel-Moreau Theorem  then establishes the absence of a duality gap \eqref{eq_C}.  Theorem \ref{t:duality} thus implies that the concave closure and the concave envelope coincide, and hence we can use the two notions interchangeably.\footnote{When $\Omega$ is finite,  this follows from Corollary 12.1.1 in \cite{rockafellar}.}

One consequence of duality in the persuasion setting is that we can provide a verification result for the persuasion problem and its dual. Within our interpretation, a feasible plan and supporting prices are optimal if and only if the cost of each operated process is equal to the income it generates.
 
\begin{cor}[Complementary Slackness]\label{cor_ver}
Distribution $\tau\in \Tau(\mu_0)$ and price $p\in \mathcal P(V)$ are optimal solutions to \eqref{primal} and \eqref{dual}, respectively,  if and only if 
\begin{equation} \label{eq_CS}
V(\mu)=\int_\Omega p(\omega)\df \mu(\omega),\quad \text{for all $\mu\in \supp(\tau)$}.\tag{C}
\end{equation}	
\end{cor}
\begin{proof}
The proof is relegated to Appendix \ref{app:complslack}.
\end{proof}

In applications, Corollary \ref{cor_ver} can be used to infer properties of solutions to the persuasion problem. However, for this approach to be applicable, we must ensure that a solution to the dual problem exists. Our final goal is to establish conditions under which dual attainment holds. Contrary to previous results, additional regularity conditions on $V$ are needed.

We say that  $\ccV$ is {\it superdifferentiable} at $\mu_0$ if there exists a continuous linear function $H$ on $M(\Omega)$ (called a {\it supporting hyperplane} of $\ccV$ at $\mu_0$) represented by $p\in \Lip(\Omega)$ (called a {\it supergradient}  of $\ccV$ at $\mu_0$) such that $\ccV(\mu_0)=H(\mu_0)$ and $\ccV(\mu)\leq H(\mu)=\int_{\Omega}p(\omega)\df \mu(\omega)$ for all $\mu\in \Delta(\Omega)$. Note that the concave closure $\ccV$ is a concave function. When $\Omega$ is finite,  a concave function on $\Delta(\Omega)$ is also continuous on the interior of the domain, and hence it is superdifferentiable at all interior points (Theorems 7.12 and 7.24 in \citealp{aliprantis2006}). Interior points in case of finite $\Omega$ correspond to priors $\mu_0$ that have full support on $\Omega$. However, when $\Omega$ is infinite, the set of probability measures $\Delta(\Omega)$ has an empty (relative) interior---any $\mu_0\in \Delta(\Omega)$ is a boundary point. As a result, the hyperplane separating $(\mu_0,\,\ccV(\mu_0))$ from the graph of $\ccV$ may be vertical, and hence the required linear function $H$ may fail to exist.\footnote{For an analogy, consider a concave and continuous function $f(x)=\sqrt{x}$ on $[0,\,1]$. This function is not superdifferentiable at the boundary point $x=0$ because the supporting hyperplane would have to be vertical.}

Following \cite{Gale1967}, we say that $\ccV$ has \textit{bounded steepness} at $\mu_0$ if there exists a constant $L$ such that 
\[
\frac{\ccV(\mu)-\ccV(\mu_{0})}{\kr{\mu-\mu_0}} \leq L,\quad  \text{for all $\mu\in\Delta(\Omega)$}.
\]
Intuitively, bounded steepness says that the marginal increase in the value of the persuasion problem is bounded above for a small perturbation of the prior. %

\begin{theorem}[Dual Attainment]\label{thm_strong}
The following statements are equivalent:
\begin{enumerate}
	\item The problem \eqref{dual} has an optimal solution.
	\item $\ccV$ is superdifferentiable at $\mu_0$.
	\item $\ccV$ has bounded steepness at $\mu_0$.
\end{enumerate} 
\end{theorem}
\begin{proof}
The proof is relegated to Appendix \ref{app:regular}.
\end{proof}

Equivalence of properties 2 and 3 is established by the Duality Theorem in \cite{Gale1967}, which we can apply because we represented the space of distributions as a normed space (by using the Kantorovich-Rubinstein norm).\footnote{\cite{holmes} and \cite{GOZ} extend \citeauthor{Gale1967}'s theorem  from normed spaces to locally convex spaces, which may be useful for future generalizations of our results.} Equivalence of properties 1 and 2 follows from the fact that continuous linear functions on $M(\Omega)$ can be identified with Lipschitz functions on $\Omega$. Intuitively, superdifferentiability of $\ccV$ at the prior means that we can find a supporting hyperplane at $\mu_0$. Due to the representation theorem, a supporting hyperplane  can be identified with a Lipschitz price function on the state space. By definition of a supporting hyperplane, this price function is feasible and touches the graph of $\ccV$ at $\mu_0$---it must therefore be optimal by weak duality (Theorem \ref{weak}). This argument shows that the optimal price function is in fact a supergradient of the concave closure $\ccV$ at the prior $\mu_0$.

Geometrically, any price function $p$ defines a hyperplane in $\Delta(\Omega)\times\mathbb{R}$ by specifying what values it takes on extreme points $(\delta_\omega,\,p(\omega))$ (as depicted by the red lines in  Figure \ref{fig1}). The price function $p$ is feasible for \eqref{dual} if the hyperplane lies above $V$ on $\Delta(\Omega)$. The dual problem is to find a hyperplane that lies above $V$ and whose value at the prior $\mu_0$ is minimized. Thus, the optimal hyperplane  supports $\ccV$ at  $\mu_0$, and the optimal price $p^\star(\omega)$ of each state $\omega$ is the value of the supporting hyperplane at the Dirac probability measure $\delta_\omega$ at $\omega$.

While Theorem \ref{thm_strong} provides a necessary and sufficient condition for dual attainment, the condition is stated in terms of a non-primitive object, the concave closure of $V$. Next, we present a useful sufficient  condition on the primitive objective function $V$. 

\begin{theorem}[Lipschitz Preservation]\label{t:Lip}
Let $V$ be Lipschitz on $\Delta(\Omega)$. Then $\widehat V$ is also Lipschitz on $\Delta(\Omega)$. Consequently, $\widehat V$ has bounded steepness at each $\mu_0\in \Delta(\Omega)$. 
\end{theorem}
\begin{proof}
The proof is relegated to Appendix \ref{a:Lip}.
\end{proof}

\begin{cor}[Strong duality]
When $V$ is Lipschitz on $\Delta(\Omega)$, strong duality  holds for the persuasion problem \eqref{primal}. 
\end{cor}

While the statement of Theorem \ref{t:Lip} may seem intuitive, its proof is quite involved in the general (infinite-dimensional) case.\footnote{Theorem \ref{t:Lip} extends Lemma 1 and Corollary 2 in \cite{GS} from the case of finite $\Omega$ to the general case. Theorem 1.17(f) in \cite{laraki} establishes a version of Theorem 4 for the total variation norm on $\Delta(\Omega)$, which is not suitable for our analysis because there is no tractable characterization of the space that is dual to $\Delta(\Omega)$ under the total variation norm.} Informally, we show that given two priors, $\mu_0$ and $\eta_0$, and an optimal distribution $\tau \in \Tau(\mu_0)$, we can find a perturbation $\eta(\mu)$ of each posterior belief $\mu \in \supp(\tau)$ such that the perturbed posteriors $\eta(\mu)$ average out to $\eta_0$ under the distribution $\tau$. Moreover, the average distance between the posteriors $\mu$ and their perturbations $\eta(\mu)$ is equal to the distance between $\mu_0$ and $\eta_0$. This implies that the value of the persuasion problem under the prior $\mu_0$ cannot exceed the value of the persuasion problem under the prior $\eta_0$ by more than $L\kr{\mu_0-\eta_0}$ when $V$ is $L$-Lipschitz. Reversing the roles of $\mu_0$ and $\eta_0$ leads to the desired conclusion.

To the best of our knowledge, Theorems \ref{thm_strong} and \ref{t:Lip} provide the first general dual attainment result for Bayesian persuasion.\footnote{At the same level of generality,  Section 8 of \cite{dworczak2019} establishes weak duality by defining a price function on the space of beliefs $\Delta(\Omega)$ and requiring it to be ``outer-convex" (a relaxation of convexity). Theorems \ref{thm_strong} and~\ref{t:Lip} demonstrate that such a price function exists when $V$ is Lipschitz, and that the price function can in fact be taken to be \textit{linear} on $\Delta(\Omega)$.} Theorem \ref{thm_strong} is mathematically more general than the existing strong duality results in the sense that it applies on a larger domain of problems; in fact, bounded steepness of the concave closure is shown to be necessary and sufficient for dual attainment so it must imply all existing sufficient conditions. However, verifying bounded steepness of the concave closure may be difficult in applications.   Our Theorem \ref{t:Lip} identifies Lipschitz continuity of $V$ as a simple \textit{sufficient} condition for strong duality; while this condition is stronger than the most permissive sufficient condition identified for one-dimensional moment persuasion (\citealp{DK}), it has the advantage of being fully universal---it applies to \textit{any} persuasion problem.

We conclude the section with an illustration of duality by studying conditions for optimality of two extreme information structures: full disclosure (distribution $\tau_F\in \Tau(\mu_0)$ uniquely characterized by attaching probability one to the set of Dirac probability measures on $\Omega$) and no disclosure (distribution $\tau_N\in \Tau(\mu_0)$ that attaches probability one to the prior $\mu_0$). We argue that strong duality makes the well-known sufficient conditions necessary.

Suppose that $\mu_0$ has full support on $\Omega$ and let $V$ be Lipschitz on $\Delta(\Omega)$ so that, by Theorems \ref{thm_strong} and \ref{t:Lip}, dual attainment holds. Then, full disclosure $\tau_F$ is optimal if and only if $V$ lies below a linear function that passes through each extreme point $(\delta_\omega,V(\delta_\omega))$:
\begin{equation}\label{eq_FD}
V(\mu)\leq \int_\Omega V(\delta_\omega)\df \mu(\omega)\text{ for all }\mu\in \Delta(\Omega). \tag{F}	
\end{equation}
No disclosure $\tau_N$ is optimal if and only if
\begin{equation}\label{eq_ND}
V \text{ is superdifferentiable at }\mu_0. \tag{N}	
\end{equation}
Theorem \ref{thm_strong} implies that the dual problem \eqref{dual} has an optimal solution. Thus, by Corollary~\ref{cor_ver}, a feasible distribution $\tau\in \Tau(\mu_0)$ is optimal if and only if the optimal price function  $p\in \mathcal P(V)$ satisfies \eqref{eq_CS}. The support of $\tau_F$ is the set of all Dirac probability measures  $\delta_\omega$ on $\Omega$, so \eqref{eq_CS} simplifies to $p(\omega)=V(\delta_\omega)$ for all $\omega\in \Omega$. Thus, $\tau_F$ is optimal if and only if $V(\delta_\omega)$, treated as a function of $\omega$, belongs to $\mathcal P(V)$---this simplifies to \eqref{eq_FD}.
Similarly, the condition for optimality of $\tau_N$ follows from the observation that feasibility of $p$ along with \eqref{eq_CS} is equivalent to $p$ being the supergradient of $V$ at the prior, yielding \eqref{eq_ND}.

Because sufficiency follows from weak duality, conditions \eqref{eq_FD} and \eqref{eq_ND} are sufficient even without the assumptions on $V$ and $\mu_0$. In Appendix B.1, we show that these intuitive conditions are no longer necessary when dual attainment fails.

\section{Moment persuasion}\label{sec:moment}
 
In this section, we apply the general duality approach developed in Section \ref{sec:duality} to a persuasion problem in which the objective function depends on the posterior belief through a finite set of moments---what we refer to as ``moment persuasion." This case arises naturally in persuasion problems in which the Sender's preferences only depend on the Receiver's action, and the Receiver's optimal action depends only on aggregate statistics such as the (potentially multivariate) mean, variance, or skewness of the posterior belief.\footnote{Even with a one-dimensional state, this nests the settings of \cite{ZZ} and \cite{NP}, as well as a separable special case of \cite{KCW}.} Multi-dimensionality  allows for applications with multiple Receivers (under public communication), potentially caring about different moments of the public belief. For another example, suppose that there are $N+1$  primitive states of the world but a Sender only observes a partially revealing signal about the primitive state. The Sender sends a signal informative about her own posterior belief to a Receiver. As long as the Receiver maximizes expectation of a utility function that depends on the primitive state---by the law of iterated expectations---her payoff will only depend on the expectation of the Sender's belief, which can be represented as an element of an $N$-dimensional simplex.\footnote{\cite{ABSY} offer an analogous interpretation of the one-dimensional moment persuasion problem.} Finally, moment persuasion captures information acquisition problems for certain well-behaved utility functions of the agent acquiring information (e.g., representing mean-variance preferences).

 Weak duality for (multi-dimensional) moment persuasion can be established directly and is often sufficient to solve instances of persuasion problems. However, our approach has two distinct advantages. First, by deriving duality for moment persuasion from the general case, we unify existing approaches (differing in the representation of the constraints in the moment persuasion problem), demonstrate how the dual variables in these alternative approaches relate to one another (Theorem \ref{thm.moment}), and extend them to the multidimensional case. More substantially---due to strong duality---we are able to derive general predictions about the \textit{structure} of solutions (Theorems \ref{t:moment}, \ref{t:Brenier2}, \ref{t:KMS}, as well as Propositions \ref{prop:KX} and \ref{t:RSsmooth} in the application in Section \ref{sec:application}). In particular, strong duality implies that the complementary slackness conditions \eqref{eq_CS} must always hold; even if the optimal $p$ is unknown, these conditions impose restrictions on the optimal persuasion scheme.

\subsection{Formulation}

We assume that, given some underlying state space $\tilde \Omega$ and prior $\tilde \mu_0$,
\[
\tilde V(\tilde \mu)=v\left(\int_{\tilde \Omega} m(\tilde \omega)\df \tilde \mu(\tilde \omega)\right), \quad \text{for all  $\tilde \mu\in \Delta(\tilde\Omega)$,}
\]
for some measurable $m:\,\tilde\Omega\to \mathbb{R}^N$ and some real-valued function $v$. It will be convenient to redefine the state space as $\Omega=m(\text{supp}(\tilde \mu_0))$ with the prior $\mu_0$ given by $\mu_0(B)=\tilde\mu_0(m^{-1}(B))$ for any measurable $B\subset \Omega$, so that
\begin{equation*}\label{def_moment_persuasion}
V(\mu)=v\left(\int_{\Omega} \omega\df \mu(\omega)\right), \quad \text{for all  $\mu\in \Delta(\Omega)$.}
\end{equation*} 
We then define the space of ``moments" $X$ as the convex hull of $\Omega$.\footnote{By redefining the state space, we have converted a general case of moment persuasion to a problem in which the objective function only depends on a multi-dimensional vector of posterior means.} We assume that $X$ is a compact convex set with non-empty interior\footnote{This is without loss of generality: As a convex set in $\mathbb{R}^N$, $X$ has a non-empty relative interior, so we can always embed $X$ in a (possibly lower-dimensional) Euclidean space such that $X$ has non-empty interior.} and that $v:\,X\to \mathbb{R}$ is Lipschitz with constant $L$.

The next lemma ensures that we can rely on dual attainment from Theorems \ref{thm_strong} and \ref{t:Lip}. 

\begin{lemma}\label{lem_Lip}
If $v$ is Lipschitz, then $V$ is also Lipschitz.
\end{lemma}

\begin{proof}
The proof is relegated to Appendix \ref{proof_thm_moment}.
\end{proof}

In moment persuasion, a distribution $\tau$ of posterior beliefs $\mu\in\Delta(\Omega)$ influences the objective only through the induced distribution of moments. By Strassen's Theorem (for example, Theorem 7.A.1 in \citealp{SS}), a distribution $\piX\in \Delta(X)$ of moments is feasible (that is, induced by some Bayes-plausible distribution of posterior beliefs) if and only if $\mu_0$ is a mean-preserving spread of $\piX$. However, anticipating our results and following \cite{Kolotilin2017}, we will formulate the moment persuasion problem as optimization over joint distributions of moments and states.  Formally, we call a distribution  $\pi\in \Delta(X\times \Omega )$ \textit{feasible}, denoted $\pi\in \Pi(\mu_0)$, if
\begin{align*}
\int_{X\times B}\df \pi(x,\omega)=\int_{B}\df \mu_0(\omega),\quad \text{for all measurable }B\subset \Omega,\\
\quad \int_{B\times \Omega} (x-\omega)\df\pi(x,\omega)=0,\quad \text{for all measurable }B\subset X.
\end{align*}
The first equation is the Bayes-plausibility constraint, which says that the marginal distribution of states induced by $\pi$ is $\mu_0$. The second equation is the martingale constraint, which says that the conditional expectation $\E_\pi [\omega|x]$ induced by $\pi$ is $x$.

We let $\piX$ denote the marginal distribution of moments induced by $\pi$. The primal problem \eqref{primal} simplifies to finding a joint distribution $\pi\in \Delta (X\times \Omega)$ to
\begin{equation}\label{primal_*}
\begin{gathered}
\text{maximize  $\int_{X} v(x) \df \piX (x)\quad$}\\
\text{subject to $\pi\in \Pi(\mu_0)$}. \tag{P$_\text{M}$}
\end{gathered}
\end{equation}
When discussing intuitions, we will sometimes refer to $\pi$ informally as a ``signal."

\subsection{Prices for moments}

Our first major result of this section derives the implications of the general duality from Section \ref{sec:duality} for the special case of moment persuasion. 

\begin{theorem}\label{thm.moment}
Suppose that $v$ is $L$-Lipschitz and fix an optimal solution $p:\,\Omega\to \mathbb{R}$ to the dual problem \eqref{dual}. There exists an extension $\bar{p}:\,X\to \mathbb{R}$ of $p$ to $X$ (i.e.,\,$p$ and $\bar{p}$ coincide on $\Omega$) such that, for any optimal solution $\pi\in \Pi(\mu_0)$ to \eqref{primal_*},
\begin{enumerate}
\item $\bar{p}$ is convex, $L$-Lipschitz, satisfies $\bar{p}\geq v$, and 
\[
\int_X v(x)\df\piX(x)=\int_\Omega \bar{p}(\omega)\df\mu_{0}(\omega);
\]
\item there exists a measurable function $q:\,X\to\mathbb{R}^{N}$ such that  $\lVert q(x)\rVert\leq L$ for all $x\in X$, 
\[ 
\bar{p}(y)=\sup_{x\in X}\left\{ v(x)+q(x)\cdot(y-x)\right\} ,\quad \text{for all $y\in X$},
\]
\[
\bar{p}(\omega)=v(x)+q(x)\cdot(\omega-x),\quad \text{for $\pi$-almost all }(x,\omega).
\]
\end{enumerate}
Conversely, if there exists a feasible $\pi\in\Pi(\mu_0)$ and a price function $\bar{p}:\,X\to \mathbb{R}$ satisfying any one of conditions \textit{1} or \textit{2}, then $\pi$ is optimal for \eqref{primal_*}. (The last claim is true under a weaker assumption that $v$ is measurable and bounded.)
\end{theorem} 

Theorem \ref{thm.moment} provides sufficient and necessary conditions for optimality of a candidate solution $\pi\in \Pi(\mu_0)$.  The main insight is that ``prices for states" can be extended to ``prices for moments."  Additionally, condition \textit{1} shows that optimal prices must be convex in moment persuasion.  To see that intuitively, note that in our interpretation of the dual problem \eqref{dual} from Section \ref{sec:model}, a measure $\mu \in \Delta(\Omega)$ of resources and one unit of resource $x=\E_\mu [\omega]$ are now equivalent for the producer. If prices failed to be convex, the producer could sell at effectively higher prices by engaging in such ``mean-preserving" transformations of the resources. Thus, the wholesaler offers convex prices to begin with.

Theorem \ref{thm.moment} recovers (under a stronger assumption) the  duality results for one-dimensional moment persuasion from \cite{Kolotilin2017}, \cite{dworczak2019}, and \cite{DK}, and establishes  strong duality for multi-dimensional moment persuasion.  By providing the two conditions \textit{1} and \textit{2} that are jointly necessary but individually sufficient, the theorem unifies two alternative approaches to moment persuasion. The price function from condition \textit{1} is a direct analog of prices for moments in \cite{dworczak2019} who derive them as a multiplier on the mean-preserving spread constraint (represented in its integral form for the one-dimensional case). 
The price function from condition \textit{2}, along with the function $q$, are analogs of the dual variables from \cite{Kolotilin2017} and \cite{KCW} who derive them as multipliers on the two constraints defining the set $\Pi(\mu_0)$ of joint distributions of moments and states. In particular, $q$ is the multiplier on the martingale constraint. Thus, the two existing duality formulations for moment persuasion are a consequence of two alternative representations of feasible distributions for the primal problem.\footnote{In Appendix B.2, we formally introduce the problem dual to \eqref{primal_*}, show that the price function $\bar{p}$ from Theorem \ref{thm.moment} is indeed a solution to that problem, and formalize the connection to previous duality formulations in Appendix B.3.} Theorem \ref{thm.moment} shows that both formulations are a special case of our general duality, and that both can be extended to the multi-dimensional case.

Next, we give an overview of the proof of Theorem \ref{thm.moment}. Because we have guaranteed dual attainment (by the assumption that $v$ is Lipschitz), there exists a solution $p$ to the dual problem \eqref{dual}, and there is no duality gap: Equality \eqref{eq_C} simplifies to
\[
\int_X v(x)\df\piX(x)=\int_\Omega  p(\omega)\df\mu_0(\omega), 
\] 
for any $\pi$ optimal for \eqref{primal_*}.
We can extend $ p$ (prices for states) from $\Omega$ to $X$ (prices for moments) using the so-called ``convex-roof" construction (\citealp{BL}):
\begin{equation}
\check p(x)\coloneqq \inf\left\{\int_\Omega  p(\omega)\df\mu(\omega):\,\mu\in \Delta(\Omega),\,\int_\Omega \omega \df\mu(\omega)=x\right\},\quad\text{for all } x\in X. \label{def_CR} \tag{R}
\end{equation}
It is easy to show that $\check p$ is convex, $\check{p}\geq v$, and hence $\check{p}$ satisfies the constraint in \eqref{dual}. Moreover, by definition, $\check p$ is pointwise smaller than $ p$ on $\Omega$. If we could show that $\check p$ is Lipschitz, then $\check p$  restricted to $\Omega$ would be a solution to the dual \eqref{dual}, and condition \textit{1} in Theorem \ref{thm.moment} would hold.

However, $\check p$ does not even have to be continuous when $N$---the dimension of the space of moments---is three or higher (even though $p$ is Lipschitz).\footnote{A careful reader might notice that this implies that some assumption of Berge's Maximum Theorem must be violated. Indeed, it turns out that the feasibility correspondence $\Phi(x)=\left\{\mu\in \Delta(\Omega):\,\int_\Omega \omega \df\mu(\omega)=x\right\}$ is not necessarily lower hemi-continuous in $\mathbb{R}^N$ for $N> 2$. However, because $\Phi$ is an upper hemi-continuous correspondence, $\check p$ is lower semi-continuous, by Lemma 17.30 in \cite{aliprantis2006}.\label{foot}} There are moment-persuasion problems in which $ p\in \Lip(\Omega)$ solves \eqref{dual} but its convex roof is discontinuous. Furthermore, for non-Lipschitz $v$, one can construct examples in which there does not exist \textit{any} convex continuous extension of optimal prices for states to prices for moments. These cases help explain why our assumptions on the objective $v$ are stronger than those imposed by \cite{dworczak2019} and \cite{DK} in the one-dimensional case.  In fact, the additional complications are a direct consequence of a multi-dimensional space of moments: It can be shown that $\check p$ is Lipschitz when $\Omega$ contains the boundary of $X$---a condition that holds trivially in the one-dimensional case.\footnote{Formal arguments supporting the claims made in this paragraph can be found in Appendix B.4.} 

To circumvent these difficulties, we prove a lemma showing that the graph of $\check p$ can be separated by a hyperplane (with a properly bounded gradient, as captured by the function $q(x)$ from condition \textit{2}) from any point $(x,\,v(x))$ on the graph of the objective function $v$. We can then define a new price function $\bar{p}:\,X\to \mathbb{R}$ that is the supremum of all such hyperplanes. The resulting price function is a convex and Lipschitz extension of $ p$ that is ``sandwiched'' between $\check{p}$ and $v$. It follows that $\bar{p}$ solves \eqref{dual} (viewed as a function on $\Omega$) and that condition \textit{1} of Theorem \ref{thm.moment} holds. Additionally, using the function $q(x)$, we can show that the complementary-slackness condition \eqref{eq_CS} takes a particularly simple form described in condition \textit{2} of Theorem \ref{thm.moment}.

In the remainder of this section, we leverage Theorem \ref{thm.moment} to derive structural properties of solutions to \eqref{primal_*}. Even though Theorem \ref{thm.moment} guarantees existence of  prices for moments, it does not provide a direct way to construct them. We show next that when $v$ is continuously differentiable, we can take $q(x)$ from condition \textit{2} of Theorem \ref{thm.moment} to be equal to the gradient of $v$ at $x$ on the support of any optimal $\piX$.

\subsection{Constructing solutions in the differentiable case}\label{s:diff}

To derive tighter implications of duality for the properties of optimal solutions, we further strengthen our assumptions on the objective function. We assume that $v$ is continuously differentiable on $X$, and thus has a continuous gradient $\nabla v$ on $X$.\footnote{We say that $v$ is differentiable at $x\in X$ if there exists a gradient $\nabla v(x)\in \R^N$ such that $f(x+h)=f(x)+\nabla v(x)\cdot h+o(\lVert h\rVert)$ for all $h\in \R^N$ such that $x+h\in X$, in which case $\nabla v(x)$ is unique. %
}  
We will show that, in this case, solving the problem \eqref{primal_*} can be reduced to finding the support of the optimal distribution of moments. 

\begin{figure}[h]  
\centering
\includegraphics[scale=0.24]{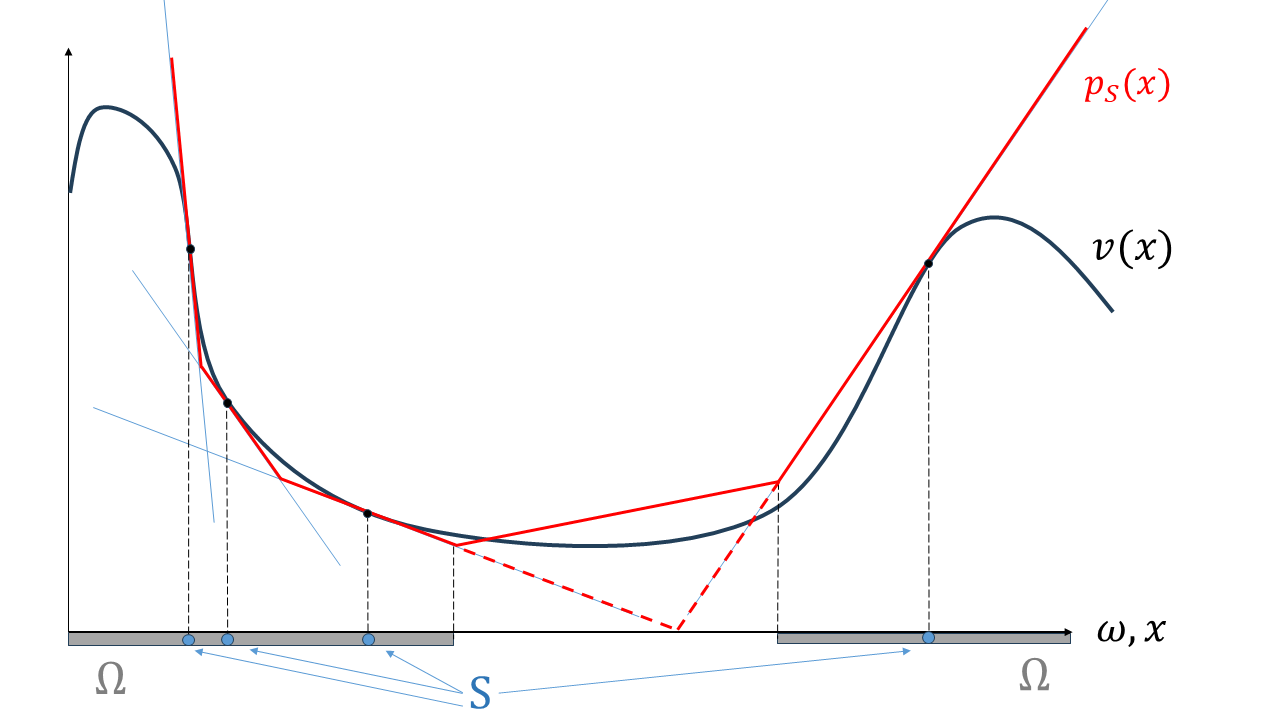} \includegraphics[scale=0.24]{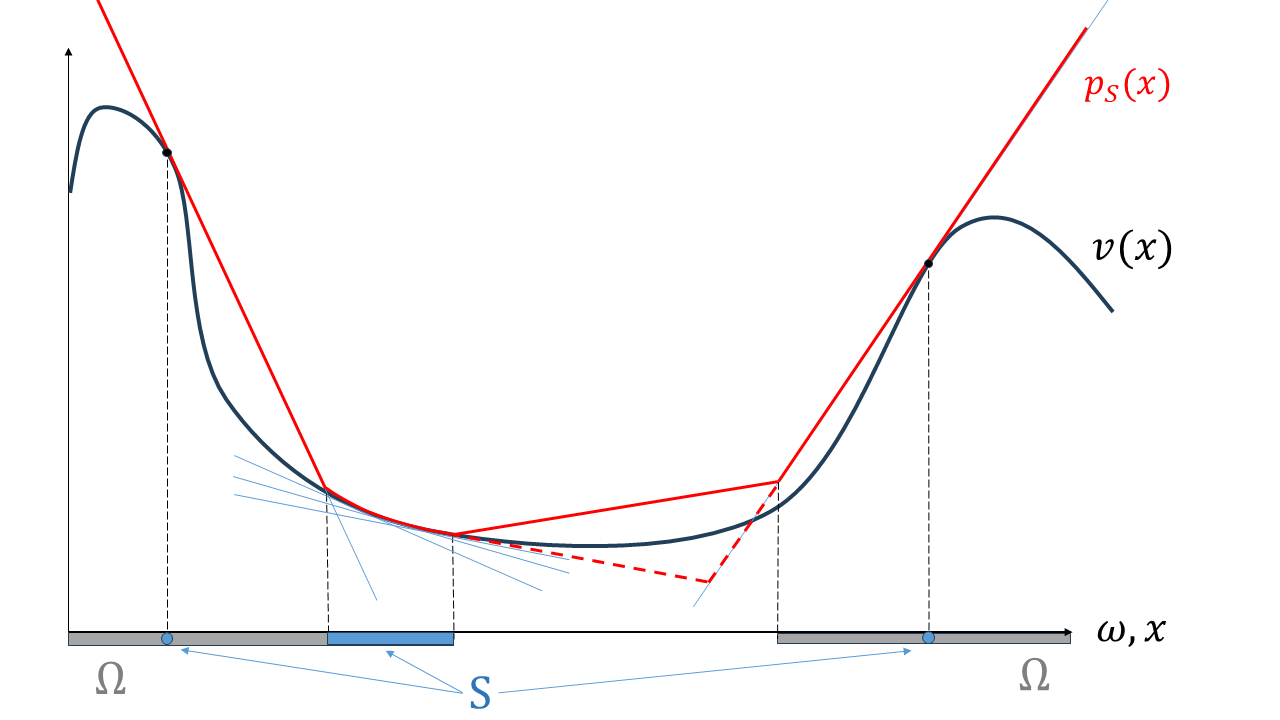}
\caption{The construction of the function $p_S(x)$ in the one-dimensional case. The gray area in the $x$-axis represents the non-convex domain $\Omega$. The left panel depicts the price function induced by a suboptimal, discrete set $S$ (indicated in blue)---the induced price function fails the first condition in \eqref{e:moment}. The right panel depicts the price function induced by a set $S$ that satisfies condition \eqref{e:moment} (for some $\pi$). The dashed red line depicts the extension of the function $p_S(\omega)$ from $\Omega$ to $X$ obtained by applying formula \eqref{e:prices} outside of $\Omega$, while the red solid line is obtained by applying the convex-roof construction.}
\label{fig3}
\end{figure}

 For any closed set $S\subset X$ (candidate support of the optimal distribution of moments), we define the function $p_S$ on $\Omega$ by
\begin{equation}\label{e:prices}
	p_S(\omega)\coloneqq\max_{x\in S} \left\{v(x)+\nabla v(x)\cdot (\omega-x)\right\},\quad \text{for all $\omega\in \Omega$}.\tag{S}
\end{equation}
In case $\Omega$ is not convex, we extend $p_S$ from $\Omega$ to $X$ using the convex-roof construction:\footnote{Note that because $p_S$ is convex on $\Omega$ by definition, it does not matter whether we use the convex roof for $x\in X\setminus \Omega$ or for all $x\in X$. The reader might be surprised that we rely on the convex roof construction after arguing that it sometimes fails to properly extend prices for states to the prices for moments. And indeed, the price function $p_S$ we construct does not necessarily satisfy all of the conditions of Theorem \ref{thm.moment}. Nevertheless, it turns out that $p_S$ satsifies the conditions that are relevant for deriving properties of optimal solutions to \eqref{primal_*} which is our ultimate goal. \label{foot_convex}}
\[
p_S(x)\coloneqq\inf\left\{\int_\Omega p_S(\omega)\df\mu(\omega):\,\mu\in \Delta(\Omega),\,\int_\Omega \omega \df\mu(\omega)=x\right\},\quad \text{for all } x\in X\setminus \Omega.
\] 
Figure \ref{fig3} illustrates these definitions: The function $p_S(\omega)$ is found  as the maximum over hyperplanes tangent to the graph of the function $v$ at points in the set $S$. The convex-roof construction extends $p_S$ from $\Omega$ to $X$ by minimizing the value achieved at any $x\in X\setminus\Omega$ of any convex combination of points belonging to the graph of $p_S$ on $\Omega$.

Finally, for any feasible $\pi\in \Pi(\mu_0)$, consider the condition:
\begin{equation}\label{e:moment} 
\begin{gathered}\tag{M}
p_S(x) \geq v(x),\quad\text{for all $x\in X$},\\
 p_S(\omega) = v(x)+\nabla v(x)\cdot (\omega - x),\quad \text{for all $(x,\omega)\in \supp(\pi)$}.
\end{gathered}	
\end{equation}
The following theorem connects condition \eqref{e:moment} to optimality of $\pi$.

\begin{theorem}\label{t:moment}
Suppose that $v$ is continuously differentiable. A joint distribution $\pi\in \Pi(\mu_0)$ is an optimal solution to \eqref{primal_*} if and only if condition \eqref{e:moment} holds with $S=\supp(\piX)$.
\end{theorem} 
\begin{proof}
The proof is relegated to Appendix \ref{proof_t:moment}.
\end{proof}

Theorem \ref{t:moment} gives rise to a ``guess and verify" procedure that can be used to identify optimal solutions to \eqref{primal_*}. The ``guess" involves conjecturing the optimal support $S$ of moments. Fixing $S$, prices $p_S$ can be computed mechanically, and then condition \eqref{e:moment} becomes necessary and sufficient for optimality of $\pi$ with support $S$.

\bigskip

In general, different solutions to \eqref{primal_*} may have different supports $S$ of  posterior moments. However, duality implies that one can define a maximal set $S^\star$ of posterior moments that can be induced by an optimal signal. In other words, any optimal signal must induce posterior moments that belong to $S^\star$. Moreover, this set $S^\star$ can be easily found as long as we have one solution to \eqref{primal_*}---we formalize this in the following remark.

\begin{remark}\label{remark1}
Suppose that $\pi^\star\in \Pi(\mu_0)$ is optimal for \eqref{primal_*}, and let \[S^\star=\{x\in X:\, p_{\supp(\piX^\star)}(x)=v(x)\}.\] Then, $\pi\in \Pi(\mu_0)$ is optimal for \eqref{primal_*} if and only if $\supp(\piX)\subset S^\star$ and condition \eqref{e:moment} holds with $S=S^\star$.\footnote{It is easy to see that $ p_S\geq v$ in this case, so only the second condition in \eqref{e:moment} is relevant.}
\end{remark}

\begin{proof}
The proof is relegated to Appendix \ref{proof_t:moment}.
\end{proof}

Remark \ref{remark1} is useful when proving uniqueness and characterizing properties of an optimal solution. We turn to these issues next. 

\subsection{Structure of solutions}\label{sec.structure}
 
In this subsection, we focus on deriving the implications of Theorem \ref{t:moment} for the structure of optimal solutions to \eqref{primal_*}. We provide a condition under which there exists a unique optimal solution $\pi$ to \eqref{primal_*} that partitions the state space into convex sets, and  pools the states in each element of the partition. This is a natural extension of the idea of monotone-partitional solutions from one-dimensional moment persuasion to the multi-dimensional case. We also generalize a result proven by \cite{ABSY} and \cite{KMS}: In the one-dimensional case, there exists an optimal signal $\pi \in \Pi(\mu_0)$ with a bi-pooling structure. We derive a multi-dimensional analog of this property. 

To simplify exposition and obtain tighter results, we  assume  that $\Omega$ is a convex set (so that $\Omega=X$). In Appendix \ref{app_ext_structure}, we extend the analysis to the general case.

\subsubsection{Optimality of convex-partitional signals}\label{s:cp}

We first address the problem of when it is without loss of optimality to restrict attention to convex-partitional signals.  Formally, we say that $\pi\in\Pi(\mu_0)$ is \textit{convex-partitional} if there is a measurable function $\chi :\Omega\rightarrow X$ such that, for all measurable sets $A\subset X$ and $B\subset \Omega$, 
\[
\pi(A,\,B)=\int_{B}\1\{\chi(\omega)\in A\} \df \mu_0(\omega),
\]
and the set $\chi^{-1}(x)$  is convex for all $x$. Intuitively, $\chi$ represents a distribution that pools all states in $\chi^{-1}(x)$ into the moment $x$. %

\begin{theorem}\label{t:Brenier2}
Suppose that $v$ is continuously differentiable and that $\mu_0$ has a density on $\Omega$ with respect to the Lebesgue measure.\footnote{The assumption that $\mu_0$ is a continuous distribution allows us to circumvent the thorny issue of how to define a convex partition when there are atoms in the distribution of states---in this case, some of the atoms may need to be split among multiple elements of the partition. } Suppose there do not exist distinct $x,\,y\in X$ with
\begin{gather*}
\nabla v(x)=\nabla v(y),\\
v(x)-\nabla v(x)\cdot x=v(y)-\nabla v(y)\cdot y,\\
\lambda v(x)+(1-\lambda)v(y)\geq  v(\lambda x+(1-\lambda)y), \quad \text{for all $\lambda\in [0,1]$}.
\end{gather*} 
Then, there is a unique optimal solution to \eqref{primal_*}, and that solution is convex-partitional. 
\end{theorem}

\begin{proof}
The proof is relegated to Appendix \ref{proof_t:Brenier2}. 
\end{proof}

Theorem \ref{t:Brenier2} gives an easy-to-verify condition on the objective function $v$ under which the optimal distribution is unique and convex-partitional. The condition can be seen as an extension of the affine-closure property from \cite{dworczak2019} that guarantees optimality of a monotone partition in the one-dimensional case.%

In Appendix B.5, we state a version of Theorem \ref{t:Brenier2} that imposes a slightly weaker sufficient condition, which  turns out to be necessary; if that weaker condition fails, then for at least some priors there exist optimal signals that are not convex-partitional. To the best of our knowledge, these results provide the most permissive conditions guaranteeing a convex-partitional signal for multi-dimensional moment persuasion. Prior to the current version of this paper, \cite{malamud} obtained a stronger sufficient  condition (requiring that $\nabla v(x)\neq \nabla v(y)$ for $x\neq y$).

In the remainder of this subsection, we give an overview of the proof of Theorem \ref{t:Brenier2}. The first part of the proof investigates the structure of optimal solutions, and does not rely on any of the assumptions of Theorem \ref{t:Brenier2}. Thus, our  goal in the overview is to present these additional results; they will be useful for subsequent analysis. The second part of the proof gives an explicit construction of the elements of the optimal convex partition from Theorem~\ref{t:Brenier2}. %

We begin by introducing some additional notation. 
Fix an optimal solution $\pi^\star\in \Pi(\mu_0)$ to \eqref{primal_*}, and define the set $S^\star$ as in Remark \ref{remark1}:
\[
S^\star:=\left\{x\in X:p_{\supp(\piX^\star)}(x)=v(x)\right\}.
\]
Recall that we can interpret $S^\star$ as the maximal set of posterior moments that can be induced by an optimal solution.
To simplify notation, let $p^\star(x)\coloneqq p_{S^\star}(x)$, for all $x\in X$.  Next, we define the set $\Gamma$ that encodes the second property in condition \eqref{e:moment}:
\[
\Gamma\coloneqq \left\{(x,\,\omega)\in S^\star\times \Omega:\,  p^\star(\omega)= v(x)+\nabla v(x) \cdot (\omega-x)\right\}.
\]
The set $\Gamma$ is called the contact set in the linear programming literature. In light of Theorem \ref{t:moment} and Remark \ref{remark1}, a feasible $\pi\in \Pi(\mu_0)$ is optimal if and only if $\supp(\pi)\subset \Gamma$. Finally, we define the $x$-section of $\Gamma$,
\[
\Gamma_x\coloneqq\{\omega\in \Omega:\,(x,\,\omega)\in \Gamma\}.
\]
Intuitively,  the set $\Gamma_x$ contains states that can appear together with $x$ in the support of an optimal solution---states in $\Gamma_x$ (and only these states) can be pooled into the moment $x$. Geometrically, $\Gamma_x$ is the projection of the face of the epigraph of $p^\star$ exposed by the direction $(-1,\nabla v(x))$ on the state space, $\Gamma_x=\arg\max_{\omega\in \Omega}\{\nabla v(x)\cdot \omega-p^\star (\omega)\}$. A more intuitive statement of this property is that states can be pooled (in an arbitrary way as long as the induced posterior moments belong to $S^\star$) within regions where the price function is affine; at the same time, the optimal solution cannot pool together states that do not belong to a region on which $p^\star$ is affine. We can thus think of $\Gamma_x$ as the ``pooling region" of moment~$x$.

The sets $\Gamma_x$ can intersect in general. If $\omega\in \Gamma_x\cap \Gamma_y$, then $\omega$ could appear in the support of $\pi$ both conditional on $x$ and conditional on $y$---this is possible when the signal is random conditional on $\omega$. However, an important consequence of the above geometric characterization is that each $\Gamma_x$ is convex, and that $\relint(\Gamma_x)\cap \relint(\Gamma_y) \neq \emptyset $ implies  $\Gamma_x=\Gamma_y$, where $\relint(\cdot)$ stands for the relative interior of a set. Thus, the set $\Gamma$ generates a partition of $\Omega$ consisting of relatively open convex components $\{\relint(\Gamma_x)\}_{x\in S^\star}$ and the set of points on the boundaries of these components: $X\setminus \bigcup_{x\in S^\star}\relint(\Gamma_x)$. If $x\neq y$ implies that $\Gamma_x\neq \Gamma_y$, then $\pi$ has a very simple structure: For any $x\in S^\star$, states in $\relint(\Gamma_x)$ are pooled together into the posterior mean $x$.

This is where the conditions of Theorem \ref{t:Brenier2} come in. When the conditions on $v$ hold, it is indeed true that $x\neq y$ implies that $\Gamma_x\neq \Gamma_y$. When $\mu_0$ has a continuous distribution, we can ignore the measure-zero set of states on the boundaries of the convex elements of the partition. Thus, a convex-partitional signal is optimal. Moreover, the optimal  $\chi:X\rightarrow X$ is uniquely determined, for $\mu_0$-almost all $\omega\in \Omega$, by 
\[\chi(\omega)=\{x\in S^\star:\,\omega\in \Gamma_x\}=\{x\in S^\star: \nabla p^\star(\omega)=\nabla v(x)\}.\] 
We illustrate this discussion with an application in the next section.
 
\subsubsection{Beyond convex-partitional signals}\label{sec_beyond}

In this subsection, we turn attention to the structure of solutions when the conditions of Theorem \ref{t:Brenier2} fail. In the one-dimensional case, the bi-pooling result of \cite{ABSY} and \cite{KMS} shows that even if no optimal signal is monotone-partitional, there still exist optimal signals with a relatively simple structure.  Namely, the state space is partitioned into intervals, and conditional on any interval, an additional binary signal may be sent.  We will derive a multi-dimensional version of this result. Our generalization is a direct consequence of duality, while \cite{ABSY} and \cite{KMS} rely on an extreme-point characterization of optimal signals.
 
For a set $A\subset X$, let $\cl(A)$ denote the closure of $A$, and $\ext(A)$ denote the set of extreme points of the closed convex hull of $A$. Fixing a solution $\pi$ to \eqref{primal_*}, %
let 
\[
S_x:=\cl(\supp(\pi_X)\cap \relint(\Gamma_x)),
\]
for any $x\in \supp(\pi_X)$. Recall that $\Gamma_x$ is the set of states that can be pooled into the posterior moment $x$ by an optimal signal. Thus, conditional on $x$ being the realized posterior moment under some optimal signal $\pi$, the set $S_x$ contains all posterior moments in the support of $\pi_X$ that could be generated by an optimal signal. For example, if the conditions of Theorem \ref{t:Brenier2} hold, then the (unique) optimal signal $\pi$ satisfies $S_x=\{x\}$ for almost all $x\in\supp(\pi_X)$. This means that any state in the support of the optimal signal conditional on $x$ must be pooled into $x$; thus, the optimal signal is deterministic (and convex-partitional  since each $\Gamma_x$ is convex). The bi-pooling result of \cite{ABSY} and \cite{KMS} in the one-dimensional case can be reformulated as stating that  there exists an optimal solution such that $S_x$ has at most two elements. That is, for any realized posterior moment $x$, there exists at most one other posterior moment $y\in \supp(\pi_X)$ such that $\Gamma_x=\Gamma_y$. In this case, states in the interval $\Gamma_x$ can be pooled into either $x$ or $y$, and we have $S_x=S_y=\{x,\,y\}$. The following result extends that conclusion to the multi-dimensional case. 

\begin{theorem}\label{t:KMS}
Suppose that $v$ is continuously differentiable and that $\mu_0$ has a density on $\Omega$ with respect to the Lebesgue measure. There exists an optimal solution $\pi\in \Pi(\mu_0)$ to \eqref{primal_*} such that $
S_x=\ext(S_x)$ for $\pi_X$-almost all $x$.
\end{theorem}

\begin{proof}
The proof is relegated to Appendix \ref{app_proofs}.
\end{proof}

The conclusion $S_x=\ext(S_x)$ means that no posterior mean in $S_x$ can be expressed as a convex combination of other posterior means in $S_x$. This generalizes the bi-pooling result of \cite{ABSY} and \cite{KMS} because in the one-dimensional case, for any set $S\subset \mathbb{R}$, $|\ext(S)|\leq 2$. In higher dimensions, Theorem \ref{t:KMS} guarantees that we can divide the state space into convex ``pooling regions" (up to a measure-zero set) and find an optimal signal that only pools states inside pooling regions; moreover, the posterior moments induced from a given pooling region form a set that only consists of extreme points (of its own convex hull). 

The proof of Theorem \ref{t:KMS} relies on the fact that $\supp(\pi)\subset \Gamma$ is both necessary and sufficient for the optimality of $\pi \in \Pi(\mu_0)$. As shown in Section \ref{s:cp}, $\Gamma$ defines (up to a measure zero set) a convex partition of the state space, with a representative element $\Gamma_x$, which could in general coincide with $\Gamma_y$ for $y\neq x$. That is, optimality of a signal requires  that states in $\Gamma_x$ are mapped only into posterior moments $y$ for which $\Gamma_y=\Gamma_x$. We can modify the solution on $\Gamma_x$ and it will remain optimal as long as we preserve the above property. Formally---to deal with the fact that sets $\Gamma_x$ may have measure zero---we introduce an auxiliary optimization problem in which we minimize the average norm of induced posterior moments subject to maintaining the condition $\supp(\pi)\subset\Gamma$. The auxiliary problem then picks an optimal solution in which $
S_x=\ext(S_x)$ must be satisfied, as otherwise the value of the auxiliary problem could be lowered by shifting probability mass towards some posterior mean $y\in S_x$ that can be expressed as a convex combination of other posterior means in $S_x$.

For one-dimensional problems, the geometric property $S_x=\ext(S_x)$ implies the cardinality restriction $|S_x|\leq 2$. This is no longer the case when the dimension $N$ of the state space is two or more. In fact, one can construct an example in which $S_x$ is infinite for any choice of optimal $\pi$.\footnote{We provide one such example in Appendix B.6.} The example implies that our result is tight if one works with the partition of the state space defined by the price function through the contact set $\Gamma$, as is implicitly assumed in our definition of $S_x$. However, that partition may sometimes be unnecessarily coarse; intuitively, the price function may be affine over a region that could be further subdivided into smaller ``pooling regions" (sets of states that are only pooled with one another but not with states from other pooling regions). \cite{obloj} and \cite{DT} show how to define the finest partition into pooling regions relying directly on the distribution of posterior moments.\footnote{In the one-dimensional case, their construction can be understood through the integral characterization of mean-preserving spreads: An element of a partition (in this case, an interval) is pinned down by two consecutive points at which the integral constraint binds. In the multi-dimensional case, the construction is significantly more complicated since there exists no convenient representation of mean-preserving spreads.} If one defines an analog of $S_x$ for the finest partition (by replacing $\Gamma_x$ in the definition of $S_x$ by the element of the finest partition containing $x$), then it might be possible to tighten the conclusion of Theorem \ref{t:KMS}, perhaps by showing that there are at most $N+1$ posterior means induced from every pooling region (as is loosely suggested by Carath\'{e}odory's Theorem). Since duality does not seem immediately useful in pursuing this direction, we leave it for future research. 
 
\section{Application: Quadratic Objective}\label{sec:application}
 
In this section, we show how our duality approach developed in the preceding section can be used to solve a class of persuasion problems in which $\mu_0$ has a density on $\Omega$ that is a compact convex set in $\R^2$ with non-empty interior (so that $\Omega=X$), the objective function depends on a pair of moments $x=(x_1,\,x_2)$, and $v(x)$ is a quadratic form: $v(x)=x \Lambda x^T$. %

Variants of this model received considerable attention in the literature. The case $v(x)=x_1 x_2$ is equivalent to the model of \cite{rayo2010optimal}, who analyzed it under the assumption that $\Omega$ is a finite set. \cite{NP} studied this problem under the assumption that $\Omega$ is a strictly convex curve. These two papers mostly focus on deriving necessary conditions for optimality.\footnote{\cite{rayo2013} and \cite{onuchic} restrict attention to monotone partitional signals in the setting of \cite{NP}.}  
\cite{tamura} considers the case where $v$ is a general quadratic form in $\R^N$ but imposes strong symmetry assumptions on the prior distribution. 
\cite{KX} consider a problem (inspired by the insider trading model of \citealp{rochetvila}) that turns out to be mathematically equivalent to a generalized version of our problem where the assumption $\Omega=X$ is not imposed---their analysis is limited in its economic predictions since their methods are designed to handle even fairly pathological distributions of the state.  Our marginal contribution is to provide a tighter characterization of optimal solutions for the well-behaved case when $\Omega$ is a compact convex set (that is, when $\Omega=X$). Relative to \cite{rayo2010optimal} and \cite{NP}, we show that a set of necessary conditions taken from these two papers become jointly sufficient for optimality in our case. Prior to the current version of this paper, \cite{malamud} provided an alternative (less explicit) characterization of solutions under weaker assumptions.

We first argue that the case of a general quadratic form can easily be reduced to the special case $v(x)=x_1 x_2$. Indeed, for any quadratic form, there exists a basis such that the quadratic form is diagonal: 
$v(x)=\lambda_1 x_1^2+\lambda_2x_2^2$. If $\lambda_1,\lambda_2\geq 0$ (respectively, $\lambda_1,\lambda_2\leq 0$), then full disclosure (respectively, no disclosure) is optimal. If $\lambda_1$ and $\lambda_2$ have opposite signs, then there exists yet another basis such that $v(x)=x_1x_2$, which we assume henceforth.

It is known from \cite{rayo2010optimal} that the posterior means induced by an optimal signal must belong to a monotone set. Using duality, we can establish a stronger claim. Formally, we will say that a set $S\subset X$ is 
\begin{itemize}
\item {\it monotone} if $(x_1- y_1)(x_2- y_2)\geq 0,\quad \text{for all } x,y\in S$;
\item {\it maximal monotone} in $X$ if it is monotone, and for each $y\in X\setminus S$, there exists $x\in S$ such that $(x_1-y_1)(x_2-y_2)<0$.
\item {\it  almost-maximal monotone} in $X$ if it is  monotone, compact, and, for each $y\in X\setminus S$, there exists $x \in S$ such that $(x_1-y_1)(x_2-y_2)\leq 0$.
\end{itemize}
Intuitively, a monotone set $S$ in $\mathbb{R}^2$ has the property that if $x\in S$, then $S$ cannot intersect the interiors of either the upper-left or the lower-right quadrants centered at $x$. A monotone set is maximal in $X$ if it is not a proper subset of any monotone set in  $X$. A maximal monotone set must be compact (when $X$ is compact, as assumed). An almost-maximal monotone set $S$ is a compact subset of a maximal monotone set $S'$ such that $S'\setminus S$ is a collection of open line segments that are either  horizontal or vertical.

\begin{prop}\label{prop:KX}
If $\pi^\star \in \Pi(\mu_0)$ is optimal, then the support of moments $\supp (\piX^\star)$ induced by  $\pi^\star $ is an almost-maximal monotone set in $X$.
\end{prop}

\begin{proof}
Suppose that $\pi^\star \in \Pi(\mu_0)$ is optimal. To simplify notation, let $p^\star:=p_{\supp (\piX^\star)}$ as defined by \eqref{e:prices}. By Remark \ref{remark1}, $p^\star\geq v$ and $\supp (\piX^\star)\subset S^\star$, where $S^\star=\{x\in X:\, p^\star(x)=v(x)\}$; moreover, $p^\star=p_{S^\star}$, and hence, since $\Omega=X$ and $v(x)=x_1x_2$, we have, for all $x\in X$, 
\[
p^\star(x)=\max_{y\in S^\star}\{x_1 y_2+x_2 y_1-y_1y_2\}.
\]
We claim that the set $S^\star$ is monotone: Otherwise, we would have $x,y\in S^\star$ such that $(x_1-y_1)(x_2-y_2)<0$, but then
\[
p^\star(x)\geq x_1x_2-(x_1-y_1)(x_2-y_2)>x_1x_2=v(x),
\]
contradicting that $x\in S^\star$. Next, we claim that the set $S^\star$ is maximal monotone in $X$. Otherwise, there would exist $x\in X\setminus S^\star$ such that $(x_1-y_1)(x_2-y_2)\geq 0$ for all $y\in S^\star$, and thus
\[
p^\star(x)=\max_{y\in S^\star}\{x_1x_2-(x_1-y_1)(x_1-y_2)\}\leq x_1x_2=v(x).
\]
But then, since $p^\star\geq v$, we would have that $p^\star(x)=v(x)$, contradicting that $x\notin S^\star$. 

Since $\supp (\piX^\star)\subset S^\star$, and we have shown that $S^\star$ is a monotone set, $\supp (\piX^\star)$ is also a monotone set. Finally, we claim that $\supp (\piX^\star)$ is almost-maximal monotone in $X$. Otherwise, there would exist $x\in X$ such that $(y_1-x_1)(y_2-x_2)> 0$ for all $y\in \supp (\piX^\star)$, which implies that (since $\supp (\piX^\star)$ is compact)
\[
p^\star(x)=\max_{y\in \supp (\piX^\star)}\{x_1x_2-(x_1-y_1)(x_1-y_2)\}<x_1x_2=v(x),
\]
contradicting that $p^\star\geq v$.
\end{proof}

\begin{figure}
\centering \includegraphics[scale=0.4]{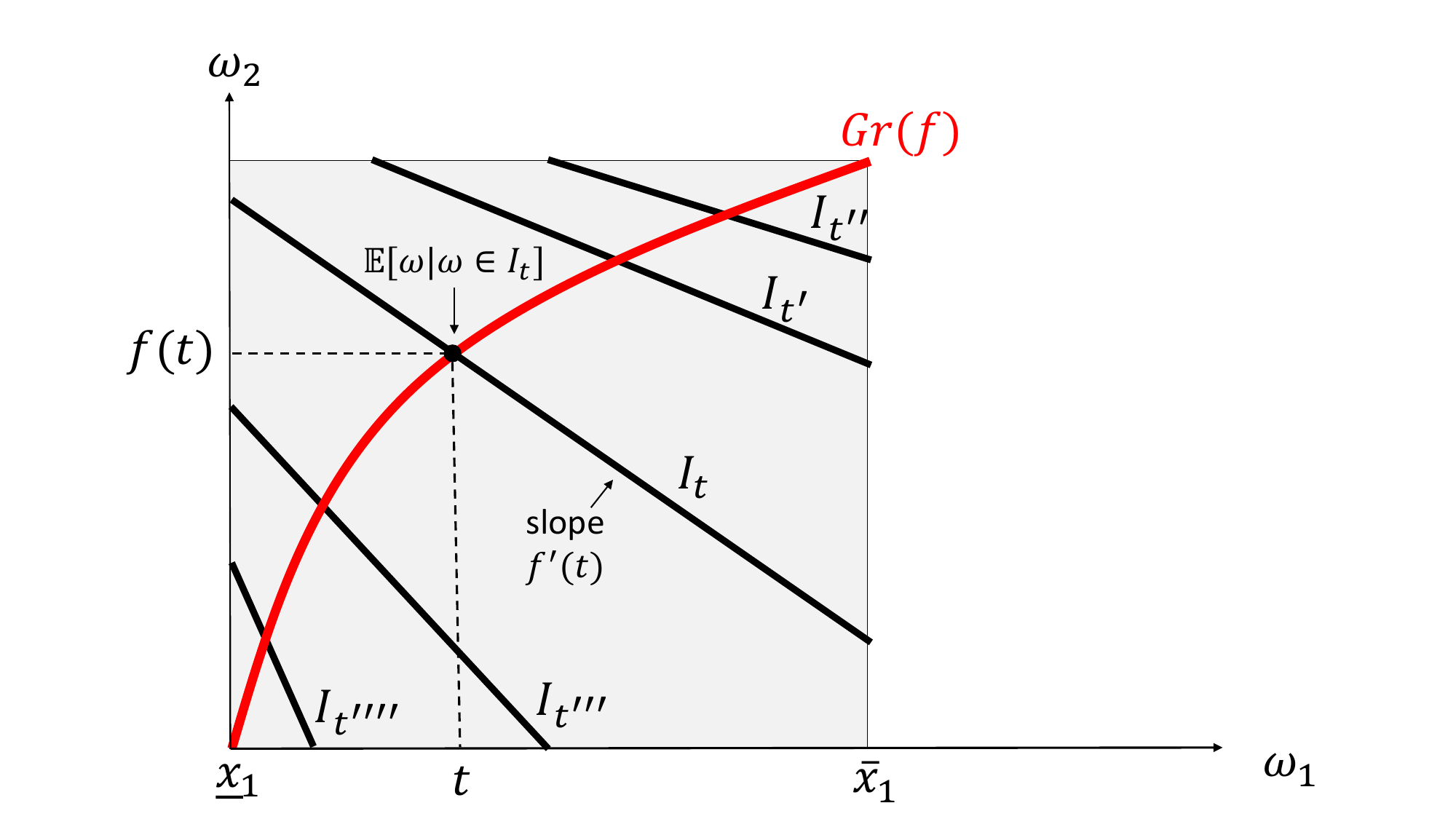}
\caption{Illustration of Proposition \ref{t:RSsmooth}: The optimal signal pools all states in each of the negatively sloped intervals $I_t$, and the resulting posterior means belong to $\Gr(f)$.}\label{fig2}
\end{figure}

In light of Remark \ref{remark1}, the proof of Proposition \ref{prop:KX} implies that the optimal price function can always be derived from some candidate support $S$ of the distribution of moments that is a maximal monotone set. A natural class of maximal monotone sets in $X$ are graphs of continuous increasing functions. %
The main result of this section describes necessary and sufficient conditions for the optimality of a solution $\pi^\star\in \Pi(\mu_0)$ with  $\supp (\piX^\star)$ equal to the graph $\Gr (f)$ of a given well-behaved function $f$. By Theorem \ref{t:Brenier2},  the unique solution is convex-partitional; the optimal partition divides $\Omega$ into negatively-sloped line segments; a line segment that induces the posterior mean $(t,\,f(t))$ has slope $-f^\prime(t)$, as illustrated in Figure \ref{fig2}. These observations are formalized in the following proposition. 

\begin{prop}\label{t:RSsmooth}
Let $f:\,[\ul x_1,\ol x_1]\rightarrow \mathbb{R}$ be a twice continuously differentiable function, with $f^\prime(t)>0$ for all $t\in [\ul x_1,\ol x_1]$, such that the graph $\Gr(f)$ is a maximal monotone subset of\,$X$. An optimal $\pi^\star\in \Pi(\mu_0)$ induces a support of moments $\supp (\piX^\star)$ equal to $\Gr(f)$ if and only if $\Omega$ can be partitioned, up to a measure zero set,\footnote{That is, $\Omega\setminus\left\{\bigcup_{t\in [\ul x_1,\ol x_1]}I_t\right\}$ has zero (Lebesgue) measure.} into a collection of disjoint open line segments $\{I_t\}_{t\in [\ul x_1,\ol x_1]}$ such that
\begin{enumerate}
\item $\mathbb{E}[\omega|\,\omega\in I_t]=(t,\,f(t))$, for almost all $t\in [\ul x_1,\ol x_1]$;\footnote{Since $I_t$ has zero measure under the prior, $\mathbb{E}[\omega|\,\omega\in I_t]$ is formally defined almost everywhere via the conditional expectation of $\omega$ conditional on a $\sigma-$algebra generated by  $\{I_t\}_{t\in [\ul x_1,\ol x_1]}$. We provide an explicit formula for the conditional expectation in Appendix B.7.}
\item $I_t=\relint\,(\{\omega\in \Omega:\,t\in \underset{s\in [\ul x_1,\ol x_1]}{\argmax}\,\{\omega_1 f(s)+\omega_2 s-sf(s)\}\})$, for all $t\in [\ul x_1,\ol x_1]$.
\end{enumerate}
Whenever the above conditions hold, the optimal signal is convex-partitional and pools the states within each $I_t$; moreover, $I_t\subseteq \{\omega\in \Omega:\,\omega_2=f(t)-f^\prime(t)(\omega_1-t)\}$, for all $t\in [\ul x_1,\ol x_1]$. 
\end{prop}

\begin{proof} 

We will prove that existence of the required partition of $\Omega$ is sufficient for optimality of the corresponding $\pi^\star$.  We relegate the more technical proof of necessity to Appendix\,\ref{proof_t:RSsmooth}.

Suppose that there exists a collection of line segments $\{I_t\}_{t\in [\ul x_1,\ol x_1]}$ satisfying properties \textit{1}-\textit{2}. We can define $\pi^\star\in \Pi(\mu_0)$ as the convex-partitional signal that pools states in each $I_t$ (it is irrelevant how the signal is defined for $\omega\in\Omega$ not belonging to any $I_t$). By the first property, the induced posterior-mean curve $\supp (\piX^\star)$ is equal to $\text{Gr}(f)$. Following Section \ref{s:diff}, define the price function
\[
p_{\text{Gr}(f)}(x)=\max_{y\in  \text{Gr}(f)}\{v(y)+\nabla v(y)\cdot(x-y)\}=\max_{t\in  [\ul x_1,\ol x_1]}\{x_1 f(t)+x_2 t-tf(t)\}.
\]
We will verify that condition \eqref{e:moment} holds; optimality of $\pi^\star$ will then follow from Theorem \ref{t:moment}.
First, we argue that  $p_{\text{Gr}(f)}(x)\geq v(x)$, for all $x\in X.$ It suffices to show that there exists a $t\in  [\ul x_1,\ol x_1]$ such that $x_1 f(t)+x_2 t-tf(t)\geq  x_1x_2$, or, equivalently, $(t-x_1) (f(t)-x_2)  \leq 0$. The claim is obvious when $x\in \text{Gr}(f)$, and follows from the fact that $\text{Gr}(f)$ is maximal monotone in $X$ when $x\in X\setminus\text{Gr}(f)$.  To complete the proof that \eqref{e:moment} holds, note that, by the second property,  for almost all $\omega \in I_t$, 
\[
p_{\text{Gr}(f)}(\omega)=v(x(t))+\nabla v(x(t))\cdot(\omega-x(t))=\omega_1 f(t)+\omega_2 t-tf(t).
\]
This shows that the equality in \eqref{e:moment} holds for all $(x,\,\omega)\in \bigcup_{t\in [\ul x_1,\ol x_1]}\left((t,f(t))\times I_t\right)$; by continuity, the equality extends to the closure of this set, which is $\supp (\pi^\star)$.  

Finally, the inclusion $I_t\subseteq \{\omega\in \Omega:\,\omega_2=f(t)-f^\prime(t)(\omega_1-t)\}$, for $t\in (\ul x_1,\ol x_1)$, follows from the observation that, by the second property, the first-order condition $(\omega_1-t) f^\prime(t)+\omega_2-f(t)=0$ must hold for all $\omega\in I_t$.\footnote{This observation shows that it would suffice to require $\mathbb{E}[\omega_1|\,\omega\in I_t]=t$ in the first property in Proposition \ref{prop:KX}. Indeed, $(\omega_1-t) f^\prime(t)+\omega_2-f(t)=0$ for all $t\in (\ul x_1,\ol x_1)$ and $\omega\in I_t$ implies that $(\mathbb{E}[\omega_1|\,\omega\in I_t]-t) f^\prime(t)+\mathbb{E}[\omega_2|\,\omega\in I_t]-f(t)=0$, from which the second required equality $\mathbb{E}[\omega_2|\,\omega\in I_t]=f(t)$ follows. \label{f:Ew1}} For $t\in \{\ul x_1,\ol x_1\}$, the proof of the inclusion is more complicated, and thus relegated to Appendix \ref{proof_t:RSsmooth}.\end{proof}

Proposition \ref{t:RSsmooth} provides a tight characterization of optimal signals under the additional regularity requirement that the induced posterior mean curve is sufficiently regular (a graph of a twice differentiable function). If an optimal signal $\pi^\star$ induces $\supp (\piX^\star)=\Gr(f)$, then it must have a simple convex-partitional structure in which only states belonging to negatively-sloped line segments $I_t$ are pooled together. Moreover, the slopes of these line segments are uniquely pinned down by $f$. The full proof in Appendix \ref{proof_t:RSsmooth} additionally reveals that the closures of these line segments can only intersect at the endpoints; the endpoints can be found by solving the optimization problem in the second property in Proposition \ref{t:RSsmooth}.

As an illustration, we provide conditions under which it is optimal to reveal only some linear combination of $\omega_1$ and $\omega_2$. A simple implication of this characterization is that it is optimal to reveal $\omega_1+\omega_2$ if the prior is symmetric around the line $\omega_2=\omega_1$.

\begin{prop} \label{thm_case} 
The joint distribution $\pi\in \Pi(\mu_0)$ induced by the disclosure of the realization of $a\omega_1 +\omega_2$, with $a>0$, is optimal if and only if $\supp(\pi_X)\subset \{(t,\,a t+b):\,t\in \mathbb{R}\}$, with $b\in \R $.
\end{prop}
\begin{proof}
\emph{If.}  Let $\pi\in\Pi(\mu_0)$ be induced by disclosure of the realization of $a \omega_1+\omega_2$, and suppose that $\supp(\pi_X)\subset \{(t,\,a t+b):\,t\in \mathbb{R}\}$. Note that $\pi$ partitions $\Omega$ into parallel open line segments $\{I_t\}_{t\in [\ul x_1,\ol x_1]}$, where $I_t=\relint\left(\{\omega\in\Omega:\,a \omega_1+\omega_2=2at+b\}\right)$, and the range $[\ul x_1,\ol x_1]$ is defined by the property that $(t,at+b)\in \Omega$. Since $\supp(\pi_X)\subset \{(t,\,a t+b):\,t\in \mathbb{R}\}$, the induced posterior mean curve is a line segment with slope $a$ that is a monotone maximal set in $\Omega$. Finally, the second property in Proposition \ref{t:RSsmooth} holds since 
\[
\{\omega\in \Omega:\,t\in \underset{s\in [\ul x_1,\ol x_1]}{\argmax}\,\{\omega_1 (as+b)+\omega_2 s-s(as+b)\}\}=\{\omega\in \Omega:\,a\omega_1+\omega_2=2at+b\},
\]
which is precisely our definition of $I_t$. Thus, Proposition \ref{t:RSsmooth} shows that $\pi$ is optimal.

\emph{Only if.} Here we prove the necessity part under a regularity condition that the support of $\pi_X$ corresponding to disclosure of the realization of $a\omega_1+\omega_2$ is a twice continuously differentiable function $f$ with $f'>0$; we relegate the complete proof (without any regularity condition) to Appendix \ref{a:linear f}. If disclosing $a \omega_1+\omega_2$ is optimal, then the open line segments $I_t$ partitioninig $\Omega$ (whose existence is guaranteed by Proposition \ref{t:RSsmooth} under the regularity condition) must be parallel and have slope $-a$. But then, we must have that $\omega_2=f(t)-f^\prime(t)(\omega_1-t)$ if and only if $\omega_2=2at+b-a\omega_1$, for some $b$, which is only possible when $f(t)=at+b$.
\end{proof}

Proposition \ref{thm_case} showcases two ways in which Proposition \ref{t:RSsmooth} can be used. First, it can be applied to verify the optimality of a conjectured posterior mean curve. Once a posterior mean curve is fixed, Proposition \ref{t:RSsmooth} allows us to construct the unique candidate solution, and then check whether it is indeed optimal. Second, Proposition \ref{t:RSsmooth} provides a way to construct the optimal signal. Suppose that we partition $\Omega$ (up to a measure-zero set) into negatively-sloped open line
segments in such a way that pooling the states within these line segments induces a posterior
mean curve that is a graph of some continuous function $f$. Then, this signal is optimal as long as the second property holds. Moreover, if $f$ is differentiable and the closures of these
line segments are disjoint, then it suffices to verify that the slope of the line segment inducing posterior mean $(t,\,f(t))$ is $-f^\prime(t)$.

Finally, we offer some intuition for our results. We can rewrite the objective function as 
\[
v(\omega)=\omega_1\omega_2=\frac{1}{a}\left[\left(\frac{a\omega_1+\omega_2}{2}\right)^2-\left(\frac{a\omega_1-\omega_2}{2}\right)^2\right].
\]
Thus, intuitively, the objective is to disclose as much information as possible about $a\omega_1+\omega_2$ while disclosing as little as possible about $a\omega_1-\omega_2$. Typically, $a\omega_1+\omega_2$ and $a\omega_1-\omega_2$ will be correlated, leading to a trade-off.  However, when $\mathbb{E}[a\omega_1-\omega_2|a\omega_1+\omega_2]=\mathbb{E}[a\omega_1-\omega_2]$, (so that disclosing $a\omega_1+\omega_2$ does not change the expectation of $a\omega_1-\omega_2$), it becomes optimal to disclose $a\omega_1+\omega_2$. The condition $\supp(\pi_X)\subset \{(t,\,a t+b):\,t\in \mathbb{R}\}$  states precisely that  $\mathbb{E}[\omega_2|a\omega_1+\omega_2]=a\mathbb{E}[\omega_1|a\omega_1+\omega_2]+b$.  Proposition \ref{thm_case} shows that this intuitive condition is not only sufficient but also necessary for the optimality of disclosing $a\omega_1+\omega_2$.  Note that no correlation between $a\omega_1-\omega_2$ and $a\omega_1+\omega_2$ requires that $a=\text{sd}(\omega_2)/\text{sd}(\omega_1)$ (where $\text{sd}$ stands for standard deviation) implying that the optimal weight equalizes the contribution of the two states to the variability of the signal. The general case, covered by Proposition \ref{t:RSsmooth}, can be understood as setting the weight $a$ locally, as captured by the condition that the slope of $I_t$ must be equal to $-f^\prime(t)$.

\section{Concluding remarks}\label{s:concl}

We conclude with a few remarks on possible extensions and connections to other problems.

\paragraph{Potential applications.} 
Several other potential applications of persuasion duality are worth mentioning. \cite{GP2} show that duality can be used to quantify the value of ``data records;" our results could thus be helpful in calculating that value. \cite{BD}, \cite{yang2022}, and \cite{KW} characterize the set of feasible distributions of posterior \textit{quantiles}; it might be interesting to study the consequences of general duality for the special case of ``quantile persuasion"---paralleling the developments for moment persuasion.   Finally, a large literature on rational inattention and costly-information acquisition studies optimization problems in which a linear objective is maximized over distributions of posterior beliefs subject to Bayes-plausibility. Our analysis applies under the assumption that the cost of information satisfies posterior-separability (see, among many others, \citealp{kaplin, caplin2015}, \citealp{gentzkow2014}, and \citealp{denti}).

\paragraph{Additional constraints in the persuasion problem.}\mbox{}  \cite{doval2018}, inspired by an earlier contribution of \cite{letreust}, observe that many persuasion problems feature additional linear constraints (such as moral-hazard, inventive-compatibility, or capacity constraints) that modify the structure of optimal persuasion schemes. Our general duality approach easily accommodates  a finite number $M$ of additional linear constraints: In this case, there are $M$ new prices that enter the objective function in the dual problem \eqref{dual} (see an earlier version of the paper \citealp{wp}, for details). 

Such an extension could be useful in analyzing problems with a privately informed Receiver (see, among others, \citealp{KMZL} and \citealp{guointerval}). \cite{candogan} point out that the one-dimensional moment persuasion problem with a privately informed Receiver reduces to the standard one-dimensional moment persuasion problem with additional linear constraints.	 It would be interesting to see if duality could be fruitfully applied to such a representation of the informed-Receiver problem.

\paragraph{Belief-based versus recommendation-based approach.} We have formulated the persuasion problem in terms of distributions of posterior beliefs. An alternative approach is to explicitly introduce a Sender and a Receiver, and maximize the Sender's utility from the realized state and the Receiver's action over joint distributions of states and recommendations, subject to Bayes-plausibility and an obedience constraints for the Receiver. 

We first note that none of these two approaches is more general---it is in fact possible to reformulate the belief-based  problem using the recommendation-based approach, and vice versa.  To illustrate this point suppose that $\Omega$ is a finite set. Consider a  problem in which the Sender's and Receiver's utility functions are $w(a,\,\omega)$ and $u(a,\,\omega)$, respectively, where $a$ is the action of the Receiver. \cite{kamenica2011} show that this problem can be analyzed through the belief-based approach by defining $V(\mu)=\E_\mu [w(a^\star(\mu),\, \omega)]$, where $a^\star(\mu)\in \argmax_{a\in A} \E_\mu [u(a,\,\omega)]$.
Conversely, the problem we introduced in Section \ref{sec:model} is equivalent to a problem in which the action space is $A=\Delta(\Omega)$, the Sender's  utility is given by $w(a,\,\omega)=V(a)$, and the Receiver's utility is $u(a,\,\omega)=2a(\omega)-\sum_{\omega'\in \Omega}a^2(\omega')$. Indeed, given a posterior $\mu$, the Receiver takes an action $a^\star(\mu)=\mu$, which maximizes his expected utility $\sum_{\omega\in \Omega} (2a(\omega)\mu(\omega) - a^2(\omega))$, and thus the objective function is $V(\mu)$.

In the context of moment persuasion, the two approaches are unified by  Theorem~\ref{thm.moment} through the lens of duality---this is because  the martingale constraint in the definition of the feasible set $\Pi(\mu_0)$ can be regarded as an obedience constraint for a Receiver with quadratic preferences who matches the action to the state (see \citealp{Kolotilin2017}).
It is interesting to ask whether duality analysis could similarly cast light on the relationship between the two approaches in more general contexts, such as a multi-dimensional version of the non-linear persuasion problem considered by \cite{KCW}.

\paragraph{Multiple Receivers.} Perhaps the biggest limitation of our setting is that it does not cover the case in which a Sender wishes to communicate privately with multiple interacting Receivers. Of course, 
our results do apply when the Sender is restricted to public signals, as in \cite{pavan}. Moreover, our duality approach could be useful in analyzing private persuasion problems in conjunction with existing results. \cite{mathevet} show how to adapt the belief-based approach to persuasion in games, by decomposing a general signal into its public and (purely) private part. Our results apply to the optimal design of the public part of the signal. Additionally, in a recent contribution, \cite{arieli22} apply transportation duality to cast light on the optimal design of the purely private signal---it is natural to ask whether our  approach and theirs could be unified. Duality may also be useful within the recommendation-based approach to information design in games (introduced by \citealp{BM2}, and \citealp{taneva}). \cite{GP} obtain strong duality under finite action and state spaces, while \cite{smolin} rely on weak duality in their analysis  of ``concave games." Obtaining conditions for strong duality in a general environment remains an open problem.

\setstretch{0.85}
\bibliographystyle{ecta}
\bibliography{bibliography}

\onehalfspace
\appendix

\section{Appendix: Proofs}

We will prove the results in Section \ref{sec:duality} in a different order than they appear in Section \ref{sec:duality}. We first deal with weak duality and primal attainment, as their proofs are standard. We then prove Theorem \ref{t:Lip}. Finally, relying on Theorem \ref{t:Lip}, we prove Theorem \ref{t:duality} and Theorem~\ref{thm_strong}.

\subsection{Proof of Theorem \ref{weak} and primal attainment}\label{app:weak_duality}

We first prove Theorem \ref{weak}. By the definition of the Lebesgue integral, $\tau$ belongs to $\Tau(\mu_0)$ if and only if for any measurable and bounded $p:\Omega\to\mathbb{R}$, 
\[
\int_{\Delta (\Omega)}\int_{\Omega} p(\omega) \df\mu(\omega) \df \tau (\mu) = \int_\Omega p(\omega) \df\mu_0(\omega).
\]
Thus, for any $\tau\in \Tau (\mu_0)$ and any such $p$ that additionally satisfies $V(\mu)\leq \int_\Omega p(\omega)\df \mu(\omega)$ for all $\mu\in \Delta(\Omega)$, we have  
\[
\int_{\Delta (\Omega)} V(\mu) \df \tau (\mu) \leq \int_{\Delta (\Omega)}\int_\Omega p(\omega)\df \mu(\omega) \df \tau (\mu)  = \int_\Omega p(\omega) \df\mu_0(\omega).
\]
Taking the supremum over $\Tau (\mu_0)$ on the left-hand side and the infimum over $\mathcal P (V)$ on the right-hand side (any $p\in \mathcal P(V)$ is measurable and bounded) yields the desired result.

Next, we prove primal attainment under the weaker assumption that $V$ is bounded only from above, because this stronger version will be used in the proof of Theorem \ref{t:KMS}. 

\begin{lemma}\label{l:popt}
Let $V:\Delta(\Omega)\rightarrow \R \cup \{-\infty\}$ be bounded from above and upper semi-continuous. Then \eqref{primal} has an optimal solution.
\end{lemma}
\begin{proof}
Because the function $\tau\to\int_{\Delta(\Omega)}\mu \df\tau(\mu)$ is continuous, the feasible set $\Tau(\mu_0)$ is compact, being a closed subset of the compact set $\Delta(\Delta(\Omega))$. Moreover, $\Tau(\mu_0)$ is  non-empty, as it contains the Dirac probability measure $\delta_{\mu_0}$ at $\mu_0$, which corresponds to no disclosure.  Since $V$ is bounded from above and upper semi-continuous, the function $\tau \rightarrow \int V(\mu)\df \tau(\mu)$ is also upper semi-continuous and thus attains its maximum on the compact set $\Tau(\mu_0)$, by the Weierstrass Theorem. Thus, an optimal solution $\tau^\star$ to the problem \eqref{primal} exists. 
\end{proof}

\subsection{Proof of Theorem \ref{t:Lip}}\label{a:Lip}
We start with a key lemma.
\begin{lemma}\label{lemma_discrete}
Let $\mu_0,\eta_0\in \Delta(\Omega)$ and $\tau\in \mathcal T(\mu_0)$. There exists a measurable function $\eta:\,\Delta(\Omega)\to \Delta(\Omega)$ such that 
\[
\int_{\Delta(\Omega)}\eta(\mu)\df\tau(\mu)=\eta_0 \quad\text{and}\quad \int_{\Delta(\Omega)} \kr{\mu-\eta(\mu) } \df \tau(\mu)=\kr{\mu_0-\eta_0}.
\]
\end{lemma}
Before proving Lemma \ref{lemma_discrete}, we show that it implies Theorem \ref{t:Lip}. First, since $V$ is Lipschitz, it is upper semi-continuous, and hence, by Lemma \ref{l:popt},  for any $\mu_0\in \Delta(\Omega)$, there exists $\tau\in \Tau(\mu_0)$ that attains the concave closure of $V$ at $\mu_0$, so that $\ccV(\mu_0)=\int_{\Delta(\Omega)} V(\mu)\df \tau(\mu).$
Next, since $V$ is Lipschitz, there exists  $L\in \R$, such that, for all $\mu_0,\eta_0\in \Delta(\Omega),$ we have $|V(\mu_0)-V(\eta_0)|\leq L\kr{\mu_0-\eta_0}.$
Then, using Lemma \ref{lemma_discrete} to define the function $\eta$, we obtain
\begin{align*}
\ccV (\mu_0) -\ccV(\eta_0) &\leq \int_{\Delta(\Omega)} V(\mu)\df \tau(\mu)- \int_{\Delta(\Omega)} V(\eta(\mu))\df \tau(\mu)  \\
&\leq \int_{\Delta(\Omega)} L\kr{\mu-\eta(\mu)}\df \tau(\mu) =L\kr{\mu_0-\eta_0}.
\end{align*}
By reversing the roles of $\mu_0$ and $\eta_0$, we conclude that $\ccV$ is Lipschitz (with constant $L$). Thus, it remains to prove Lemma \ref{lemma_discrete}.

\begin{proof}[Proof of Lemma \ref{lemma_discrete}]
 The idea behind the proof is to ``perturb" each posterior belief $\mu \in \Delta(\Omega)$ (with $\eta$ describing the perturbation function) in such a way that perturbed posteriors $\eta(\mu)$ average out to $\eta_0$, and the average distance between each posterior $\mu$ and its perturbation $\eta(\mu)$ is  the same as the distance between the ``priors" $\mu_0$ and $\eta_0$. A naive approach would be to perturb each posterior $\mu$ by the same magnitude and in the same direction $\eta_0-\mu_0$. However, this could easily take the perturbed beliefs outside the set $\Delta(\Omega)$. Thus, our construction is more complicated: We rely on a property of the Kantorovich-Rubinstein norm to find the transportation plan $\lambda\in \Delta(\Omega\times\Omega)$ that defines the distance between $\mu_0$ and $\eta_0$; we then define the perturbation by conditioning on each realized posterior belief $\mu\in \Delta(\Omega)$, and using a properly constructed conditional transportation plan.

Since every norm is convex, we have $\kr{\mu_0-\eta_0}\leq \int_{\Delta(\Omega)} \kr{\mu-\eta(\mu) } \df \tau(\mu)$ for any measurable function $\eta:\,\Delta(\Omega)\to \Delta(\Omega)$ such that $\int_{\Delta(\Omega)}\eta(\mu)\df\tau(\mu)=\eta_0$. Thus, it suffices to show the existence of a measurable function $\eta:\,\Delta(\Omega)\to \Delta(\Omega)$ satisfying the reverse inequality.

By the Kantorovich-Rubinstein theorem (Theorem 8.10.45 in \citealp{bogachev2007}), 
\[
\kr{\mu_0-\eta_0}=\min_{\lambda\in \Lambda(\mu_0,\eta_0)}\int_{\Omega\times \Omega} \rho(\omega,\omega')\df \lambda(\omega,\omega'),
\]
where $\Lambda(\mu_0,\eta_0)$ is the set of probability measures $\lambda \in \Delta(\Omega\times \Omega)$ such that
$\lambda(A\times\Omega)=\mu_0(A)$ and $\lambda (\Omega\times B)=\eta_0(B)$ for all measurable sets $A,\,B\subset \Omega$. In particular, the minimum is attained at some $\lambda\in \Lambda(\mu_0,\eta_0)$, which we fix for the remainder of the proof.

Define a probability measure $\sigma\in \Delta(\Delta(\Omega)\times \Omega)$ by $\sigma(M,\, A)=\int_{M}\mu(A) \df \tau(\mu)$ for all measurable $M\subset \Delta(\Omega)$ and $A\subset \Omega$. For all measurable $A\subset \Omega$, we have $\sigma(\Delta(\Omega),\,A)=\mu_0(A)$ because $\int_{\Delta(\Omega)} \mu \df \tau(\mu)=\mu_0$. For any probability measure on a product of two compact metric spaces we can define its conditional measures (Theorem 10.4.5 in \citealp{bogachev2007}). Since $\Omega$ and $\Delta(\Omega)$ are compact, there exists a measurable function $\omega\to \sigma(\cdot|\omega)$, from $\Omega$ into $\Delta(\Delta(\Omega))$, such that $\sigma(M,\, A)= \int_A \sigma(M| \omega)\df \mu_0(\omega)$ for all measurable $A\subset \Omega$ and $M\subset \Delta(\Omega)$. Similarly, there exists a measurable function $\omega \to \lambda(\cdot|\omega)$, from $\Omega$ into $\Delta(\Omega)$, such that $\lambda(A,\,B)= \int_A \lambda(B| \omega)\df \mu_0(\omega)$ for all measurable $A,\,B\subset \Omega$.

Define a probability measure $\zeta\in \Delta(\Omega\times\Omega\times\Delta(\Omega))$ by $\zeta (A,\,B,\,M)=\int_{A} \lambda(B|\omega)\sigma(M|\omega)\df \mu_0(\omega)$ for all measurable $A,\,B\subset \Omega$, and $M\subset \Delta(\Omega)$. For all measurable  $A,\,B\subset \Omega$, and $M\subset \Delta(\Omega)$, we have $\zeta (A,\,B,\,\Delta(\Omega))=\lambda(A,\,B)$ and $\zeta (A,\Omega,\,M)=\sigma(M,\,A)$, by construction.  Since $\Omega\times\Omega$ and $\Delta(\Omega)$ are compact, there exists a measurable function $\mu\to \zeta (\cdot|\mu)$, from $\Delta(\Omega)$ into $\Delta(\Omega\times \Omega)$, such that $\zeta (A,\, B,\, M)=\int_M \zeta (A,B|\mu)\df \tau (\mu)$ for all measurable $A,\,B\subset \Omega$, and $M\subset \Delta(\Omega)$. 

Finally, define a measurable function $\mu\to \eta(\mu)$, from $\Delta(\Omega)$ into $\Delta(\Omega)$, by $\eta(\mu)(B)=\zeta (\Omega,\,B|\mu)$ for all $\mu\in \Delta(\Omega)$ and all measurable $B\subset \Omega$.
Notice that the conditional measure $\zeta(\cdot\,,\,\cdot\,|\mu)$ on $\Omega\times\Omega$ is a feasible transportation plan between $\mu$ and $\eta(\mu)$, for $\tau$-almost all $\mu\in \Delta(\Omega)$. Indeed, $\zeta (\Omega,\,\cdot\,|\mu)=\eta(\mu)(\cdot)$ by construction, and for any measurable $A\subset \Omega$ and $M\subset \Delta(\Omega)$,
\[
\int_M\zeta(A,\,\Omega|\mu)\df \tau(\mu)=\zeta(A,\,\Omega,\,M)=\sigma(M,\,A)=\int_M\mu(A)\df \tau(\mu),
\]
establishing that $\zeta(\cdot\,,\Omega|\mu)=\mu(\cdot)$ for $\tau$-almost all $\mu\in \Delta(\Omega)$.

To show that the constructed function $\eta$ satisfies the required properties, note first that, for any measurable $B\subset \Omega$,
\[
\int_{\Delta(\Omega)} \eta(\mu)(B)\df \tau(\mu)=\int_{\Delta(\Omega)} \zeta (\Omega,\,B|\mu)\df \tau(\mu)=\zeta(\Omega,\,B,\,\Delta(\Omega))=\lambda(\Omega,\,B)=\eta_0(B), 
\]
and hence $\int_{\Delta(\Omega)} \eta(\mu)\df \tau(\mu)=\eta_0$. Moreover,
\begin{align*}
\int_{\Delta(\Omega)} \kr{\mu-\eta(\mu) } \df \tau(\mu) \leq 
\int_{\Delta(\Omega)}\left[\int_{\Omega\times \Omega} \rho(\omega,\,\omega')\df \zeta(\omega,\,\omega'|\mu)\right]\df \tau(\mu)\\
=	\int_{\Omega\times \Omega} \rho(\omega,\,\omega')\df \lambda(\omega,\,\omega')=\kr{\mu_0-\eta_0},
\end{align*}
where the inequality follows from the Kantorovich-Rubinstein theorem and the fact that the conditional measure $\zeta(\cdot\,,\,\cdot\,|\mu)$ on $\Omega\times\Omega$ is a feasible transportation plan between $\mu$ and $\eta(\mu)$, for  $\tau$-almost all $\mu \in \Delta(\Omega)$, as shown above. 
\end{proof}

\subsection{Proof of Theorem \ref{t:duality}}\label{app:weak}

Existence of an optimal solution to the primal problem follows from Lemma \ref{l:popt}.

To prove the rest of the theorem, we introduce some basic tools from convex analysis, used in the proof of the next lemma.\footnote{See Chapter 1.4 in \cite{brezis} for further details.} Let $E$ be a normed vector space and $E^\star$ its topological dual space, that is, the space of all continuous linear functions on $E$. Let $\varphi:E\rightarrow \mathbb R\cup \{+\infty\}$ be an extended-valued function that is not identically $\{+\infty\}$. The Legendre transform of $\varphi$ is the function $\varphi^\star:E^\star\rightarrow \R \cup\{+\infty\}$ given by
\[\varphi^\star(z^\star)=\sup_{z\in E}\left\{\langle z^\star,z\rangle -\varphi (z)\right\}\quad \text{for all $z^\star\in E^\star$}, \]
where $\langle\cdot,\cdot \rangle$ is the duality product between $E$ and $E^\star$.
It is easy to verify that $\varphi^\star$ is convex, lower semi-continuous, and not identically $\{+\infty\}$. %
Next, define the function $\varphi^{\star\star}:E\rightarrow \R \cup \{+\infty\}$ as the Legendre transform of $\varphi^\star$, restricted from $E^{\star\star}$ to $E$,
\[\varphi^{\star\star}(z)=\sup_{z^\star\in E^\star}\left\{\langle z^\star,z\rangle  -\varphi^\star (z^\star)\right\}\quad \text{for all $z\in E$}. \]
Clearly, $\varphi^{\star\star}$ is a convex and lower semi-continuous function satisfying $\varphi^{\star\star}(z)\leq \varphi(z)$ for all $z\in E$. The Fenchel-Moreau Theorem states that if $\varphi:E\rightarrow \R \cup \{+\infty\}$ is convex and lower semi-continuous, and not identically $\{+\infty \}$, then $\varphi^{\star\star}=\varphi$. We remark that the Fenchel-Moreau Theorem is a consequence of an appropriate hyperplane separation theorem.\footnote{Indeed, an earlier version of this paper \cite{wp} contained a proof of strong duality that directly relied on a hyperplane separation theorem.}

We prove the theorem in two steps. First, we show the conclusion for Lipschitz objective functions. Here, we rely on the (already proven) Theorem \ref{t:Lip}. Second, we use an approximation argument to extend the conclusion to all bounded and upper semi-continuous objectives. 

\begin{lemma}\label{l:Vcont}
Let $V\in \Lip(\Delta(\Omega))$. Then \eqref{eq_C} holds.
\end{lemma}
\begin{proof} 
Let $E=(M(\Omega),\kr {\cdot})$; then, as argued in the main text, $E^\star=\Lip(\Omega)$. 
Define the function $\varphi$ on  $M(\Omega)$ as
\[
\varphi(\eta)=
\begin{cases}
-\sup_{\tau\in \Tau(\eta)} \int_{\Delta(\Omega)}V(\mu)\df\tau(\mu), &\eta \in \Delta(\Omega),\\
+\infty, &\eta \notin \Delta(\Omega).
\end{cases}
\]

First, we note that $\varphi$ is convex. Indeed, let $\eta_1,\eta_2\in M(\Omega)$ and $\lambda \in (0,1).$ If $\eta_1,\eta_2\in \Delta(\Omega)$, then, by Lemma \ref{l:popt}, there exist $\tau_1\in \Tau(\eta_1)$ and $\tau_2\in \Tau (\eta_2)$ such that
\[
\varphi (\eta_1)=-\int_{\Delta(\Omega)}V(\mu)\df\tau_1(\mu)\in \R\quad \text{and}\quad  \varphi(\eta_2)=-\int_{\Delta(\Omega)}V(\mu)\df\tau_2(\mu)\in \R.
\]
By the definition of $\Tau$,
\[
\lambda \tau_1+(1-\lambda)\tau_2 \in \mathcal \Tau(\lambda \eta_1+(1-\lambda)\eta_2)
\]
and hence, by the definition of $\varphi$,
\begin{align*}
\varphi(\lambda\eta_1+(1-\lambda)\eta_2)&\leq -\int_{\Delta(\Omega)}V(\mu)\df(\lambda\tau_1+(1-\lambda)\tau_2)
= \lambda \varphi(\eta_1)+(1-\lambda)\varphi(\eta_2).
\end{align*}
If $\eta_1\notin \Delta (\Omega)$ or $\eta_2\notin \Delta(\Omega)$, then, trivially,
\[
\varphi(\lambda\eta_1+(1-\lambda)\eta_2)\leq  \lambda\varphi(\eta_1)+(1-\lambda)\varphi(\eta_2)=+\infty.
\]
Second, we note that $\varphi:M(\Omega)\rightarrow \R\cup \{+\infty\}$ is lower semi-continuous, because $\varphi$ is Lipschitz on the compact set $\Delta (\Omega)$, by Theorem \ref{t:Lip}.

Let us compute the Legendre transform of $\varphi$. For each $g\in \Lip(\Omega)$, 
\begin{align*}
	\varphi^\star(g) &=\sup_{\eta\in M(\Omega)} \left\{ \int_\Omega g(\omega)\df\eta (\omega) - \varphi(\eta) \right\}\\
	&= \sup_{\eta\in \Delta(\Omega),\tau\in \Tau(\eta)}\left\{ \int_\Omega g(\omega)\df\eta (\omega) + \int_{\Delta(\Omega)} V(\mu)\df \tau (\mu) \right\}\\
	&= \sup_{\eta\in \Delta(\Omega),\tau\in \Tau(\eta)}\left\{ \int_{\Delta(\Omega)} \left(\int_\Omega g(\omega)\df\mu (\omega) +  V(\mu)\right)\df \tau (\mu) \right\}\\
	&= \sup_{\eta\in \Delta(\Omega)}\left\{  \int_\Omega g(\omega)\df\eta (\omega) +  V(\eta) \right\},
\end{align*}
where the last equality follows from the fact that by treating $\widetilde{V}(\mu):=\int_\Omega g(\omega)\df\mu (\omega) +  V(\mu)$ as an objective function, we obtain a persuasion problem in which we choose both a prior $\eta$ and a distribution $\tau$ of posteriors, which  averages out to the prior, so it is optimal to choose a prior $\eta\in \argmax_{\mu\in \Delta(\Omega)}\widetilde V(\mu)$ and a degenerate distribution $\tau=\delta_\eta$.\footnote{This observation is also made in the proof of Theorem 2 in  \cite{dworczak2020}.}

Let us, finally, compute $\varphi^{\star\star}(\mu_0)$,
{\allowdisplaybreaks
\begin{align*}
\varphi^{\star\star} (\mu_0)&= \sup_{p \in \Lip(\Omega)}  \left \{ \int_{\Omega} p(\omega) \df\mu_0(\omega) -\varphi^\star(p)\right\}\\
&= \sup_{p \in \Lip(\Omega)}  \left \{ \int_{\Omega} p(\omega) \df\mu_0(\omega) -\sup_{\eta\in \Delta(\Omega)}\left\{  \int_\Omega p(\omega)\df\eta (\omega) +  V(\eta) \right\}\right\}\\
&= -\inf_{p \in \Lip(\Omega)}  \left \{ \int_{\Omega} p(\omega) \df\mu_0(\omega) +\sup_{\eta\in \Delta(\Omega)}\left\{    V(\eta)- \int_\Omega p(\omega)\df\eta (\omega) \right\}\right\}\\
&=- \inf_{p \in \Lip(\Omega)}  \left \{ \int_{\Omega} p(\omega) \df\mu_0(\omega):\,\sup_{\eta\in \Delta(\Omega)}\left\{    V(\eta)- \int_\Omega p(\omega)\df\eta (\omega) \right\}=0\right\}\\
&=
- \inf_{p \in \mathcal P(V)}\left \{ \int_{\Omega} p(\omega) \df\mu_0(\omega)\right\},
\end{align*}}
where the third equality follows from substituting $p$ for $-p$ as the optimization variable, and the fourth follows because, for any fixed $\eta$, adding a constant to $p$ does not change the value of the outer infimum---it is thus without loss of generality to normalize $p$ by insisting that the inner supremum is equal to $0$ (note that the inner supremum is attained and finite at each $p\in \Lip(\Omega)$).  
The Fenchel-Moreau Theorem implies that $\varphi=\varphi^{\star\star}$, so \eqref{eq_C} follows from  $\varphi(\mu_0)=\varphi^{\star\star}(\mu_0)$.
\end{proof}

\begin{lemma}\label{l:Vupp}
Let $V$ be bounded and upper semi-continuous. Then, \eqref{eq_C} holds.
\end{lemma}
\begin{proof}
This follows from a standard approximation argument, as, for example, in the proof of Theorem 1.3 in \cite{villani2003}. By Baire's Theorem (see, for example, Box 1.5 in \citealp{santambrogio2015}), there exists a non-increasing sequence of Lipschitz functions $V_k\in \Lip(\Delta(\Omega))$ converging pointwise to $V$. That is, $V_k(\mu)\geq V_{k+1} (\mu)$ for all $\mu\in \Delta(\Omega)$ and $k\in \mathbb N$, and $\lim_{k\rightarrow \infty}V_k(\mu)=V(\mu)$ for all $\mu\in \Delta(\Omega)$. Let $\tau^\star_k$ denote an optimal solution to \eqref{primal} with the objective function $V_k$. For each $k\in \mathbb N$, we have
\begin{align*}
\int_{\Delta (\Omega)} V(\mu) \df \tau^\star (\mu) &\leq \inf_{p\in \mathcal P(V)}\int_{\Omega} p (\omega) \df \mu_0 (\omega)\leq \inf_{p\in \mathcal P(V_k)}\int_{\Omega} p (\omega) \df \mu_0 (\omega)=\int_{\Delta (\Omega)} V_k(\mu) \df \tau^\star_k (\mu),
\end{align*}
where the first inequality holds by Theorem \ref{weak}, the second inequality holds by $\mathcal P(V_k)\subset\mathcal P(V)$ for $V_k\geq V$, and the equality holds by Lemma \ref{l:Vcont} for Lipschitz $V_k$. To establish \eqref{eq_C} for upper semi-continuous $V$, it is thus sufficient to show that
\[
\lim_{k\rightarrow \infty } \int_{\Delta (\Omega)} V_k(\mu) \df \tau^\star_k (\mu)\leq \int_{\Delta (\Omega)} V(\mu) \df \tau^\star (\mu).
\]
Thanks to compactness of $\Tau(\mu_0)$, up to extraction of a subsequence, we can suppose that $\tau^\star_k$ converges weakly to some $\ol \tau \in \Tau(\mu_0)$. Then, for each $j\in \mathbb N$, we have
\[
\lim_{k\rightarrow \infty }\int_{\Delta (\Omega)} V_k(\mu) \df \tau^\star_k (\mu)\leq \lim_{k\rightarrow \infty}  \int_{\Delta (\Omega)} V_j(\mu) \df \tau^\star_k (\mu) =\int_{\Delta (\Omega)} V_j(\mu) \df \overline \tau (\mu) ,
\]
where the first inequality holds because $V_k\leq V_j$ for $k\geq j$, and the equality holds because $V_j$ is (Lipschitz) continuous and $\tau^\star_k \rightarrow \overline \tau$. Then, letting $j$ go to infinity and invoking the monotone convergence theorem, 
\[
\lim_{j\to \infty}\int_{\Delta(\Omega)}V_j(\mu)\df\ol\tau(\mu)=\int_{\Delta(\Omega)}V(\mu)\df\ol \tau(\mu),
\]
we obtain
\[
\lim_{k\rightarrow \infty }\int_{\Delta (\Omega)} V_k(\mu) \df \tau^\star_k (\mu)\leq \int_{\Delta(\Omega)}V(\mu)\df\ol \tau(\mu)\leq \int_{\Delta (\Omega)} V(\mu) \df \tau^\star (\mu),
\]
where the last inequality holds because $\tau^\star$ is an optimal solution to \eqref{primal}. This establishes \eqref{eq_C} for upper semi-continuous $V$. As a by-product, it also shows the optimality of $\ol \tau$.\footnote{In the persuasion literature, a similar argument appears in the proof of Theorem 1 in \cite{DK} for the special case of one-dimensional moment persuasion.}
\end{proof}

\subsection{Proof of Corollary \ref{cor_ver}}\label{app:complslack}

By Theorem \ref{t:duality}, $\tau\in \Tau(\mu_0)$ and $p\in \mathcal P(V)$ are optimal solutions to \eqref{primal} and \eqref{dual} if and only if 
\begin{gather*}
\int_{\Delta (\Omega)} V(\mu) \df \tau (\mu)  = \int_\Omega p(\omega) \df\mu_0(\omega)\iff \int_{\Delta(\Omega)}\left(V(\mu)-\int_\Omega p(\omega)\df\mu(\omega)\right)\df\tau(\mu)  =0.
\end{gather*}
Since the term in parenthesis is non-positive for $p\in \mathcal P(V)$, it follows that $\tau(\Lambda)=1$ where
\[
\Lambda=\left\{\mu\in \Delta(\Omega):\, V(\mu)=\int_\Omega p(\omega)\df \mu(\omega)\right\}=\left\{\mu\in \Delta(\Omega):\, V(\mu)\geq \int_\Omega p(\omega)\df \mu (\omega)\right\}.
\]
The set $\Lambda$ is closed because $V(\mu)$ is upper semi-continuous in $\mu$ and $\int_\Omega p(\omega)\df \mu (\omega)$ is continuous in $\mu$, given that each $p\in \mathcal P(V)$ is Lipschitz continuous. Thus, $\supp(\tau)\subset \Lambda$ and \eqref{eq_CS} follows, since $\supp(\tau)$ is defined as the smallest closed set on which $\tau$ is concentrated.

\subsection{Proof of Theorem \ref{thm_strong}}\label{app:regular}
The Duality Theorem in \cite{Gale1967} shows that $\ccV$ is superdifferentiable at $\mu_0$ if and only if
	 $\ccV$  has bounded steepness at $\mu_0$. Thus, Theorem \ref{thm_strong} follows from the following lemma.

\begin{lemma}\label{l:sd}
There exists an optimal solution $p\in \mathcal P(V)$ to \eqref{dual} if and only if $\ccV$ is superdifferentiable at $\mu_0$.
\end{lemma}

\begin{proof}
If $\widehat V$ is superdifferentiable at $\mu_0$, then, by the fact that $(M(\Omega),\kr {\cdot})^\star=\Lip(\Omega)$, there exists $p\in \Lip(\Omega)$ such that
\[
\widehat V(\mu_0)=\int_\Omega p(\omega)\df \mu_0(\omega)  \quad\text{and}\quad   \widehat V(\mu)\leq \int_{\Omega} p(\omega)\df \mu(\omega),\quad \text{for all $\mu\in \Delta(\Omega)$}. 
\]
Thus,
\[
V(\mu)\leq \widehat V(\mu)\leq \int_\Omega p(\omega)\df \mu(\omega),\quad \text{for all $\mu\in \Delta(\Omega)$},
\]
so $p\in \mathcal P(V)$ is an optimal solution to \eqref{dual},  by Theorem \ref{weak}.

Conversely, if $p\in \mathcal P(V)$ is optimal, then we have $p\in \Lip(\Omega)$,
\[
\ceV(\mu_0)=\int_\Omega p(\omega)\df \mu_0(\omega), 
\quad \text{and}\quad 
V(\mu)\leq \int_\Omega p(\omega)\df \mu(\omega),\quad \text{for all $\mu \in \Delta(\Omega)$}.
\]
By the definition of the concave envelope,
\[ 
\ceV(\mu)\leq \int_\Omega p(\omega)\df \mu(\omega),\quad \text{for all $\mu \in \Delta(\Omega)$}.
\]
Therefore, by Theorem \ref{t:duality},
\[
\ccV(\mu_0)=\int_\Omega p(\omega)\df \mu_0(\omega), 
\quad \text{and}\quad 
\ccV(\mu)\leq \int_\Omega p(\omega)\df \mu(\omega),\quad \text{for all $\mu \in \Delta(\Omega)$}.
\]
Thus, $p$ is a supergradient of $\ccV$ at $\mu_0$, and thus $\widehat V$ is superdifferentiable at $\mu_0$ (simply define $H(\mu)=\int_\Omega p(\omega)\df \mu(\omega)$, which is a continuous linear function on $M(\Omega)$ because $p\in \Lip(\Omega)$).
\end{proof}

\subsection{Proof of Lemma \ref{lem_Lip}}\label{proof_thm_moment}

Suppose that $v$ is $L$-Lipschitz on $X\subset \mathbb{R}^N$. Since all norms are equivalent in an $N$-dimensional Euclidean space, without loss of generality, we endow $\R^N$ with the Euclidean norm, 
\[\lVert x \rVert=\sqrt{\textstyle\sum_{i=1}^N x_i^2},\quad \text{for all $x\in \R^N$}. \] 
For any $\mu,\,\eta\in \Delta(\Omega)$, with $\mu\neq \eta$, 
\begin{align*}
\frac{|V(\mu)-V(\eta)|}{\kr{\mu-\eta}}=\frac{|v(\mathbb{E}_\mu[\omega])-v(\mathbb{E}_\eta[\omega])|}{\lVert \mathbb{E}_\mu[\omega]-\mathbb{E}_\eta[\omega]\rVert} \frac{\lVert \mathbb{E}_\mu[\omega]-\mathbb{E}_\eta[\omega]\rVert}{\kr{\mu-\eta}}\leq L \frac{\lVert \mathbb{E}_\mu[\omega]-\mathbb{E}_\eta[\omega]\rVert}{\kr{\mu-\eta}}.
\end{align*}
Because the function $f(\omega)=\omega_i$ is 1-Lipschitz,
\[
\big|\mathbb{E}_\mu[\omega_i]-\mathbb{E}_\eta[\omega_i]\big|=\left|\int_\Omega \omega_i \df(\mu-\eta)(\omega)\right|\leq \kr{\mu-\eta},
\]
and thus
\[
\lVert\mathbb{E}_\mu[\omega]-\mathbb{E}_\eta[\omega]\rVert=\sqrt{\textstyle \sum_{i=1}^N\left(\mathbb{E}_\mu[\omega_i]-\mathbb{E}_\eta[\omega_i]\right)^2}\leq \sqrt{N}\kr{\mu-\eta},
\]
showing that $V$ is $L\sqrt N$-Lipschitz.

\subsection{Proof of Theorem \ref{thm.moment}}\label{proof_thm.moment}

By Lemma \ref{lem_Lip}, we know that $V:\,\Delta(\Omega)\to \mathbb{R}$ is Lipschitz, since $v$ is Lipschitz. It follows from Theorems \ref{t:duality}, \ref{thm_strong}, and \ref{t:Lip} that there exists a solution $ p\in \Lip(\Omega)$ to the dual problem \eqref{dual}; moreover, since \eqref{primal_*} is a special case of the general problem \eqref{primal}, $\pi\in \Pi(\mu_0)$ is then optimal for \eqref{primal_*} if and only if  
\begin{equation*}\label{no_dual_gap}
\int_X v(x)\df \piX(x)=\int_\Omega  p(\omega)\df \mu_0(\omega).
\end{equation*}
Let $\check p$ be the convex roof extension of $ p$ from $\Omega$ to $X$, defined in the main text. By construction, $\check  p\leq  p$ on $\Omega$.
Moreover, the infimum in the definition of $\check{p}$ is attained because $ p$ is (Lipschitz) continuous on $\Omega$ and the set of feasible distributions is compact. Hence, for any $x\in X$, we can write $\check{p}(x)=\int_\Omega  p(\omega)\df\mu_x(\omega)$ for some $\mu_x\in \Delta(\Omega)$ with $\int_\Omega \omega \df \mu_x(\omega)=x$. By the definition of $\check p$, for any $x,\,y\in X$, $\lambda\in (0,\,1)$, we have 
\[
\lambda \check{p}(x)+(1-\lambda)\check{p}(y)=\int_\Omega  p(\omega)\df(\lambda \mu_x+(1-\lambda)\mu_y)(\omega)\geq \check{p}(\lambda x+(1-\lambda)y),
\]
showing that $\check{p}$ is convex. Moreover, by feasibility of $ p$, for any $x\in X$, 
\begin{equation*}
\check p(x) = \int_{\Omega}  p(\omega) \df \mu_{x} (\omega) \geq V(\mu_x) = v(x).
\end{equation*}
Next, we prove a key lemma. 

\begin{lemma}\label{lemma_q(x)}
Let $v$ be $L$-Lipschitz and $\check p\geq v$. There exists a measurable function $q:\,X\to \mathbb{R}^N$ such that $\lVert q(x)\rVert\leq L$ for all $x\in X$, and 
\[
\check p(y)\geq v(x)+q(x)\cdot(y-x),\quad \text{for all $y,\,x\in  X$}.
\]
\end{lemma}
\begin{proof}
Define 
\[
F(x)\coloneqq\{r\in \R^N:\check p(y)\geq v(x)+r\cdot (y-x),\quad \text{for all $y\in X$}\},
\]
and let 
\[
q(x)\coloneqq\argmin_{r\in F(x)} \lVert r \rVert,\quad \text{for all $x\in X$}.
\]
Note that $F(x)$ is closed-valued and convex-valued. Thus, if $F(x)$ is non-empty, then $q(x)$ exists and is unique because $q(x)$ is the projection of $0$ onto the  non-empty closed convex set $F(x)$. If we can additionally prove that $\lVert q(x)\rVert\leq L$ for all $x\in X$, then $q$ will be measurable by the measurable maximum theorem (Theorem 18.19 in \citealp{aliprantis2006}). To see that, note that the definition of $q$ will not change if we additionally require that $\lVert r\rVert \leq L$, so that the correspondence $x\rightrightarrows F(x)\cap \{r\in \R^N:\, \lVert r \rVert\leq L \}$ is compact-valued and upper hemi-continuous (given that $\check p$ is lower semi-continuous and $v$ is continuous), and thus measurable, by Theorem 18.20 in \cite{aliprantis2006}.

We deal with some easy cases first. If $0\in F(x)$, then  $q(x)=0$ and $0=\lVert q(x)\rVert \leq L$. 
Next, if $0\notin F(x)$ but $\check p(x)=v(x)$, then we have, for any $y\in X$, 
\[
\check p(y)-\check p(x)\geq v(y)-v(x)\geq -L\lVert y-x\rVert,
\]
because $\check{p}\geq v$ and $v$ is $L$-Lipschitz. By the Duality Theorem in \cite{Gale1967}, $q(x)$ is well defined and 
\[
\lVert q(x)\rVert = - \inf _{y\in X} \frac{\check p(y)-\check p(x)}{\lVert y-x\rVert}\leq L.
\]
Thus,  for the rest of the proof, we fix an arbitrary $x\in X$ such that $0\notin F(x)$ and $\check p(x)>v(x)$.

We first show that $F(x)$ is non-empty. Because, $\check p(x)>v(x)$, the point $(x,\, v(x))$ does not belong to the epigraph of $\check{p}$, defined as $\epi (\check p):=\{(y, t)\in X\times \R:\, t\geq \check p(y) \}$. Note that $\epi (\check p)$ is closed and convex, because $\check p$ is  lower semi-continuous (see footnote \ref{foot}) and convex. By the separation theorem (for example, Corollary 11.4.1 in \citealp{rockafellar}), there exists $(\alpha,\beta)\in \R^N\times \R$ such that, for all $y\in X$ and $t\geq \check p(y)$, 
\[
\alpha \cdot y +\beta t > \alpha \cdot x +\beta v(x).
\]
Clearly, $\beta \geq 0$; otherwise, the inequality would be violated for sufficiently large $t$. Moreover, $\beta \neq 0$; otherwise, the inequality would be violated for $(y,\, t)=(x,\, \check p(x))$. Thus, evaluating the inequality for $t=\check p(y)$, for all $y\in X$, proves that $-\alpha/\beta$ belongs to $F(x)$. Thus,  $F(x)$ is indeed non-empty (and hence $q(x)$ is well-defined). 

We now show that $\lVert q(x)\rVert \leq L$. 
Define the set 
\[
Y\coloneqq\{y\in X:\check p(y)=v(x)+q(x)\cdot (y-x)\}.
\]
Note that $Y$ is non-empty: If there is no $y\in X$ such that $\check p(y)=v(x)+q(x)\cdot (y-x)$, then the constraint in the definition of $F(x)$ is slack, so it is possible to reduce $\lVert r\rVert$, contradicting that $q(x)$ is a minimizer (this step uses the fact that $\check p$ is lower semi-continuous). Since $\check p$ is convex, the set $Y$ is convex. Since $\check p(x)>v(x)$, the set $Y$ cannot contain $x.$ Also, let
\[
E:=\{e\in \R^N:  e\cdot q(x)<0\}.
\]
We will prove that there exists  $y^\star\in Y$ such that $e\cdot (y^\star-x)\geq 0$ for all $e\in E$. Suppose that such $y^\star$ does  not exist. Since any such $y^\star$ must satisfy $y^\star-x=-tq(x)$ for some $t\geq 0$, we conclude that the compact convex set $Y-x\coloneqq\{y-x:\,y\in Y\}$ and the closed convex cone $\{-tq(x):t\geq 0\}$ must be disjoint. By the separation theorem (for example, Corollary 11.4.1 in \citealp{rockafellar}), there exists $e\in \R^N$ such that
\[\max_{y\in Y} e\cdot (y-x)<\inf_{t\geq 0} e\cdot (-tq(x)).\]
Notice that we must have $e\cdot q(x)\leq 0$, as otherwise the right-hand side is $-\infty$ and the inequality cannot hold. In fact, there exists $e\in \R^N$ such that $e\cdot q(x)<0$, because we can always replace $e$ with $e-\varepsilon q(x)$ for a sufficiently small $\varepsilon>0$ without violating the above inequality, given that $Y$ is compact. Since there is $e\in E$ such that $e\cdot (y-x) <0$ for all $y\in Y$, there is $\delta >0$ such that for all $z$ in the $\delta$-neighborhood of $Y$, we have $e\cdot (z-x)<0$, and thus for all $\varepsilon>0$,
\[
v(x)+(q(x)+\varepsilon e) \cdot (z-x) <v(x)+q(x)\cdot (z-x).
\]
Since $\check p(z)>v(x)+q(x) \cdot (z-x)$ for $z\notin Y$, and $\check p$ is convex and lower semi-continuous, there exists $\gamma>0$ such that for all $z\in X$ outside the $\delta$-neighborhood of $Y$, we have
\[
\check p(z)>v(x)+q(x)\cdot (z-x) +\gamma.
\]
Consequently, there exists a sufficiently small $\varepsilon>0$ such that, for all $z\in X$,
\[
\check p(z) > v(x)+(q(x)+\varepsilon e)\cdot (z-x).
\]
This is a contradiction with the definition of $q(x)$. Indeed, the above inequality shows that  $q(x)+\varepsilon e\in F(x)$ and, by the fact that $e\in E$ and $q(x)\neq 0$, we have $\lVert q(x)+\varepsilon e \rVert< \lVert q(x)\rVert$ for sufficiently small $\varepsilon>0$.

We have thus proven that there exists  $y^\star\in Y$ such that $e\cdot (y^\star-x)\geq 0$ for all $e\in E$. Since $e\cdot(y^\star-x)\geq 0$ for all $e\in E$ and $Y$ does not contain $x$, it follows that there exists $t>0$ such that $x-y^\star =tq$. Thus, 
\[
q(x)\cdot(x-y^\star)=\lVert q(x)\rVert \,\lVert x-y^\star\rVert.
\]
And since $y^\star\in Y$, we have that 
\[
v(x)-\check{p}(y^\star)=q(x)\cdot(x-y^\star).
\]
Putting these two equalities together, we conclude that
\[
\lVert q(x)\rVert \lVert x-y^\star\rVert = v(x)-\check p(y^\star) \leq v(x) -v(y^\star) \leq L\lVert x-y^\star\rVert,
\]
showing that $\lVert q(x)\rVert \leq L$. 
\end{proof}

Fixing $q(x)$ from Lemma \ref{lemma_q(x)}, we define
\[
\bar{p}(y)\coloneqq\sup_{x\in X} \{v(x)+q(x)\cdot(y-x)\},\quad \text{for all $y\in X$}.
\]
Note that $\bar{p}$ is convex as a pointwise supremum of affine functions. It lies everywhere above $v$, by definition. Finally, we show that $\bar{p}$ is $L$-Lipschitz. Take any $y,\,z\in X$. Let $x_n$ be a sequence of points in $X$ that generate the supremum in the definition of $\bar{p}(y)$. Because $X$ is compact and $q$ is bounded, we can assume that $x_n$ and $q(x_n)$ converge. Then, we have that
\begin{align*}
\bar{p}(y)-\bar{p}(z)&=\lim_{n\to \infty}\{v(x_n)+q(x_n)\cdot(y-x_n)\}-\bar{p}(z)\\
 &\leq \lim_{n\to \infty}\{v(x_n)+q(x_n)\cdot(y-x_n)-v(x_n)-q(x_n)\cdot(z-x_n)\}\\
 &=\lim_{n\to \infty}\{ q(x_n)\}\cdot (y-z)\leq L\lVert y-z\rVert.
\end{align*}
Because $y$ and $z$ were arbitrary, this proves that $\bar{p}$ is $L$-Lipschitz.

Finally, notice that $\bar{p}\leq \check{p}$, by Lemma \ref{lemma_q(x)}. Therefore, on $\Omega$, we have that 
\[
\bar{p}\leq \check{p}\leq  p.
\]
Since $\bar{p}$ is Lipschitz, $\bar{p}\geq v$ and $\bar{p}$ is convex, it follows that $\bar{p}$ (restricted to $\Omega$) is feasible for the dual \eqref{dual}; indeed, for any $\mu\in \Delta(\Omega)$, 
\[
\int_\Omega \bar{p}(\omega)\df \mu(\omega)\geq \bar{p}\left( \int_\Omega \omega\df \mu(\omega) \right)\geq  v\left( \int_\Omega \omega\df \mu(\omega) \right)=V(\mu).
\]
But since $ p$ solves the dual problem \eqref{dual}, we must have that $ p=\bar{p}$ almost surely on $\Omega$. Since both these function are (Lipschitz) continuous, we can conclude that $ p$ and $\bar{p}$ coincide on $\Omega$. In particular, we have shown that $\bar{p}$ is convex and solves \eqref{dual} when restricted to $\Omega$.%

Next, we prove that if $\pi\in \Pi(\mu_0)$ is optimal for \eqref{primal_*}, then conditions \textit{1} and \textit{2} hold. We have already shown that $\bar{p}$ is convex, Lipschitz, and satisfies $\bar{p}\geq v$. To finish the proof that condition \textit{1} holds, note that 
\[
\int_X v(x)\df\piX(x)=\int_\Omega  p(\omega)\df\mu_0(\omega)=\int_\Omega \bar{p}(\omega)\df\mu_0(\omega),
\]
where the first equality is due to the absence of a  duality gap (Theorem \ref{t:duality}) and the second is by the fact that $ p=\bar{p}$ on $\Omega$.  We can also prove that condition \textit{2} holds: $\bar{p}$ satisfies the required equality by definition when $q$ is defined by Lemma \ref{lemma_q(x)}; moreover, 
\[
\int_{X\times \Omega} (v(x)+q(x)\cdot (\omega-x))\df\pi(x,\omega)=\int_X v(x)\df\piX(x)=\int_\Omega \bar{p}(\omega)\df\mu_0(\omega)=\int_{X\times \Omega} \bar{p}(\omega)\df\pi(x,\omega),
\]
where the first and last equality follow from  the feasibility of $\pi$, and the second equality was established above. Because, by definition, $\bar{p}(\omega)\geq v(x)+q(x)\cdot (\omega-x)$ for all $(x,\omega)$, we must have that for $\pi$-almost all $(x,\omega)$, \[
v(x)+q(x)\cdot(\omega-x)=\bar{p}(\omega).
\]
It remains to show that any one of conditions \textit{1} or \textit{2} imply optimality of $\pi\in \Pi(\mu_0)$. Note that we will not use the assumption that $v$ is Lipschitz in that part of the proof. 

Assume that condition \textit{1} holds. Note that, under these assumptions, $\bar{p}$ is feasible for the dual \eqref{dual} when viewed as a function on $\Omega$ (in particular, as shown previously, convexity and $\bar{p}\geq v$ imply that $\int_\Omega \bar{p}(\omega)\df \mu(\omega)\geq V(\mu)$, for all $\mu\in \Delta(\Omega)$). But then the fact that $\piX$ achieves no duality gap means that $\pi$ must be optimal.

Assume that condition \textit{2} holds. Note that under these assumptions, we have shown previously (using only the definition of $\bar{p}$ and the property that $q$ is measurable with $\lVert q(x)\rVert\leq L$ for all $x\in X$) that $\bar{p}$ is feasible for the dual \eqref{dual} on $\Omega$. Moreover, by the last equation of condition \textit{2},
\[
\int_\Omega \bar{p}(\omega)\df\mu_0 (\omega)=\int_{X\times \Omega} (v(x)+q(x)\cdot(\omega-x))\df\pi(x,\omega)=\int_X v(x)\df\piX(x),
\]
showing that $\bar{p}$ and $\piX$ achieve no duality gap, and hence $\pi$ is optimal.

\subsection{Proof of Theorem \ref{t:moment} and Remark \ref{remark1}}\label{proof_t:moment} 
 
Since $v$ is continuously differentiable on the compact set $X$, it is $L$-Lipschitz on $X$ where
\[
L:=\max_{x\in X}\lVert \nabla v(x)\rVert <\infty,
\]
so all previous results apply. We now prove the two implications of the equivalence separately.
 
\emph{If.} 
Fix $\pi\in \Pi(\mu_0)$, and let $S=\supp(\piX)$. The function $p_S$ is convex (see footnote \ref{foot_convex}). Moreover, by condition \eqref{e:moment}, $p_S\geq v$. Thus, there exists a function $q$ as in Lemma \ref{lemma_q(x)}. Then, for any feasible $\tilde\pi\in \Pi (\mu_0)$, we have 
\begin{align*}
\int_{X\times\Omega} v(x)\df\tilde \pi(x,\omega)& =\int_{X\times\Omega} (v(x)+q(x)\cdot (\omega-x))\df\tilde \pi(x,\omega)\\
&\leq \int_{X\times\Omega} p_S(\omega)\df\tilde \pi(x,\omega)=\int_{\Omega } p_S(\omega) \df \mu_0(\omega)= \int_{X\times\Omega}  p_S(\omega)\df  \pi(x,\omega)\\
&= \int_{X\times\Omega} (v(x)+\nabla v(x)\cdot (\omega-x))\df \pi(x,\omega)=\int_{X\times\Omega} v(x)\df \pi(x,\omega),
\end{align*}
showing that $\pi$ is optimal. The inequality follows from Lemma \ref{lemma_q(x)}. The second to last equality holds by condition \eqref{e:moment}. The remaining equalities follow from the feasibility of $\tilde \pi$ and $\pi$.

\emph{Only if.} Fix an optimal distribution $\pi\in \Pi(\mu_0)$. By Theorem \ref{thm.moment}, there exists an optimal solution $ p$ to \eqref{dual} and it is convex on $\Omega$. Define the convex roof extension $\check{p}$ of $ p$ from $\Omega$ to $X$, as in formula \eqref{def_CR}. For each $x\in X$, the infimum in the definition of $\check p(x)$ is attained at some $\mu_x\in \Delta(\Omega)$. By feasibility of $ p$, for any $x\in X$,
\[
\check p(x)=\int_\Omega  p(\omega) \df \mu_x (\omega) \geq V(\mu_x)=v(x).
\]
Consequently,
\[
\int_X v(x)\df \piX(x) \leq \int _X \check p(x) \df \piX(x) \leq \int_\Omega \check p(\omega)\df \mu_0(\omega)=\int_\Omega  p(\omega)\df \mu_0(\omega),
\]
where the first inequality holds because $\check p\geq v$, the second inequality holds because $\check p$ is convex and $\mu_0$ is a mean-preserving spread of $\piX$, and the equality holds because $\check p$ coincides with $ p$ on $\Omega$, given that $ p$ is convex on $\Omega$. Hence condition \textit{1} in Theorem \ref{thm.moment} implies that all inequalities hold with equality, 
\[
\int_X v(x)\df \piX(x) = \int _X \check p(x) \df \piX(x)=\int_\Omega \check p(\omega)\df \mu_0(\omega).
\]
Thus, $\piX(\check S)=1$, where $\check S=\left\{x\in X:v(x)=\check p(x)\right\}$.
Since $X$ is closed, $v$ is continuous, $\check p$ is lower semi-continuous (see footnote \ref{foot}), and the set $\check S$ can be equivalently written as $\check S=\left\{x\in X:v(x)\geq \check p(x)\right\}$, it follows that the set $\check S$ is closed. Thus, $\supp(\piX)\subset \check S$. 

Taking into account that $v$ is continuously differentiable and $\check p$ is convex and satisfies $\check p\geq v$, we obtain that $\check p$ has a subgradient $\nabla v(x)$ at each $x\in \check S$, so, for all $y\in X$,
\[
\check p(y)\geq \check p(x)+\nabla v(x)\cdot (y-x)=v(x)+\nabla v(x)\cdot (y-x).
\]
Indeed, for $x\in \check S$, $y\in X$, and $\varepsilon>0$, we have
\[
\check p(y)-\check p(x)\geq \frac {1}{\varepsilon}(\check p(x+\varepsilon(y-x))-\check p(x))\geq\frac 1 \varepsilon (v(x+\varepsilon(y-x))-v(x)),
\]
where the first inequality is by convexity of $\check p$, and the second inequality is by $\check p\geq v$ and $\check p (x)=v(x)$. Taking $\varepsilon\downarrow 0$ yields that $\nabla v(x)$ is a subgradient of $\check p$ at $x\in \check S$.

Thus, since $\pi\in \Pi(\mu_0)$ and $ p=\check p$ on $\Omega$, we have
\begin{align*}
\int_{\Omega}   p(\omega)\df \mu_0(\omega)\geq \int_{X\times \Omega} (v(x)+\nabla v(x) \cdot (\omega-x))\df \pi(x,\omega) =\int_{X\times \Omega} v(x)\df \pi(x,\omega).
\end{align*}
As shown above, the inequality holds with equality, so $\pi(\check\Gamma)=1$, where 
\[\check \Gamma=\left\{(x,\omega)\in \check S\times \Omega:\,  \check p(\omega)= v(x)+\nabla v(x) \cdot (\omega-x)\right\}.\] 
Note that the set $\check \Gamma$ is closed, given that $\check S$ and $\Omega$ are closed and $\nabla v$ and $\check p$ are continuous on $X$ and $\Omega$, respectively. Thus, $\supp(\pi)\subset \check \Gamma$. But then we have that, for all $\omega\in\Omega$, 
\[
p_{\supp(\piX)}(\omega)=\max_{x\in \supp(\piX)}\{v(x)+\nabla v(x)\cdot (\omega-x)\}=\check p(\omega),
\]
where the first equality is by the definition of $p_S$, and the second equality is by $\supp(\pi)\subset \check \Gamma$. This shows that $p_{\supp(\piX)}(\omega)=\check p(\omega)= p(\omega)$ for  $\omega\in \Omega$, and hence also that $p_{\supp(\piX)}(x)=\check p(x)$ for $x\in X$. Thus, we have shown that  $p_{\supp(\piX)}$ satisfies condition \eqref{e:moment}, which finishes the proof of the theorem. 

Finally, we explain why the above proof also implies Remark \ref{remark1}. First, note that in the ``only if" part of the proof  we established $p_{\supp(\piX)}\equiv\check p$ for an arbitrary optimal $\pi$. It follows that $S^\star$, as defined in Remark \ref{remark1}, is equal to $\check S$ in the proof (note that $\check S$ does not depend on which optimal solution $\pi$ we consider). Thus, we also have that $p_{S^\star}\equiv\check p$.

Fix a feasible $\pi\in\Pi(\mu_0)$. 
Suppose that $\pi$ is optimal for \eqref{primal_*}. Then, the ``only if" part of the above proof shows that $\supp(\piX)\subset \check S$ and $\supp(\pi)\subset \check \Gamma$. As argued in the previous paragraph, we can replace $\check S$ with $S^\star$ and $\check p$ with $p_{S^\star}$ and hence condition \eqref{e:moment} holds with $S=S^\star$. Conversely, if $\supp(\piX)\subset S^\star$ and condition \eqref{e:moment} holds with $S=S^\star$, then the ``if" part of the proof shows that $\pi$ is optimal for \eqref{primal_*}.

\subsection{Generalized analysis for Section \ref{sec.structure}}\label{app_ext_structure}

In this appendix, we set up generalized notation that agrees with the notation defined in Section \ref{sec.structure} in the special case of convex $\Omega$ but may differ in the general case of non-convex $\Omega$. In Appendix \ref{app_proofs}, we use this generalized notation to prove Theorems \ref{t:Brenier2} and  \ref{t:KMS} without assuming that $\Omega$ is convex.

It will be convenient to consider solutions $\pi\in \Pi(\mu_0)$ on the extended space $X\times X$ even though $\supp(\pi)\subseteq X\times \Omega$. To make our notation more intuitive, we will use the symbols  $x,\,y,\,z\in X$ to refer to moments, and $\omega\in X$ to refer to the ``extended states."

For a closed set $S\subset X$, let $p_S:\,X\to \R$ be defined as in Section \ref{s:diff}. Let $S^\star$ be defined as in Remark \ref{remark1}. Specifically, $S^\star$ is the closed subset of $X$ such that \[S^\star =\{x\in X:\, p_{S^\star} (x)=v(x)\},\] and condition \eqref{e:moment} holds with $S=S^\star$ (for any optimal solution $\pi$). Define the function $p^\star:\, X\rightarrow \R$,
\[
p^\star(\omega)\coloneqq \max_{x\in S^\star } \left\{ v(x) +\nabla v(x)\cdot (\omega-x)\right\},\quad \text{for all $\omega\in X$}.
\]
Note that this definition agrees with the one introduced in Section \ref{sec.structure} when $\Omega=X$ because $p^\star$ and $p_{S^\star}$ coincide on $\Omega$; however, $p^\star$ and $p_{S^\star}$  may differ on $X\setminus \Omega$.

Define the contact set $\Gamma\subset X \times X$,
\[
\Gamma\coloneqq \left\{(x,\,\omega)\in S^\star\times X:\,  p^\star(\omega)= v(x)+\nabla v(x) \cdot (\omega-x)\right\},
\]
and its $x$-section,
\[
\Gamma_x\coloneqq \left \{\omega\in X:\, (x,\,\omega)\in \Gamma \right\}, \quad\text{for all $x\in S^\star$}.
\] 
To extend Theorem \ref{t:Brenier2}, we must first define convex-partitional signals for the case when $\Omega$ is not necessarily a convex set. To circumvent this difficulty, we define the partition on the convex hull of $\Omega$ (that is, on $X$), and we require each element of the partition of $X$ to be convex.\footnote{\label{f:Y}To understand why we adopt this convention, consider the distribution $\pi$ induced by no disclosure. Intuitively, pooling all states should correspond to a convex-partitional signal. However, the support of this distribution over states conditional on the induced moment is equal to $\Omega$, and is hence not convex when $\Omega$ is not convex. We circumvent this by defining the partition on $X$; then, the unique element of that partition corresponding to no disclosure is $X$ itself, a convex set. And of course, this partition restricted to $\Omega$ still represents no disclosure.} Formally, we say that $\pi\in\Pi(\mu_0)$ is \textit{convex-partitional} if there is a measurable function $\chi :X\rightarrow X$ such that, for all measurable sets $A\subset X$ and $B\subset \Omega$, 
\[
\pi(A,\,B)=\int_{B}\1\{\chi(\omega)\in A\} \df \mu_0(\omega),
\]
and, for all $x\in X$, the set $\chi^{-1}(x)$  is convex.

\subsection{Proof of Theorems \ref{t:Brenier2} and \ref{t:KMS}}\label{app_proofs}

In this appendix, we rely on the general notation set up in Appendix \ref{app_ext_structure}.

Before proceeding to the proofs of Theorems \ref{t:Brenier2} and \ref{t:KMS}, we state and prove a key lemma.  Define the correspondence $\mathcal X:\, X\rightrightarrows X$ by
\[
\mathcal X(\omega):=\argmax_{x\in S^\star} \{v(x)+\nabla v(x)\cdot (\omega-x)\},\quad \text{for all $\omega\in X$},
\]
and fix any measurable selection $\chi:\, X\to X$ from $\mathcal X$, which exists by the measurable maximum theorem (Theorem 18.19 in \citealp{aliprantis2006}). We start with a key lemma that we will be using throughout.

\begin{lemma}\label{l:face} \hfill
\begin{enumerate}
	\item The function $p^\star$ is convex and Lipschitz on $X$. Moreover, $p^\star$ is differentiable at any $\omega\in \interior (X)$ if and only if the set $\{\nabla v(x):\, x\in \mathcal X(\omega)\}$ is a singleton, and in that case $\nabla p^\star(\omega)=\nabla v(x)$ for all $x\in \mathcal X(\omega)$. 
	\item The set $\Gamma\subseteq X\times X$ is closed. Its projection along the first coordinate is $S^\star$, and its projection along the second coordinate is $X$. For each $x\in S^\star$, $\Gamma_x$ is  a compact convex set such that $x\in \Gamma_x$ and
\begin{align*}
	\Gamma_x=\argmin_{\omega\in X}\left\{ p^\star (\omega) -\nabla v(x)\cdot \omega\right\}.
\end{align*}
Moreover, for any $x,\,y\in S^\star$, we have:
\begin{enumerate}
\item  $\nabla v(x)=\nabla v(y)\implies \Gamma_x=\Gamma_y$;
\item  $\relint (\Gamma_x)\cap \relint (\Gamma_y)\neq \emptyset\implies\Gamma_x=\Gamma_y$;
\item  $\relint (\Gamma_x )\cap \Gamma_y \neq \emptyset\implies\Gamma_x\subset \Gamma_y$.
\end{enumerate}\end{enumerate} 
\end{lemma}
\begin{proof}
\textit{1.} Clearly, $p^\star$ is convex on $X$ as a pointwise maximum of affine functions. Moreover, it is Lipschitz on $X$ because, for any $\omega,\,\omega'\in X$,
\begin{align*}
p^\star(\omega)-p^\star(\omega') &\leq  v(\chi (\omega))+ \nabla v(\chi (\omega)) \cdot (\omega-\chi(\omega))-v(\chi(\omega))-\nabla v(\chi (\omega))\cdot 	(\omega'-\chi (\omega))\\
&= \nabla v(\chi (\omega))\cdot (\omega-\omega')\leq L\lVert \omega-\omega'\rVert,
\end{align*}
with $L$ defined (as in Appendix \ref{proof_t:moment}) as the maximal value of the norm of the gradient of $v$ on $X$.

The remainder of part \textit{1} is a consequence of the envelope theorem. For $N=1$, this follows immediately from Corollary 4 in \cite{MS}. Below, we extend their analysis to the general case $N\geq 1$. 

Suppose, by contradiction, that $p^\star$ is differentiable at $\omega\in \interior (X)$ but there exist $x,\, y\in \mathcal X(\omega)$ such that $\nabla v(x)\neq \nabla v(y)$. Denote $u\coloneqq \nabla v(x)-\nabla v(y)$, so that $\nabla v(x)\cdot u> \nabla v(y)\cdot u$. 
Since $\omega\in \interior (X)$, we have $\omega\pm hu\in X$ for small enough $h>0$. Moreover, by the definitions of $p^\star$ and $\mathcal X$,
\begin{align*}
	\frac{p^\star(\omega+ hu)-p^\star(\omega)}{h} \geq \nabla v(x)\cdot u\quad \text{and}\quad \frac{p^\star(\omega-hu)-p^\star(\omega)}{h} \geq -\nabla v(y)\cdot u,
\end{align*}
and thus
\[
-\lim_{h\downarrow 0}\frac{p^\star(\omega-hu)-p^\star(\omega)}{h}\leq \nabla v(y)\cdot u< \nabla v(x)\cdot u\leq\lim_{h\downarrow 0} \frac{p^\star(\omega+ hu)-p^\star(\omega)}{h},
\]
showing that $p^\star$ is not differentiable at $\omega$.

Conversely, suppose that $\omega\in \interior (X)$ and $\{\nabla v(x):\, x\in \mathcal X(y)\}$ is a singleton. Fix any $u\in \R^N$ and small enough $h''>h'>0$, so that $\omega+h'u$ and $\omega+h''u$ are both in\ $X$. By the definition of $p^\star$,
\begin{align*}
	{\nabla v (\chi(\omega+h'u))\cdot u} \leq \frac{p^\star(\omega+h''u)-p^\star(\omega+h'u)}{h''-h'}\leq \nabla v(\chi (\omega+h''u))\cdot u.
\end{align*}
Taking the limit superior in this inequality as $h'\downarrow 0$ yields
\begin{align*}
\limsup _{h'\downarrow 0}	{\nabla v (\chi(\omega+h'u))\cdot u} \leq \frac{p^\star(\omega+h''u)-p^\star(y)}{h''}\leq \nabla v(\chi(\omega+h''u))\cdot u.
\end{align*}
Taking the limit inferior in the resulting inequality as $h''\downarrow 0$ yields
\begin{align*}
\limsup _{h'\downarrow 0}	{\nabla v (\chi(\omega+h'u))\cdot u} \leq \lim_{h''\downarrow 0} \frac{p^\star(\omega+h''u)-p^\star (\omega)}{h''}\leq \liminf _{h''\downarrow 0}\nabla v(\chi(\omega+h''u))\cdot u.
\end{align*}
Since the limit superior is never smaller than the limit inferior, we conclude that the two limits coincide, and hence
\[
\lim_{h\downarrow 0} \frac{p^\star(\omega+hu)-p^\star(\omega)}{h}= \lim _{h\downarrow 0}	{\nabla v (\chi(\omega+hu))\cdot u}.
\]
Since the correspondence $\mathcal X:\, X\rightrightarrows X$ is upper hemicontinuous, a version of Berge's Maximum Theorem (see Lemma 17.30 in \citealp{aliprantis2006}) yields
\[
\lim_{h\downarrow 0} \frac{p^\star(\omega+hu)-p^\star(\omega)}{h}=\lim _{h\downarrow 0}	{\nabla v (\chi(\omega+hu))\cdot u}\leq \max_{x\in \mathcal X(\omega)}\nabla v(x)\cdot u.
\]
Since $\{\nabla v(x):x\in \mathcal X (\omega)\}$ is a singleton, we have $\max_{x\in \mathcal X(\omega)}\nabla v(x)\cdot u=\nabla v(x)\cdot u$ for all $x\in \mathcal X(\omega)$.
Finally, taking into account that, by the definition of $p^\star$, for any $x\in \mathcal X(\omega)$ and any small enough $h>0$, we have
\[
\nabla v(x)\cdot u\leq \frac{p^\star(\omega+hu)-p^\star(\omega)}{h},
\] 
it follows that
\[
\lim_{h\downarrow 0} \frac{p^\star(\omega+hu)-p^\star(\omega)}{h}=\nabla v(x)\cdot u,\quad \text{for all $x\in \mathcal X(\omega)$},
\]
showing that $p^\star$ is differentiable at $y$ and $\nabla p^\star(\omega)=\nabla v(x)$ for all $x\in \mathcal X(\omega)$.

\textit{2.} The set $\Gamma$ is closed, because the function $p^\star(\omega)- v(x)-\nabla v(x) \cdot (\omega-x)$ is continuous in $(x,\, \omega)$ on $X\times X$. The projection of $\Gamma$ along the second coordinate is $X$, because $(\chi (\omega),\, \omega)\in \Gamma $ for each $\omega\in X$. The projection of $\Gamma$ along the first coordinate is $S^\star$ by the definition of $S^\star$ and the fact that $\Gamma_x$ is non-empty, for any $x\in S^\star$, which is shown in the next paragraph. 

Fix any $x\in S^\star$. We have
\begin{align*}
\Gamma_x &=\left\{\omega\in X:\, p^\star (\omega)= v(x)+\nabla v(x)\cdot (\omega-x)\right\}	=\left\{\omega\in X:\, p^\star (\omega)\leq v(x)+\nabla v(x)\cdot (\omega-x)\right\},
\end{align*}
where the first equality is by the definition of $\Gamma$ and $\Gamma_x$, and the second equality is by the definition of $p^\star$, which, in particular, implies that
\[
p^\star (\omega) \geq v(x)+\nabla v(x)\cdot (\omega-x),\quad \text{for all $\omega\in X$}.
\]
Thus, the set $\Gamma_x$ is compact and convex, as it is a sublevel set of the (Lipschitz) continuous and convex function  $p^\star (\omega)-v(x)-\nabla v(x)\cdot (\omega-x)$ (viewed as a function of $\omega$). Moreover, we have $x\in \Gamma_x$, because 
\[
v(x)= p_{S^\star}(x)\geq p^\star(x)\geq v(x),
\]
where the equality is by $x\in S^\star$, the first inequality is by the definition of $p_{S^\star}$, and the last inequality is by the definition of $p^\star$ and the fact that $x\in S^\star$. Since $p^\star (x)=v(x)$, we have
\[
p^\star (\omega)\geq p^\star (x)+\nabla v(x)\cdot (\omega-x),\quad \text{for all $\omega\in X$},
\]
and thus
\begin{align*}
	\Gamma_x=\arg\max_{\omega\in X}\left\{  \nabla v(x)\cdot \omega - p^\star (\omega)\right\}.
\end{align*}
We have thus shown that $\Gamma_x$ is the projection along the first coordinate of the face of the epigraph of $p^\star$ exposed by the direction $(-1,\nabla v(x))$. Then, implication (a) is immediate, whereas implications (b) and (c) follow from Corollary 18.1.2 and Theorem 18.1 in \cite{rockafellar}. For completeness, we provide short self-contained proofs of (b) and (c). To show (c), let $\omega\in \relint( \Gamma_x)\cap \Gamma_y$. Since $\Gamma_x$ is convex, for any $\omega'\in \Gamma_x$ with $\omega'\neq \omega$, there exists $\omega''\in \Gamma_x$ and $\lambda \in (0,1)$ such that $\omega=\lambda \omega'+(1-\lambda )\omega''$. Next, by the definition of $p^\star$, we have  
\[
p^\star (\omega')\geq v(y)+\nabla v(y)\cdot (\omega'-y) \text{ and } p^\star (\omega'')\geq v(y)+\nabla v(y)\cdot (\omega''-y).
\]
Both inequalities must hold with equality, as otherwise we would have 
\[
p^\star(\omega)\geq \lambda p^\star (\omega')+(1-\lambda)p^\star (\omega'')>v(y)+\nabla v(y)\cdot (\omega-y),
 \]
contradicting that $\omega\in \Gamma_y$. Since $\omega'$ is arbitrary, we get $\Gamma_x\subset \Gamma_y$, proving (c). To prove (b), notice that if $\relint (\Gamma_x)\cap \relint (\Gamma_y)\neq \emptyset$, then $\relint (\Gamma_x)\cap \Gamma_y\neq \emptyset$ and $\relint (\Gamma_y)\cap \Gamma_x\neq \emptyset$, implying that $\Gamma_x\subset \Gamma_y$ and $\Gamma_y\subset \Gamma_x$, and thus $\Gamma_x=\Gamma_y$. 
\end{proof}

In the remainder, we complete the proofs of Theorems  \ref{t:Brenier2} and \ref{t:KMS}. To deal with the general case in which $\Omega\neq X$, we follow Appendix \ref{app_ext_structure} and consider solutions defined on the larger space $X$ rather than on $\Omega$. All the notation used in the following proof completions is then defined as in Appendix \ref{app_ext_structure}, and becomes consistent with the notation used in the main text under the assumption that $\Omega$ is convex (so that $\Omega=X$). 

\subsubsection*{Completion of the proof of Theorem \ref{t:Brenier2}}\label{proof_t:Brenier2}

Let $\tilde X$ be the set of interior points of $X$ where $p^\star$ is differentiable. The set of boundary points of the convex set $X$ is Lebesgue-negligible, by Theorem 1 in \cite{Lang}. The set of interior points of $X$ where $p^\star$ is not differentiable is Lebesgue-negligible by Rademacher's Theorem (Theorem 10.8 in \citealp{villani2009}). Thus, taking into account that $\mu_0$ has a density on $X$, the set $\tilde X$ has full measure under $\mu_0$: $\mu_0(\tilde X)=1$.

Fix $\omega\in \tilde X$. We claim that $|\mathcal X(\omega)|=1$. Suppose, by contradiction, that there exist distinct $x,\, y\in \mathcal X(\omega)$.  Since $\omega\in \interior (X)$ and $p^\star $ is differentiable at $\omega$, part \textit{1} of Lemma \ref{l:face} yields 
\[\nabla p^\star(\omega)=\nabla v(x)=\nabla v(y).\]
In turn, part \textit{2} of Lemma \ref{l:face} yields $x\in \Gamma_x$, $y\in \Gamma_y$, and $\Gamma_x=\Gamma_y$, and thus, given that $p^\star$ is affine on $\Gamma_x$ by the definition of $\Gamma_x$, we have $p^\star(y)=p^\star(x)+\nabla p^\star(\omega)\cdot (y-x)$ or, equivalently,
\[
v(x)-\nabla v(x)\cdot x=v(y)-\nabla v(y)\cdot y.
\]
Next, for all $\lambda \in [0,1]$, we have $p_{S^\star} (\lambda x+(1-\lambda)y) =\lambda v(x)+(1-\lambda) v(y)$ as follows from
\begin{align*}
\lambda v(x) +(1-\lambda) v(y)&=\lambda p^\star(x)+(1-\lambda)p^\star (y)= p^\star (\lambda x +(1-\lambda)y)\\
&\leq p_{S^\star} (\lambda x+(1-\lambda) y)\leq \lambda p_{S^\star} (x)+(1-\lambda )p_{S^\star} (y)=\lambda v (x)+(1-\lambda)v (y),
\end{align*}
where the first equality is by $x\in \Gamma_x$ and $y\in \Gamma_y$, the second equality is by affinity of $p^\star$ on the convex set $\Gamma_x=\Gamma_y$, the first and second inequality follow from the definition of $p_{S^\star}$, and the last equality is by $p_{S^\star}=v$ on $S^\star$. Thus, since $p_{S^\star}\geq v$ on $X$, we get
\[
\lambda v (x)+(1-\lambda)v (y)\geq v(\lambda x+(1-\lambda)y),\quad \text{for all $\lambda\in [0,1]$}.
\]
This contradicts the conditions of the theorem. Thus, $\mathcal X(\omega)$ is a singleton $\{\chi(\omega)\}$ for each $\omega\in \tilde X$, where $\chi (\omega)$ is determined by
\[
\{\chi (\omega)\}=\{x\in S^\star:\, \omega\in \Gamma_x\}=\{x\in S^\star:\,\nabla p^\star(\omega)=\nabla v(x)\}.
\]
The first equality is by the definition of $\mathcal X$, and the second is by part \textit{1} of Lemma \ref{l:face}. 

Finally, for any optimal $\pi\in \Pi(\mu_0)$, we have
\[
1=\pi(\Gamma)=\pi\left(\cup_{\omega\in \tilde X}\left(\{\chi(\omega)\}\times\{\omega\}\right)\right),
\]
where the first equality is by Remark \ref{remark1}, and the second equality is by $\Gamma=\cup_{\omega\in X} \left(\mathcal X (\omega)\times \{\omega\}\right)$, $\mathcal X(\omega)=\{\chi(\omega)\}$ for $\omega\in \tilde X$, and $\mu_0(\tilde X)=1.$ Since $\chi(\omega)$ is determined by $p^\star$ for $\mu_0$-almost all $\omega\in X$, and $p^\star$ is independent of $\pi$, we conclude that $\pi$ is uniquely determined by
\[
\pi(A,\, B)=\int_B \1 \{\chi (\omega)\in A\}\df \mu_0(\omega),\quad \text{for all measurable $A\subset X$ and $B\subset X$}.
\]
\vspace{-2em}

\subsubsection*{Completion of the proof of Theorem \ref{t:KMS}}\label{proof_t:KMS}

Fixing any solution to the primal problem \eqref{primal_*} and the corresponding price function, define the set $S^\star$, the contact set $\Gamma$, and the sets $\Gamma_x$ as in Appendix \ref{app_ext_structure}. Recall that $S_x=\cl(\supp (\pi_X)\cap \relint(\Gamma_x))$. By Theorem 1 in \cite{larman}, $X$ can be partitioned (up to a measure zero set) into a collection of disjoint (relatively) open sets $\Xi =\{\relint(\Gamma_x) \}_{x\in S^\star}$ (where we ignore duplicates whenever $\Gamma_x=\Gamma_y$ for $x\neq y$).

Consider an auxiliary problem of finding a joint distribution $\pi\in \Pi(\mu_0)$ to maximize $\int_{X\times X} w(x,\omega)\df \pi (x,\omega)$, where
\[
w(x,\omega)\coloneqq
\begin{cases}
-\lVert x\rVert^2, &(x,\omega)\in \Gamma,\\
-\infty, &(x,\omega)\in (X\times X)\setminus \Gamma.
\end{cases}
\]
Note that $\int_{X\times X} w(x,\omega)\df \pi (x,\omega)$ is finite for $\pi\in \Pi(\mu_0)$ if and only if $\supp (\pi)\subset \Gamma$, which in turn is equivalent to optimality of $\pi \in \Pi(\mu_0)$ for the primary problem. Since $w$ is upper semi-continuous and bounded from above, by Lemma \ref{l:popt}, there exists an optimal solution $\pi\in \Pi(\mu_0)$ to the auxiliary problem, which is also optimal for the primal problem \eqref{primal_*}. We fix such $\pi\in\Pi(\mu_0)$.

Intuitively, the auxiliary problem selects a solution to the primal problem \eqref{primal_*} that minimizes the average norm of the induced posterior means. 
The rest of the proof shows that if the set $S_x$ induced by $\pi$ differs from $\ext(S_x)$ on a positive measure set of $x\in\supp(\pi_X)$, 
we would obtain a contradiction with $\pi$ solving the auxiliary problem. While this conclusion is intuitive, the details of the proof are complicated by the fact that the selection induced by the auxiliary problem may be ``local" in the sense that it affects the structure of the solution on uncountably many measure-zero sets. Our strategy is to decompose the distribution $\pi$ into conditional distributions conditional on each induced $\relint(\Gamma_x)$.

Note that we can treat the set $\Xi$ as a measurable space, endowing it with the Borel $\sigma-$ algebra generated by the Hausdorff metric. We can then define $\pi_\Xi$ to be the probability distribution over $\Xi$ induced by $\pi$: For any measurable subset $A\subset \Xi$,
\[
\pi_\Xi(A)\coloneqq\pi\left(\{(x,\,\omega)\in X\times X:\,\omega \in \relint(\Gamma_x),\,\relint(\Gamma_x)\in A\}\right).
\]
By the disintegration theorem (e.g., Theorem 2.3 in \citealp{CD}), there exists a measurable function $\xi \mapsto \pi(\cdot|\xi)$ from $\Xi$ to $\Delta(X\times X)$ such that for every ``test function'' $h\in C(X\times X)$, we have 
\[
\int_{X\times X} h(x,\omega)\df \pi (x,\omega) =\int_\Xi \int_{X\times X} h(x,\omega)\df \pi(x,\omega|\xi) \df \pi_\Xi (\xi). 
\]
Let $\pi_X(\cdot|\xi)$ and $\pi_\Omega(\cdot|\xi)$, for $\xi\in \Xi$, denote the marginal distributions of $x$ and $\omega$ (that is, the first and second coordinate, respectively) induced by $\pi(\cdot|\xi)$. Then, for $\pi_X$-almost all $x\in X$, we have 
\begin{gather}
\supp (\pi_\Omega(\cdot|\relint (\Gamma_x)))\subset  \cl( \relint(\Gamma_x)),\label{eq_inc}\\
\supp (\pi_X(\cdot|\relint (\Gamma_x)))= S_x,\label{eq_equal}\\
\int_{A\times X} (\omega-x)\df \pi(x,\omega|\relint (\Gamma_x))=0, \, \text{for all measurable } A\subset X,\label{eq_mart}\\
\int w(x,\omega)\df \pi(x,\omega|\relint (\Gamma_x))\geq \int w(x,\omega)\df \tilde  \pi (x,\omega),\,\text{for all } \tilde \pi \in \Pi(\pi_\Omega(\cdot|\relint(\Gamma_x))),\label{eq_max}
\end{gather}
where the first three properties follow from definitions, and the last inequality must be true because otherwise we would have a contradiction with the definition of $\pi$ as the solution to the auxiliary problem. 

Toward a contradiction, suppose that there exists a $\pi_X$-positive-measure set of points $x$ such that $S_x\neq \ext(S_x)$; that is,  there exist distinct $x^0,x^1,\dots, x^n\in S_x$ such that $x^0=\lambda^1 x^1+\dots +\lambda^n x^n$, where $\lambda^1,\dots, \lambda^n>0$ and $\lambda^1+\dots +\lambda^n=1$. (We suppress the dependence of these variables on $x$.) By condition \eqref{eq_equal}, since $x^1,\dots,x^n\in S_x$, for all $i=1,\dots,\,n$, and $\delta>0$, we have $\pi_X(B_\delta(x^i)|\relint (\Gamma_x))>0$, where $B_\delta(x^i)$ denotes an open ball with radius $\delta$ centered at $x^i$. To simplify notation, let $\pi^i_\delta(\cdot)$ denote the conditional probability measure on $X$ induced from $\pi_X(\cdot|\relint(\Gamma_x))$ by conditioning on the event $B_\delta(x^i)$. There exists a sufficiently small $\delta$ such that for some $\lambda^1_\delta,\dots, \lambda^n_\delta>0$ with $\lambda^1_\delta+\dots +\lambda^n_\delta=1$, we have $x^0=\lambda^1_\delta x^1_\delta+\dots +\lambda^n_\delta x^n_\delta$ where $x^i_\delta =\int_X x \df\pi^i_\delta(x)$. Finally, by condition \eqref{eq_mart}, for some sufficiently small $\epsilon>0$, there exists $\tilde \pi\in \Pi(\pi_\Omega(\cdot|\relint(\Gamma_x)))$ such that, for all measurable $A\subset X$, 
\[
\tilde \pi_X(A)=\pi_X(A|\relint(\Gamma_x)) +\varepsilon \delta_{x^0}-\varepsilon \sum_i \lambda^i_\delta \pi^i_\delta(A),
\]
where $\delta_{x^0}$ denotes the Dirac measure at $x^0$. Intuitively, $\tilde{\pi}_X$ modifies $\pi_X(\cdot|\relint(\Gamma_x))$ by transferring some mass from the neighborhoods of points $x^i$ into $x^0$.  But then, by Jensen's inequality, and relying on conditions \eqref{eq_inc} and \eqref{eq_equal} to ensure that $\supp(\pi(\cdot|\relint(\Gamma_x) )\subset \Gamma$ and $\supp(\tilde{\pi})\subset \Gamma$, we have
\begin{align*}
\int_{X\times X} w(x,\omega)\df \tilde  \pi(x,\omega)&- \int_{X\times X} w(x,\omega)\df \pi(x,\omega|\relint (\Gamma_x))\\
&= \varepsilon \left(\sum_i \lambda^i_\delta \int_{X}x^2\df \pi_\delta^i(x)-(x^0)^2\right)\geq \varepsilon \left(\sum_i \lambda^i_\delta (x^i_\delta)^2-(x^0)^2\right)>0,
\end{align*}
yielding a contradiction with \eqref{eq_max}.

\subsection{Proof of Proposition \ref{t:RSsmooth}}\label{proof_t:RSsmooth}  
In this appendix, we prove the necessity part of Proposition \ref{t:RSsmooth}.
Fix an optimal $\pi^\star\in \Pi(\mu_0)$. Since $\mu_0$ has a density and $\nabla v(x)=(x_2,x_1)\neq (y_2,y_1)=\nabla v(y)$ for $x\neq y$, Theorem \ref{t:Brenier2} implies that $\pi^\star$ is the unique optimal signal, and that it is convex-partitional. Suppose that $\supp (\piX^\star)$ is the graph of the function $f$, as described in the proposition.

By the definition of $\Gamma_x$ from Section \ref{sec.structure}, for each $t\in [\ul x_1,\ol x_1]$,
\[
\Gamma_{(t,f(t))}=\{\omega\in \Omega:\,t\in \underset{s\in [\ul x_1,\ol x_1]}{\argmax}\,\{\omega_1 f(s)+\omega_2 s-sf(s)\}.
\]

First, consider $t\in (\ul x_1,\ol x_1)$. The necessary first-order condition yields $\omega_2=f(t)-f^\prime(t)(\omega_1-t)$ for all $\omega \in \Gamma_{(t,f(t))}$. Define, for all $t\in [\ul x_1,\ol x_1]$, 
\begin{gather*}
\ul l_t:= \min_{\omega\in X} \left\{\omega_1-t\right\},\\
\text{subject to } \omega_2=f(t)-f^\prime(t)(\omega_1-t),\\
\omega_2+\frac{(t-\omega_1)(f(t)-\omega_2)}{s-\omega_1} \leq f(s),\quad \text{for all $s\in (\omega_1,\ol x_1]$},
\end{gather*}
and
\begin{gather*}
\ol l_t:= \max_{\omega \in X} \left\{\omega_1-t\right\},\\
\text{subject to } \omega_2=f(t)-f^\prime(t)(\omega_1-t),\\
\omega_2+\frac{(t-\omega_1)(f(t)-\omega_2)}{s-\omega_1} \geq f(s),\quad \text{for all $s\in [\ul x_1,\omega_1)$}.
\end{gather*}
Notice that $(t+\ul l_{t},\,f(t)-f^\prime(t)\ul l_t)$ and $(t+\ol l_t,\,f(t)-f^\prime(t)\ol l_t)$ are the points  in $\Gamma_{(t,f(t))}$ with the lowest and highest first coordinate. To see this, consider $\omega\in \Gamma_{(t,f(t))}$ with $t>\omega_1$ (and thus $f(t)-\omega_2=-f^\prime(t)(t-\omega_1)<0$) and notice that, for $s\leq \omega_1$, we have $f(s)\leq f(t)< \omega_2$; thus,
\[
(t-\omega_1)(f(t)-\omega_2)<0\leq (s-\omega_1)(f(s)-\omega_2).
\]
Consequently, $\omega\in \Omega$ with $\omega_1<t$ belongs to $\Gamma_{(t,f(t))}$ if and only if
\begin{gather*}
\omega_2=f(t)-f^\prime(t)(\omega_1-t),\\
\omega_2+\frac{(t-\omega_1)(f(t)-\omega_2)}{s-\omega_1} \leq f(s),\quad \text{for all $s\in (\omega_1,\ol x_1]$}.
\end{gather*}
Since ${(t,f(t))}\in \Gamma_{(t,f(t))}$, it follows that $(t+\ul l_t,\,f(t)-f^\prime(t)\ul l_t)$ is indeed  the point in $\Gamma_{(t,f(t))}$ with the lowest first coordinate. An analogous argument shows that $(t+\ol l_t,\,f(t)-f^\prime(t)\ol l_t)$ is the point in $\Gamma_{(t,f(t))}$ with the highest first coordinate. Finally, since, by Lemma \ref{l:face}, $\Gamma_{(t,f(t))}$ is convex, it follows that 
\[
\Gamma_{(t,f(t))}=\cl(I_t):=\{\omega\in \Omega:\,\omega_1=x_1+l,\, \omega_2=f(x_1)-f^\prime(x_1)l,\, l\in [\ul l(x_1),\ol l(x_1)]\}.
\]
It turns out that the above property also holds for $x\in \supp(\piX^\star)$ with $x_1\in \{\ul x_1,\ol x_1\}$. However, the proof of that fact is significantly more complicated.
\begin{lemma}\label{lemma_help}
$\Gamma_{(t,f(t))}=\cl(I_t)$ for $t\in \{\ul x_1,\ol x_1\}$.
\end{lemma}
\begin{proof}
See Appendix \ref{proof_lemma_help}.
\end{proof}
By Lemma \ref{lemma_help}, we can conclude that $\Gamma_{(t,f(t))}=\cl(I_t)$ for each $t\in [\ul x_1,\ol x_1]$. Since the projection of the contact set $\Gamma$ along the second coordinate is $X=\Omega$, it follows that $\Omega=\bigcup _{t\in [\ul x_1,\ol x_1]} \cl(I_t)$. Define $I_t=\text{relint}(\cl(I_t))$, for $t\in [\ul x_1,\ol x_1]$.\footnote{Note that $I_t$ is a point when $\cl(I_t)$ is degenerate, since a point is a relatively open set.} By part 2(b) in Lemma \ref{l:face}, for $t\neq s$, the open line segments $I_t$ and $I_s$ do not intersect. In fact, part 2(c) in Lemma \ref{l:face} yields a stronger conclusion that, for $t\neq s$, the closed line segments $\cl(I_t)$ and $\cl(I_s)$ can intersect only at a common endpoint. 
Thus, as in the proof of Theorem \ref{t:Brenier2}, invoking Theorem 1 in \cite{larman}, we conclude that $\Omega\setminus\left\{\bigcup_{t\in [\ul x_1,\ol x_1]}I_t\right\}$ has zero (Lebesgue) measure. In sum, we have established that there exists a collection $\{I_t\}_{t\in [\ul x_1,\ol x_1]}$ of open disjoint line segments that partition $\Omega$, up to a measure-zero set.

The first property then follows directly from the above characterization of the optimal signal $\pi^\star$ and the assumption that $\supp(\piX^\star)=\Gr (f)$. The second property follows from the definition of $\cl(I_t)$. Moreover, the inclusion $I_t\subseteq \{\omega\in \Omega:\,\omega_2=f(t)-f^\prime(t)(\omega_1-t)\}$, for $t\in [\ul x_1,\ol x_1]$, follows directly from the fact that $\Gamma_{(t,f(t))}=\cl(I_t)$ for each $t\in [\ul x_1,\ol x_1]$. This finishes the proof of the proposition.

\subsubsection{Proof of Lemma \ref{lemma_help}}\label{proof_lemma_help}

We start by proving yet another lemma.

\begin{lemma}\label{l:parabola<h}
There exists $\varepsilon>0$ such that 
\[
\omega_2+\frac{(x_1-\omega_1)(f(x_1)-\omega_2)}{y_1-\omega_1} < f(y_1),
\]
for all $x_1\in [\ul x_1,\ol x_1]$, $\omega_1\in [x_1-\varepsilon ,x_1]$, $y_1\in (\omega_1, x_1)\cup (x_1, \ol x_1]$, and $\omega_2=f(x_1)-f'(x_1)(\omega_1-x_1)$.
\end{lemma}
\begin{proof}
Since $f'$ and $f''$ are continuous and $f'>0$ on the compact set $[\ul x_1,\ol x_1]$, we have $\ul f'=\min_{\tilde x_1\in [\ul x_1,\ol x_1]}f'(\tilde x_1)>0$ and $\ul f''=\min_{\tilde x_1\in [\ul x_1,\ol x_1]} f''(\tilde x_1)\in \R$. Thus, there exists $\varepsilon>0$ such that $2\ul f' +\varepsilon\ul f''>0$. Fix such $\varepsilon$.
By direct calculation,
\[
\omega_2+\frac{(x_1-\omega_1)(f(x_1)-\omega_2)}{y_1-\omega_1}< f(x_1)+f'(x_1)\varepsilon -\frac{f'(x_1)\varepsilon^2}{y_1-x_1+\varepsilon},
\]
for all $x_1\in [\ul x_1,\ol x_1]$, $\omega_1\in (x_1-\varepsilon ,x_1]$, $y_1\in(\omega_1,x_1)\cup (x_1, \ol x_1]$, and $\omega_2=f(x_1)-f'(x_1)(\omega_1-x_1)$.
Thus, it suffices to show that
\[
f(x_1)+f'(x_1)\varepsilon -\frac{f'(x_1)\varepsilon^2}{y_1-x_1+\varepsilon}< f(y_1),
\]
for all $x_1\in [\ul x_1,\ol x_1]$ and $y_1\in(x_1-\varepsilon,x_1)\cup (x_1, \ol x_1]$.

 If $\ul f''\geq 0$, the inequality holds because the right-hand side $f(y_1)$ is convex in $y_1$ with derivative $f'(x_1)$ at $x_1$, while the left-hand side is strictly concave in $y_1$ with derivative $f'(x_1)$. So assume that $\ul f''<0$ and denote
\[
\hat y_1=x_1+\frac{f'(x_1)-\ul f'}{-\ul f''}.
\]
Since $f''(y_1)\geq \ul f''$ and $f'(y_1)\geq \ul f'$, for all $y_1\in [\ul x_1,\ol x_1]$, we have $f(y_1)\geq \ul f(y_1)$, where
\[
\ul f(y_1)=
\begin{cases}
f(x_1)+f'(x_1)(y_1-x_1)+\frac{\ul f''}{2}(y_1-x_1)^2, &y_1\leq \hat y_1,\\
f(x_1)+f'(x_1)(\hat y_1-x_1)+\frac{\ul f''}{2}(\hat y_1-x_1)^2+\ul f' \left(y_1-\hat y_1\right), &y_1> \hat y_1.
\end{cases}
\]
Thus, it suffices to show that 
\begin{equation}\label{eq_show}
f(x_1)+f'(x_1)\varepsilon -\frac{f'(x_1)\varepsilon^2}{y_1-x_1+\varepsilon}<\ul f(y_1).
\end{equation}
By direct calculation, for $y_1\in (x_1-\varepsilon, x_1)\cup (x_1,\hat y_1]$, inequality \eqref{eq_show} is equivalent to $2f'(x_1)+\ul f'' (y_1-x_1+\varepsilon)>0$, which holds if and only if it holds at $\hat y_1$. At $\hat y_1$, \eqref{eq_show} simplifies to $f'(x_1)+\ul f' +\ul f'' \varepsilon>0$, which holds because $2\ul f' +\varepsilon \ul f''>0$. Again, by direct calculation, for $y_1> \hat y_1$, inequality \eqref{eq_show}
is equivalent to
\[
\frac {(f'(x_1)-\ul f')^2}{2(-\ul f'')}(y_1-x_1+\varepsilon) +\ul f' (y_1-x_1)(y_1-x_1+\varepsilon)-f'(x_1)(y_1-x_1)\varepsilon>0,
\]
where the left-hand side is quadratic and convex in $y_1$. Moreover, the derivative at $y_1=\hat y_1$ is positive because $3\ul f'+f'(x_1)+2\ul f'' \varepsilon>0$, as follows from $2\ul f' +\varepsilon \ul f''>0$. Thus, the left-hand side is increasing in $y_1$ and inequality \eqref{eq_show} holds for $y_1>\hat y_1$, because it holds for $y_1=\hat y_1$, as shown above.
\end{proof}

We are now ready to prove Lemma \ref{lemma_help}. We will focus on the case  $t= \ul x_1$ since the other case is fully analogous. The necessary Kuhn-Tucker condition yields $\omega_2\leq f(\ul x_1)-f^\prime(\ul x_1)(\omega_1-\ul x_1)$ for all $\omega \in \Gamma_{(\ul x_1,f(\ul x_1))}$. We claim that $\omega_2\geq  f(\ul x_1)-f^\prime(\ul x_1)(\omega_1-\ul x_1)$ for all $\omega \in X$, and thus $\omega_2= f(\ul x_1)-f^\prime(\ul x_1)(\omega_1-\ul x_1)$ for all $\omega \in \Gamma_{(\ul x_1,f(\ul x_1))}$, so $\Gamma_{(\ul x_1,f(\ul x_1))}=\cl(I_{\ul x_1})$, by the same argument as previously. Towards a contradiction, suppose that there exists $z\in X$ such that $z_2<  f(\ul x_1)-f'(\ul x_1)(z_1-\ul x_1)$ and $z_1<x_1$ (the case $z_1>x_1$ is analogous and omitted). Since $X$ is convex and the graph of $f$ is a maximal monotone set in $X$, it follows that $z_2> f(\ul x_1)$ and that there exists $\varepsilon>0$ such that, for all $\omega_1\in (\ul x_1-\varepsilon,\ul x_1)$, points $(\omega_1,\,f(\ul x_1)-f'(\ul x_1)(\omega_1-\ul x_1))$ and $(\omega_1,\,f(\ul x_1)-\iota (\omega_1-\ul x_1))$ with $\iota=(z_2-f(\ul x_1))/(x_1-z_1)\in (0,f'(\ul x_1))$ belong to $X$. It is easy to see that, for all $\omega_1<\ul x_1$ and $y_1>x_1$, we have
\[
\omega_2-\iota (\omega_1-x_1) -\iota\frac{(\ul x_1-\omega_1)^2}{y_1-\omega_1}<\omega_2-f'(x_1) (\omega_1-x_1) -f'(x_1)\frac{(\ul x_1-\omega_1)^2}{y_1-\omega_1}.
\]
Thus, by Lemma \ref{l:parabola<h}, for sufficiently small $\varepsilon>0$, points $(\omega_1,\,f(\ul x_1)-f'(\ul x_1)(\omega_1-\ul x_1))$ and $(\omega_1,\,f(\ul x_1)-\iota(\omega_1-\ul x_1))$ belong to $\Gamma_x$. But then $\Gamma_x$ has a non-empty interior, and all points in the interior belong only to $\Gamma_x$, by Lemma \ref{l:face}. Consequently, since $\mu_0$ has full support density on $X$,
\[
\int_{\Gamma_x} (\omega_2-f(x_1)-f'(x_1)(\omega_1-x_1))\df \mu_0(\omega)=\int_{\interior (\Gamma_{x})} (\omega_2-f(x_1)-f'(x_1)(\omega_1-x_1))\df \mu_0(\omega)<0,
\] 
as the boundary of the convex set $\Gamma_x$ has zero Lebesgue measure, by Theorem 1 in \cite{Lang}, and the integrand is strictly negative on the interior of $\Gamma_x$, as implied by the Kuhn-Tucker condition. This shows that any $\pi$ supported on $\Gamma$ cannot be in $\Pi(\mu_0)$, as it violates the second constraint in the definition of $\Pi(\mu_0)$. A contradiction.

\subsection{Proof of Proposition \ref{thm_case}}\label{a:linear f}

Suppose that  $\pi\in \Pi(\mu_0)$, induced by the disclosure of the realization of $a\omega_1+\omega_2$, is optimal.
Define $\Theta=\{\theta=a\omega_1+\omega_2:\omega\in \Omega\}$. Since $\Omega$ is a compact convex set with a non-empty interior, we have $\Theta=[\ul \theta,\ol \theta]$ for some $\ul\theta<\ol \theta$. By Proposition \ref{prop:KX}, $\supp(\pi_X)$ is a monotone set. Thus, since $\mu_0$ has full-support density on $\Omega$, we have $\supp (\pi_X)=\{(x_1(\theta),x_2(\theta)):\, \theta\in \Theta\}$ for some non-decreasing functions $x_1$ and $x_2$ satisfying $ax_1(\theta)+x_2(\theta)=\theta$ for all $\theta\in \Theta$ and $(x_1(\theta),x_2(\theta))\in \interior (\Omega)$ for almost all $\theta\in \Theta$. Note that $x_1$ is $1/a$-Lipschitz and $x_2$ is $1$-Lipschitz, and thus $\tilde\Theta=\{\theta\in\Theta:\, (x_1(\theta),x_2(\theta))\in \interior(\Omega) \}$ is an open set of full measure.

\begin{lemma}\label{l:line}
For each $\theta\in \tilde \Theta$, there exists $\delta>0$ such that, for all $\theta'\in (\theta-\delta,\theta+\delta)$,
\[
a(x_1(\theta')-x_1(\theta))=x_2(\theta')-x_2(\theta)=\frac 12 (\theta'-\theta).
\]
\end{lemma}
\begin{proof} 
Since $\theta\in \tilde \Theta$, there exists $\varepsilon>0$ such that $\omega\in \interior(\Omega)$ for all $\omega\in \R^2$ such that $\omega_1\in (x_1(\theta)-\varepsilon,x_1(\theta)+\varepsilon)$ and $\omega_2\in (x_2(\theta)-\varepsilon,x_2(\theta)+\varepsilon)$. Fix $\delta=\min \{\varepsilon/2,a\varepsilon/2\}$. We claim that for all $\theta'\in (\theta-\delta ,\theta+\delta)$ and all $\omega'\in \R^2$ such that $\omega_2'\in (x_2(\theta')-\delta,x_2(\theta')+\delta)$ and $a\omega_1'+\omega_2'=\theta'$, we have $x(\theta')\in \interior(\Omega)$ and $\omega'\in \interior(\Omega)$. 
Indeed, since $x_1$ and $x_2$ are non-decreasing and satisfy $ax_1(\theta')+x_2(\theta')=\theta'$, we have $x_1(\theta')\in (x_1(\theta)-\delta/a,x_1(\theta)+\delta/a)$ and $x_2(\theta')\in (x_2(\theta)-\delta,x_2(\theta)+\delta)$, so $x(\theta')\in \interior (\Omega)$. Next, since $a\omega_1'+\omega_2'=\theta'=ax_1(\theta')+x_2(\theta')$ and $\omega'_2\in (x_2(\theta')-\delta,x_2(\theta')+\delta)$, we have $\omega_2'\in (x_2(\theta')-\delta,x_2(\theta')+\delta)\subset (x_2(\theta)-2\delta,x_2(\theta)+2\delta)$ and 
$\omega_1'\in (x_1(\theta')-\delta/a,x_1(\theta')+\delta/a)\subset (x_1(\theta)-2\delta/a,x_1(\theta)+2\delta/a)$, so $\omega'\in \interior (\Omega)$.

Fix $\theta'\in (\theta-\delta,\theta+\delta)$ and an integer $n>0$. For $i\in \{0,\dots,n\}$, define $\theta^i=\theta+(\theta'-\theta) i/n$, $\omega^{Li}=(x_1(\theta^i)-\delta/a,x_2(\theta^i)+\delta)$, and $\omega^{Ri}=(x_1(\theta^i)+\delta/a,x_2(\theta^i)-\delta)$. As shown in the previous paragraph, we have $x(\theta^i),\omega^{Li},\omega^{Ri}\in \interior(\Omega)$ for all $i$. Next, by Theorem \ref{t:moment}, we have, for all $i\in \{0,\dots,n\}$,
\begin{gather*}
(x_1(\theta^i)-\omega^{Li}_1)(x_2(\theta^i)-\omega_2^{Li})\leq (x_1(\theta^{i+1})-\omega^{Li}_1)(x_2(\theta^{i+1})-\omega_2^{Li})	\\
\iff x_2(\theta^{i+1})-x_2(\theta^i)\geq\frac{a(x_1(\theta^{i+1})-x_1(\theta^i))}{1+\frac{a}{\delta}(x_1(\theta^{i+1})-x_1(\theta^i))},
\end{gather*}
and
\begin{gather*}
(x_1(\theta^i)-\omega^{Ri}_1)(x_2(\theta^i)-\omega_2^{Ri})\leq (x_1(\theta^{i+1})-\omega^{Ri}_1)(x_2(\theta^{i+1})-\omega_2^{Ri})	\\
\iff x_2(\theta^{i+1})-x_2(\theta^i)\leq\frac{a(x_1(\theta^{i+1})-x_1(\theta^i))}{1-\frac{a}{\delta}(x_1(\theta^{i+1})-x_1(\theta^i))}.
\end{gather*}
Since $x_1$ is $1/a$-Lipschitz, we have, for all $i\in \{0,\dots,n\}$,
\[
\frac{a(x_1(\theta^{i+1})-x_1(\theta^i))}{1+\frac{1}{n\delta}(\theta'-\theta)}\leq x_2(\theta^{i+1})-x_2(\theta^i)\leq \frac{a(x_1(\theta^{i+1})-x_1(\theta^i))}{1-\frac{1}{n\delta}(\theta'-\theta)}.
\]
Summing over $i\in \{0,\dots,n-1\}$ gives
\[
\frac{a(x_1(\theta')-x_1(\theta))}{1+\frac{1}{n\delta}(\theta'-\theta)}\leq x_2(\theta')-x_2(\theta)\leq \frac{a(x_1(\theta')-x_1(\theta))}{1-\frac{1}{n\delta}(\theta'-\theta)}.
\]
Since $n$ is arbitrary, we have $x_2(\theta')-x_2(\theta)=a(x_1(\theta')-x_1(\theta))$. Taking into account that $ax_1(\theta)+x_2(\theta)=\theta$ and $ax_1(\theta')+x_2(\theta')=\theta'$ completes the proof of the lemma.
\end{proof}
Since $\tilde \Theta$ is an open set in $\R$, it is the union of at most countably many disjoint open intervals $(\ul \theta{}^i,\ol \theta{}^i)$. Lemma \ref{l:line} implies that
\[
a(x_1(\theta')-x_1(\theta))=x_2(\theta')-x_2(\theta)=\frac 12 (\theta'-\theta), \quad \text{for all $\theta',\theta\in (\ul \theta^i,\ol \theta^i)$}.
\]
Since $\tilde \Theta$ has full Lebesgue measure, it follows that $\cl (\tilde \Theta)=\Theta$. Since $x_1$ and $x_2$ are (Lipschitz) continuous, we have
\[
a(x_1(\theta')-x_1(\theta))=x_2(\theta')-x_2(\theta)=\frac 12 (\theta'-\theta), \quad \text{for all $\theta',\theta\in \Theta$},
\] 
and thus $x_2(\theta)=ax_1(\theta)+b$ for all $\theta\in \Theta$ and some $b\in \R$.

\theoremstyle{definition} \newtheorem{OAdefn}{\protect\definitionname}
\theoremstyle{plain} \newtheorem{OAthm}{\protect\theoremname}
\theoremstyle{plain} \newtheorem{OAcor}{\protect\corollaryname}
\theoremstyle{plain} \newtheorem{OAprop}{\protect\propositionname}
\theoremstyle{plain} \newtheorem{OAlem}{\protect\lemmaname}
\theoremstyle{definition} \newtheorem{OAexample}{\protect\examplename}
\theoremstyle{plain} \newtheorem{OAclaim}{\protect\claimname}
\theoremstyle{definition} \newtheorem{OAobs}{\protect\observationname}
\theoremstyle{plain} \newtheorem{OAresult}{\protect\resultname}

\providecommand{\claimname}{Claim} \providecommand{\corollaryname}{Corollary}
\providecommand{\definitionname}{Definition} \providecommand{\observationname}{Observation}
\providecommand{\examplename}{Example} \providecommand{\propositionname}{Proposition}
\providecommand{\resultname}{Result} \providecommand{\lemmaname}{Lemma}
\providecommand{\theoremname}{Theorem}

\global\long\def\thesection{B.\arabic{section}}%
\global\long\def\thesubsection{B.\arabic{subsection}}%
\global\long\def\theequation{B.\arabic{equation}}%
\global\long\def\theOAthm{B.\arabic{OAthm}}%
\global\long\def\theOAprop{B.\arabic{OAprop}}%
\global\long\def\theOAclaim{B.\arabic{OAclaim}}%
\global\long\def\theOAexample{B.\arabic{OAexample}}%
\global\long\def\theOAcor{B.\arabic{OAcor}}%
\global\long\def\theOAlem{B.\arabic{OAlem}}%
\global\long\def\theOAdefn{B.\arabic{OAdefn}}%
\global\long\def\theOAobs{B.\arabic{OAobs}}%
\global\long\def\theOAresult{B.\arabic{OAresult}}%
\setcounter{prop}{0} \setcounter{exampl}{0} %
\setcounter{claim}{0} \setcounter{defn}{0} %
\setcounter{section}{0}

\section{Supplementary Material}

\subsection{An example with no dual attainment for Section \ref{sec:duality}}\label{s:illustration}

Let $\mu_0$ be the Lebesgue measure on $\Omega=[0,1]$ and let $V(\mu)=\1_{\{\mu=\delta_0/2+\delta_1/2\}}$. Since $\mu_0(\{0,1\})=0$, there does not exist $\tau \in \Tau(\mu_0)$ with $\tau(\delta_0/2+\delta_1/2)>0$, so each feasible distribution $\tau\in \Tau (\mu_0)$ is optimal. However, the  conditions for optimality of $\tau_F$ and $\tau_N$ both fail. In particular, \eqref{eq_FD} does not hold at $\mu=\delta_0/2+\delta_1/2$. As for condition \eqref{eq_ND}, suppose that $p\in \mathcal{P}(V)$ is its supergradient. Then, we would need $\int_0^1 p(\omega)\df\omega=0$, so that the supporting hyperplane defined by $p$ touches $V$ at the prior. But since $p$ is Lipschitz and non-negative, this implies that $p$ is identically $0$; hence, the hyperplane defined by $p$ does not lie above the graph of $V$ at $\delta_0/2+\delta_1/2$. 

The above arguments indirectly show that the dual problem does not have an optimal solution. Indeed, the dual problem is to find a non-negative Lipschitz function $p$ satisfying $p(0)/2+p(1)/2\geq 1$ that minimizes $\int_0^1 p(\omega)\df\omega$. We know from Theorem \ref{t:duality}, that the infimum is $0$. Clearly, the infimum is not attained: It is approximated by  a sequence of Lipschitz functions that take value $1$ at $\omega=0$ and $\omega=1$, and converge to $0$ on $(0,\,1)$.

\subsection{Two duality formulations for moment persuasion}\label{app_two_duals}

In this appendix, we complement the analysis of Section \ref{sec:moment} by formulating the dual problem for moment persuasion. We also introduce an alternative formulation of the dual problem, and show that the price function from Theorem \ref{thm.moment} solves both of these problems. This in turn allows us to sharpen the connection between our results and existing duality methods in the next section.

The problem dual to \eqref{primal_*} is to find functions $p:\,X\to \mathbb{R}$ and $q:\,X\to \mathbb{R}^N$  to 
\begin{equation}\label{dual_*2}
\begin{gathered}\tag{D$_\text{M}$}
	\text{minimize} \int_{\Omega} p(\omega)\df \mu_0(\omega)\\
	\text{subject to } p(y)\geq v(x) + q(x) \cdot (y-x)\text{ for all } x,\,y\in X,
	\\
	\text{$p$ is Lipschitz on $X$,\,$q$ is measurable and bounded on $X$.} 
\end{gathered}
\end{equation}

This duality formulation is a consequence of the fact that in our primal problem we represent feasible solutions as joint distributions of moments and states (similarly to \citealp{Kolotilin2017} and \citealp{KCW}). The dual variable $p$ is a multiplier on the Bayes-plausibility constraint, while the dual variable $q$ is a multiplier on the martingale constraint.  

When, instead, feasibility for the primal problem is described in terms of marginal distributions of moments using a mean-preserving spread constraint (as in \citealp{dworczak2019} and \citealp{DK}), we can write the dual problem as finding a function $p:\,X\to\mathbb{R}$ to
\begin{equation}\label{dual_*}
\begin{gathered}
	\text{minimize } \int_\Omega p(\omega) \df \mu_0 (\omega)\tag{D$'_\text{M}$}\\
	\text{subject to $p(x)\geq v(x)$ for all $x\in X$,}\\
	\text{$p$ is convex and Lipschitz on $X$.} 
\end{gathered}
\end{equation}

In both dual formulations, only prices for states in $\Omega$ matter. However, as Theorem \ref{thm.moment} formally shows, we can always extend these prices to the space of moments $X$, which provides additional insights about the structure of the solution to the primal problem \eqref{primal_*}.

We now show that these problems  can both be treated as duals to \eqref{primal_*} in the sense that their values provide the relevant upper bound on the value of \eqref{primal_*} that is tight and attained by the price function identified in Theorem \ref{thm.moment}.

\begin{OAthm}\label{prop_two_duals}
\
\begin{enumerate}
\item \textbf{Weak duality:} If $v$ is measurable and bounded, then for any $\pi$ feasible for \eqref{primal_*} and any $p$ feasible for either \eqref{dual_*2} or \eqref{dual_*}, $\int_X v(x)\df\piX(x)\leq \int_\Omega p(\omega)\df\mu_0(\omega)$.
\item  \textbf{No duality gap and primal attainment:} If $v$ is bounded and upper semi-continuous, then there exists an optimal solution to \eqref{primal_*}, and the problems \eqref{primal_*}, \eqref{dual_*2}, \eqref{dual_*} all have the same value.
\item \textbf{Dual attainment:} If $v$ is Lipschitz, then the price function $\bar{p}$ from Theorem \ref{thm.moment} solves \eqref{dual_*}, and together with the function $q$ from condition \textit{2} of Theorem \ref{thm.moment}  solves \eqref{dual_*2}.
\end{enumerate}
\end{OAthm}

\begin{proof}

\noindent \emph{Weak duality.} Suppose that $\pi$ is feasible for \eqref{primal_*}. If $(p,\,q)$ is feasible for \eqref{dual_*2}, then
\[
\int_X v(x)\df \piX(x)=\int_{X\times \Omega} (v(x)+q(x)\cdot (\omega -x)) \df \pi(x,\,\omega)\leq \int_{X\times \Omega} p(\omega)\df \pi(x,\,\omega)= \int_\Omega p(\omega)\df\mu_0(\omega). 
\]
If instead $p$ is feasible for \eqref{dual_*},  then 
\[
\int_X v(x)\df \piX(x)\leq \int_X p(x)\df \piX(x)\leq \int_\Omega p(\omega)\df\mu_0(\omega). 
\]
\newline
\noindent \emph{No duality gap and primal attainment.} When $v$ is bounded and upper semi-continuous on $X$, the corresponding $V$ is also bounded and upper semi-continuous on $\Delta (\Omega)$, and hence, by Lemma \ref{l:popt}, the problem \eqref{primal_*} has an optimal solution $\pi^\star\in \Pi(\mu_0)$. 

Thus, weak duality above implies that $\max\eqref{primal_*}\leq \inf \eqref{dual_*2}$. Moreover, if $p$ is feasible for \eqref{dual_*}, then, by Corollary 13.3.3 in \cite{rockafellar}, $p$ has a bounded subgradient (which we denote $q$), so that, for all $x,\, y\in X$,
\[
p(y)\geq p(x)+q(x)\cdot (y-x)\geq v(x)+q(x)\cdot (y-x),
\]
showing that $(p,\,q)$ is feasible for \eqref{dual_*2} and hence $\max\eqref{primal_*}\leq\inf \eqref{dual_*2}\leq \inf \eqref{dual_*}$.

Thus, it suffices to show that $\max\eqref{primal_*}=\inf \eqref{dual_*}$. The proof is essentially the same as the proof of Lemma \ref{l:Vupp}. Let $\mathcal P_M(v)$ denote the set of functions $p:\,X\to \R$ feasible for \eqref{dual_*}.
 By Baire's Theorem, there exists a non-increasing sequence of Lipschitz functions $v_k$ converging pointwise to $v$. Let $\pi^\star_k$ denote an optimal solution to \eqref{primal_*} with the objective function $v_k$. For each $k\in \mathbb N$, we have
\begin{align*}
\int_{X\times \Omega} v(x) \df \pi^\star (x,\omega) &\leq \inf_{p\in \mathcal P_M(v)}\int_{\Omega} p (\omega) \df \mu_0 (\omega)\leq \min_{p\in \mathcal P_M(v_k)}\int_{\Omega} p (\omega) \df \mu_0 (\omega)=\int_{X\times \Omega} v_k(x) \df \pi^\star_k (x,\omega),
\end{align*}
where the first inequality holds by  $\max\eqref{primal_*}\leq \inf \eqref{dual_*}$, the second inequality holds by $\mathcal P_M(v_k)\subset\mathcal P_M(v)$ for $v_k\geq v$, and the equality holds by Theorem \ref{thm.moment}. It is thus sufficient to show that
\[
\lim_{k\rightarrow \infty } \int_{X\times \Omega} v_k(x) \df \pi^\star_k (x,\omega)\leq \int_{X\times \Omega} v(x) \df \pi^\star (x,\omega).
\]
Thanks to compactness of $\Pi(\mu_0)$, up to extraction of a subsequence, we can suppose that $\pi^\star_k$ converges weakly to some $\ol \pi \in \Pi(\mu_0)$. Then for each $j\in \mathbb N$, we have
\[
\lim_{k\rightarrow \infty }\int_{X\times \Omega} v_k(x) \df \pi^\star_k (x,\omega)\leq \lim_{k\rightarrow \infty}  \int_{X\times \Omega} v_j(x) \df \pi^\star_k (x,\omega) =\int_{X\times \Omega} v_j(x) \df \overline \pi (x,\omega) ,
\]
where the first inequality holds because $v_k\leq v_j$ for $k\geq j$, and the equality holds because $v_j$ is (Lipschitz) continuous and $\pi^\star_k \rightarrow \overline \pi$. Then letting $j$ go to infinity and invoking the monotone convergence theorem, 
\[
\lim_{j\to \infty}\int_{X\times \Omega}v_j(x)\df\ol\pi(x,\omega)=\int_{X\times \Omega}v(x)\df\ol \pi(x,\omega),
\]
we obtain
\[
\lim_{k\rightarrow \infty }\int_{X\times \Omega} v_k(x) \df \pi^\star_k (x,\omega)\leq \int_{X\times \Omega}v(x)\df\ol \pi(x,\omega)\leq \int_{X\times \Omega} v(x) \df \pi^\star (x,\omega),
\]
where the last inequality holds because $\pi^\star$ is an optimal solution to \eqref{primal_*}. This establishes that $\max\eqref{primal_*}=\inf \eqref{dual_*2}=\inf \eqref{dual_*}$. \newline
\noindent\emph{Dual attainment.} When $v$ is Lipschitz, Theorem \ref{thm.moment} guarantees existence of $\bar{p}$ and $q$ with all required properties, and such that for any $\pi$ optimal for \eqref{primal_*},
\[
\int_X v(x)\df\piX(x)=\int_\Omega \bar{p}(\omega)\df\mu_0(\omega).
\]
It follows that $\bar{p}$ solves \eqref{dual_*} and $(\bar{p},\,q)$ solve \eqref{dual_*2}.
\end{proof} 

Theorem \ref{prop_two_duals} formalizes our claim from Section \ref{sec:moment} that the two conditions in Theorem \ref{thm.moment} correspond to two alternative formulations of the problem dual to \eqref{primal_*}. At the same time, the proposition shows that these two problems have the same solution, at least under the conditions of Theorem \ref{thm.moment}. This observation allows us to describe the exact connection between our general duality result and existing duality approaches to moment persuasions.

\subsection{Detailed relationship to other duality methods}\label{app.lit}

The one-dimensional moment persuasion problem  has received special attention (see, for example, \citealp{GK-RS}, \citealp{KMZL}, \citealp{Kolotilin2017},  \citealp{dworczak2019}, and \citealp{DK}). When the objective function is Lipschitz, Theorem \ref{thm.moment} generalizes Theorems 1 and 2 in \cite{dworczak2019}: By a simple transformation,  condition \textit{1} of  Theorem \ref{thm.moment} establishes existence of a convex and (Lipschitz) continuous function $p^\star$ and a cumulative distribution function  $G^\star$ of moments (a mean-preserving contraction of $F_0$) such that 
\begin{gather*}
p^\star\geq v,\\
\supp(G^\star)\subseteq\{x\in X:\,p^\star(x)=v(x)\},
\\
\int_\Omega p^\star(x)\df F_0(x)=\int_\Omega p^\star(x)\df G^\star(x).
\end{gather*}
Moreover, the theorem resolves (positively) the conjecture of \citeauthor{dworczak2019} that if the objective function $V$  is measurable with respect to a moment $m(\omega)$, then so is the corresponding price function.

It is worth noting that we impose stronger regularity conditions on the price function compared to \citeauthor{dworczak2019}. In our dual formulation \eqref{dual_*}, we assume that prices $p$ are Lipschitz continuous, while \citeauthor{dworczak2019} only assume continuity. The general trade-off is that stronger regularity conditions on the dual variable make it more difficult to prove that the dual problem has a solution in the assumed class, but---conditional on proving existence---impose tighter structure on the solutions to the primal problem. We impose a stronger condition on $p$ because Lipschitz continuity is directly implied by Theorem \ref{thm.moment}. 

\cite{DK} prove, under weaker assumptions on the objective function,  that the prices that solve the dual problem of \citeauthor{dworczak2019} are in fact Lipschitz. \citeauthor{DK} show that the dual problem in one-dimensional moment persuasion has an optimal solution by demonstrating that feasible solutions can be restricted to a compact set of uniformly Lipschitz functions. Our proof strategy is different: We construct the optimal solution (a price function on the space of moments) from the supergradient of the concave closure of $V$.

Although in most economically relevant cases imposing Lipschitz continuity of prices in the dual is without significant loss of generality, it it is easy to come up with examples where the dual problem \eqref{dual_*} has a solution in the class of continuous functions but not in the class of Lipschitz functions. For instance, when $\mu_0$ is fully supported on $\Omega=[0,\,1]$ and  $v(x)=-\sqrt{x}$, $p(x)=-\sqrt{x}$ is continuous  and achieves the lower bound in \eqref{dual_*}, but a Lipschitz solution does not exist.

\cite{Kolotilin2017}, \cite{GP2}, and \cite{KCW} use an alternative approach to the persuasion problem. Instead of working with an  objective function $V:\,\Delta(\Omega)\rightarrow \R$, they consider a Sender and a Receiver whose utility functions are $w:\,A\times \Omega\rightarrow \R$ and $u:\,A\times\Omega \rightarrow \R$ where $A$ is the space of the Receiver's actions. The Sender chooses a joint distribution $\pi\in \Delta (A\times\Omega)$ of the recommended action $a$ and the state $\omega$. On top of the Bayes plausibility constraint, $\pi$ must satisfy the obedience constraint, which requires each recommended action to be incentive-compatible for the Receiver given the beliefs it induces. As noted in Section \ref{s:concl}, it is possible to reformulate the alternative problem as our problem, and vice versa.

By setting $A=X$ in the model of \cite{Kolotilin2017}, we can draw a tight connection between the two duality approaches.
For $w(a,\omega)=v(a)$ and $u(a,\omega)=-(a-\omega)^2$, the dual problem in \cite{Kolotilin2017} is to find a continuous function $p:\,\Omega\to \R$ and a bounded measurable function $q:\,A\to \mathbb{R}$ to
\begin{equation}\label{dual_alt}
\begin{gathered}
	\text{minimize } \int_{\Omega} p (\omega)  \df \mu_0 (\omega)\tag{D$_\text{A}$}\\
	\text{subject to }
	p (\omega) + q(a) (a-\omega)  \geq v(a)\text{ for all $(a,\,\omega)\in A\times\Omega$},
\end{gathered}
\end{equation}
where $p$ and $q$ are multipliers for the Bayes plausibility and obedience constraints. Thus, the problem \eqref{dual_alt} corresponds to our dual problem \eqref{dual_*2}, and condition \textit{2} of Theorem \ref{thm.moment} establishes that this problem is solved by the price function $\bar{p}$ derived from our general duality results from Section \ref{sec:duality}.

\subsection{Comments on the convex-roof construction}\label{app.convex_roof}

In this appendix, we further investigate the properties of the convex-roof construction that underlies the proof of Theorem \ref{thm.moment}. Our goal is twofold: On one hand, we are interested in regularity conditions under which the convex roof is (Lipschitz) continuous, guaranteeing that it can be used as the price function $\bar{p}$ satisfying conditions \textit{1} and \textit{2} of Theorem \ref{thm.moment} (and hence as a solution to the dual problems \eqref{dual_*} and \eqref{dual_*2}). On the other hand, we show (by means of examples) that the convex roof can behave in surprisingly pathological ways when the space of moments is multi-dimensional, explaining why we need stronger assumptions to extend existing duality methods to the multi-dimensional case.

The main result in this appendix shows that if the support of the prior contains the boundary of its convex hull, then the convex roof preserves the Lipschitz constant of the objective function, and hence the convex roof could be used as a solution to problems \eqref{dual_*2} and \eqref{dual_*}.

\begin{OAprop}\label{prop_convex_roof_L}
Let $v$ be $L$-Lipschitz on $X$, and let $\Omega$ contain the boundary of $X$. Then $\check p$ is $L$-Lipschitz on $X$.	
\end{OAprop} 
\begin{proof}
By the proof of Theorem \ref{thm.moment}, there exists a price function $\bar{p}:\,X\to \mathbb{R}$ that is convex and  $L$-Lipschitz. Moreover, for each $z\in X$, we have $\check p(z)\geq \bar{p}(z)$ and, for each $y\in \Omega$, there exists a sequence $x_n\in X$ converging to some $x\in X$ such that $q(x_n)$ converges to some $r(y)\in \R^N$, with $\lVert r(y)\rVert \leq L$, and
\[
\check p(y)=\bar{p}(y)= \lim_{n\rightarrow \infty} \{ v(x_n) +q(x_n)\cdot (y-x_n) \}.
\]
Thus, for each $z\in X$ and each $y\in \bd X\subset\Omega$, we have
\[
\check p(z)-\check p(y)\geq \lim_{n\rightarrow \infty} \{v(x_n)+q(x_n)\cdot (z-x_n) -v(x_n) -q(x_n)\cdot (y-x_n)\} =r(y)\cdot (z-y),
\]
showing that $r(y)$, with $\lVert r(y)\rVert \leq L$, is a subgradient of $\check p$ at $y\in \bd X$.

By Theorem 7.12 in \cite{aliprantis2006}, at each $z\in \interior (X)$, the convex roof $\check p$ has a subgradient $r(z)\in \R^N$. We claim that $\lVert r(z)\rVert\leq L$. Suppose that $r(z)\neq 0$, as otherwise the claim is trivial. Since $z\in \interior (X)$ and $\lVert r(z)\rVert >0$, there exists $t>0$ such that $y:=z+tr(z)\in \bd (X)\subset \Omega$. Hence,
\begin{align*}
L\left\lVert y-z \right\rVert &\geq \bar{p} (y)-\bar{p}(z)=\check p(y)-\bar{p}(z)\geq \check p(y)-\check p(z)\\
&\geq \check p(z)+r(z)\cdot (y-z)- \check p(z)= r(z)\cdot (y-z)= \lVert r(z)\rVert\lVert y-z\rVert,
\end{align*}
showing that $\lVert r(z)\rVert\leq L$.

Thus, for each $z,y\in X$, we have
\[
\check p(z)-\check p(y)\leq r(y)\cdot(z-y)\leq \lVert r(y)\rVert\lVert z-y\rVert\leq L\lVert y-z\rVert, 
\]
showing that $\check p$ is $L$-Lipschitz on $X$.
\end{proof}

Next, we construct an example showing that the assumptions of Proposition \ref{prop_convex_roof_L} are not redundant:  $\check p$ does not always preserve the Lipschitz constant of $v$ even when $N=2$ and $\Omega$ is finite.
\begin{OAexample}
Let $\Omega=\{(-l,0),(0,1),(l,0)\}$ with $l>1$ and $v(x)=|x_1|$ for $x\in X$, which is $1$-Lipschitz. We can apply Corollary \ref{cor_ver} to show that full disclosure is optimal and thus $ p$ that coincides with $v$ on $\Omega$ solves \eqref{dual}. Indeed, condition \eqref{eq_CS} holds, and, by Jensen's inequality,
\[
V(\mu)=\left \lvert \int_\Omega \omega_1\df \mu(\omega) \right \rvert \leq \int_\Omega \lvert\omega_1\rvert \df\mu(\omega) =\int_\Omega  p(\omega) \df \mu(\omega)\text{ for all $\mu\in \Delta(\Omega)$}.
\]
It is easy to see that $\check p$ is given by $\check p(x)= l (1-x_2)$ for all $x\in X$, so the Lipschitz constant of $\check p$ is $l>1$. Of course, by Theorem \ref{thm.moment}, there exists a different convex  extension $\bar{p}$ of $ p$ from $\Omega$ to $X$ (for example, consider $ \bar{p}=v$ on $X$) that is convex, $1$-Lipschitz, and satisfies $\bar{p}\geq v$. \hfill $\blacksquare$
\end{OAexample}

The next example demonstrates the additional difficulties that arise when the dimension of the space of moments is three (or higher). In this case, even when the objective function is continuously differentiable, and the set of extreme points of $X$ is compact, the convex roof may be discontinuous. 

\begin{OAexample}
The example is adapted from Example 5.1 in \cite{BL}. Let 
\[
K=\left\{(x_1,x_2,x_3):\,x_1=-1,\,x_2^2+x_3^2=1 \right\}\cup\left\{(x_1,x_2,x_3):\,x_1=1,\,x_2^2+x_3^2=1  \right\},
\]
and $\omega^\star=(0,0,1).$ Define $\Omega=K\cup \{\omega^\star$\}, and note that its convex hull $X$ is a cylinder:
\[
X=\left\{ (x_1,x_2,x_3):\,-1\leq x_1\leq 1,\,x_2^2+x_3^2\leq 1 \right\}.
 \]
Define the objective function as $v(x)=x_1^2$ for $x\in X$, which is Lipschitz. We can again apply Corollary \ref{cor_ver} to show that  $ p$ that coincides with $v$ on $\Omega$ solves \eqref{dual}. 

We will now show that the convex roof $\check p$ is discontinuous at $\omega^\star$.  On any line segment $\{(x_1,x_2,x_3):\,-1\leq x_1\leq 1,\,x_2=y,\,x_3=z)\}$ with $y\neq 0$ and $y^2+z^2=1$, the convex roof $\check p$ must be identically 1. This shows that $\check p$ is discontinuous  at $\omega^\star=(0,0,1)$, because $\check p(\omega^\star)=0$ yet $\check p(\omega^n)=1$ for the sequence $\omega^n=(0,1/n,\sqrt{1-1/n^2)}$ that converges to $\omega^\star$, as $n\rightarrow \infty$.

By Theorem \ref{thm.moment}, there exists a convex, Lipschitz extension $\bar{p}$ (for example, $\bar{p}=v$). \hfill $\blacksquare$
\end{OAexample}

Finally, we construct an instance of moment persuasion (with a discontinuous objective function) in which there exists an optimal convex and Lipschitz price function on $\Omega$ solving the original dual \eqref{dual}, but the price function  cannot be extended to a convex and continuous function on $X$. This example, unlike the previous ones, goes beyond indicating a problem with the convex-roof construction; it shows that---beyond the case of a Lipschitz $v$--- requiring the price function to be (Lipschitz) continuous on $X$ in the multi-dimensional moment persuasion problem may be too demanding.

\begin{OAexample}
The example is adapted from Example 5.4 in \cite{BL}. Let
\[
K=\left\{(x_1,x_2,x_3):\,-1\leq x_1\leq -x_3,\,x_2^2+x_3^2=1 \right\}\cup\left\{(x_1,x_2,x_3):\,x_1=1,\,x_2^2+x_3^2=1  \right\},
\]
and $\omega^\star=(0,0,1).$ Define $\Omega=K\cup \{\omega^\star$\}, and note that its convex hull $X$ is the same cylinder:
\[
X=\left\{ (x_1,x_2,x_3):\,-1\leq x_1\leq 1,\,x_2^2+x_3^2\leq 1 \right\}.
 \]
Define the objective function
\[
v(x)=\begin{cases}
1, & x\in K, \\
0, & x= \omega^\star, \\
-1, &  x\notin K\cup \{\omega^\star\}.
\end{cases}
\]
Because the sets $K$ and $\{\omega^\star\}$ are closed and disjoint, the function $v$ is upper semi-continuous.
 
We claim that full disclosure is optimal in this instance of moment persuasion. We can again apply Corollary \ref{cor_ver} by defining $ p=v$ on $\Omega$. Then, $ p$ is trivially Lipschitz, and condition \eqref{eq_CS} holds, so all we have to check is that for all $x\in X$, and $\mu\in \Delta(\Omega)$ such that $\int_\Omega \omega \df \mu(\omega)=x$, $\int_\Omega  p(\omega)\df \mu(\omega)\geq v(x).$ When $x\notin K$, this is trivial because $ p\geq 0$. When $x\in K$, the conclusion is trivial for all $\mu$ with support in $K$. So the only case we have to check is when $x\in K$ but $\supp(\mu)$ contains the point $\omega^\star$. We will prove that this case cannot arise. Indeed, since $\omega^\star$ is an isolated point of $\Omega$, it would have to be that $\mu(\omega^\star)>0$ and
\[
x=\mu(\omega^\star)\omega^\star+\int_{K}\omega \df \mu(\omega).
\]
But $x\in K$ implies that, for almost all $\omega\in \supp(\mu)$, $\omega_2=0$ and $\omega_3=1$ (as otherwise $x_2^2+x_3^2< 1$). But the only points in $K$ with that property are $(-1,0,1)$ and $(1,0,1)$. This is a contradiction with $\mu(\omega^\star)>0$, because $\mu(\omega^\star)>0$ implies that $x_1\in (-1,1)$. 

We will now show that there does not exist a convex and continuous extension of $ p$ to $X$. On any line segment $\{(x_1,x_2,x_3):\,-1\leq x_1\leq 1,\,x_2=y,\,x_3=z)\}$ with $y\neq 0$ and $y^2+z^2=1$, the function $ p$ takes the value $1$ for $x_1\in [-1,-z]\cup \{1\}$. Hence, any convex extension $\bar{p}$ of $  p$ must be identically equal to $1$ on such a line segment. This, however, means that such $\bar{p}$ must be discontinuous at $\omega^\star=(0,0,1)$. Indeed, $\bar{p}(\omega^\star) =0$, but $\bar{p}(\omega^n)=1$ for the sequence $\omega^n=(0,1/n,\sqrt{1-1/n^2)}$ that converges to $\omega^\star$.  \hfill  $\blacksquare$
\end{OAexample}

\subsection{A version of Theorem \ref{t:Brenier2} with a converse}

In this appendix, we state a version of Theorem \ref{t:Brenier2} with a weaker sufficient condition for optimality of a convex-partitional signal, and then show that this condition is necessary for unique optimality of a convex-partitional signal regardless of the prior.

\begin{OAthm}\label{t:Brenier1}
Suppose that $\Omega$ is a convex set and that $\mu_0$ has a density on $\Omega$ with respect to the Lebesgue measure. Moreover, suppose there do not exist distinct $x,\, y\in X$ and $\varepsilon>0$ such that
\begin{equation*}\label{e:ac}
\begin{gathered}
\nabla v(x)=\nabla v(y),\\
\lambda x+(1-\lambda)y\in X, \quad \text{for all $\lambda \in [-\varepsilon,1+\varepsilon]$},\\
\lambda v(x)+(1-\lambda)v(y)\geq v(\lambda x+(1-\lambda)y), \quad \text{for all $\lambda \in [-\varepsilon,1+\varepsilon]$}.
\end{gathered}
\end{equation*}
Then, there is a unique optimal solution, and that solution is convex-partitional.

Conversely, if there exist distinct $x,\, y\in X$ and $\varepsilon>0$ satisfying the above conditions, then there exists a prior with a density on a convex subset of $X$ such that there are (multiple) optimal solutions that are not convex-partitional.
\end{OAthm}

\begin{proof}
First, we introduce additional notation and results that we will need to prove Theorem \ref{t:Brenier1}. For $x\in S^\star$, let $\mathcal U_x$ be the collection of all relative interiors of faces of $\Gamma_x$. By Theorem 18.2 in \citet{rockafellar}, $\mathcal U_x$ is a partition of $\Gamma_x$. Since, by part \textit{2} of Lemma \ref{l:face}, $x\in \Gamma_x$, there exists a unique face $F_x\in \mathcal U_x$ such that $x\in \relint (F_x)$. Define the set $\Gamma^\star$ by letting its $x$-section be given by
\[
\Gamma^\star_x=
\begin{cases}
F_x, &x\in S^\star,\\
\emptyset, &x\notin S^\star, 	
\end{cases} 
\quad \text{for all $x\in X$}.
\]
The key property of $\Gamma^\star$ is that $x\in \relint (\Gamma_x^\star)$ for all $x\in S^\star$.
Note that the projection of $\Gamma^\star$ along the first coordinate is still $S^\star$, which is compact. Note that each $\Gamma^\star_x$ is also compact, because each face of a compact convex set is compact, by Corollary 18.1.1 in \cite{rockafellar}. However, the projection of $\Gamma^\star$ along the second coordinate $X^\star=\cup_{x\in S^\star}\Gamma_x^\star$ is not necessarily compact, so $\pi(\Gamma^\star)=1$ does not imply that $\supp (\pi)\subset \Gamma^\star$.
\begin{OAlem}\label{l:regular}
Let $\pi\in \Pi(\mu_0)$ satisfy  $\supp(\pi)\subset \Gamma$. Then, for $\pi_X$-almost all $x$, a conditional probability $\pi(\omega|x)$ of $\omega$ given $x$ induced by $\pi$ satisfies  $\int_\Omega (\omega-x) \df \pi(\omega|x)=0$ and $\supp (\pi(\cdot|x))\subset \Gamma_x^\star$, so $\pi (\Gamma^\star)=1$ and $\mu_0(X^\star)=1$.
\end{OAlem}
\begin{proof}
Since $\supp(\pi)\subset \Gamma$, by Remark \ref{remark1}, we have $\supp(\pi_X)\subset S^\star$. Let $\pi(\omega|x)$ be any version of the conditional probability of $\omega$ given $x$ induced by $\pi$. Since $\pi \in \Pi (\mu_0)$ and $\supp(\pi)\subset  \Gamma$, we have
\[
\int_\Omega (\omega-x)\df \pi (\omega|x)=0\quad \text{and}\quad \supp(\pi(\cdot|x))\subset\Gamma_x,\quad \text{for $\pi_X$-almost all $x$}.
\]
Moreover, $\int_\Omega (\omega-x)\df \pi(\omega|x)=0$ implies that $x$ is in the convex hull of $\supp (\pi(\cdot|x))$. We claim that $x$ is actually in the relative interior of the convex hull of $\supp (\pi(\cdot|x))$. Let $k$ denote the dimension of the affine hull of $\supp (\pi(\cdot|x))$. Suppose, on the contrary, that $x$ belongs to the relative boundary of the convex hull of $\supp (\pi(\cdot|x))$. Thus, by Theorem 11.2 in \cite{rockafellar}, there exists some $(k-1)$-dimensional hyperplane $H$ that (weakly) separates $x$ from the convex hull of $\supp (\pi(\cdot|x))$. This is possible only if  $\omega \in H$ for all $\omega$ in the convex hull of $\supp (\pi(\cdot|x))$, but then $H$ contains the convex hull of $\supp (\pi(\cdot|x))$, which contradicts the assumption that $k$ is the dimension of the affine hull of $ \supp (\pi(\cdot|x))$.

By Theorem 18.2 in \citet{rockafellar}, every relatively open convex subset of $\Gamma_x$ is contained in one of the sets in $\mathcal U_x$. Since $\mathcal U_x$ is a partition of $\Gamma_x$, and since $x$ is in the relative interior of the convex hull of $ \supp (\pi(\cdot|x))$ and in the relative interior of $F_x\in \mathcal U_x$, it follows that the relative interior of the convex hull of $ \supp (\pi(\cdot|x))$ is contained in $\relint (F_x)$ and thus $\supp (\pi(\cdot|x))\subset \Gamma^\star_x$. Since $\supp (\pi(\cdot|x))\subset \Gamma^\star_x$ for $\pi_X$-almost all $x$, it follows that $\pi(\Gamma^\star)=1$. Thus, $\mu_0(X^\star)=1$ follows from $\mu_0(X^\star)= \pi(X\times X^\star)\geq \pi(\Gamma^\star)=1$, where the first equality is by $\pi\in \Pi(\mu_0)$ and the inequality is by $\Gamma^\star\subset X\times X^\star$.
\end{proof}

We are now ready to prove Theorem \ref{t:Brenier1}. Let $S^\star_+=\{z\in S^\star:\dim (\Gamma_z)\geq 1\}$ and $X^L=\interior (X)\setminus (\cup_{z\in S^\star_+} \rbd (\Gamma_z))$, where, for a set $A\subset X$,  $\dim (A)$ denotes the dimension of the affine hull of $A$ and $\rbd(A)$ denotes the relative boundary of $A$. The set $X^L$ has full Lebesgue measure by Theorem  1 in \citet{larman}. Let $X^R$ be the subset of $X$ where $p^\star$ is differentiable. Since $p^\star$ is Lipschitz, the set $X^R$ has full Lebesuge measure by Rademacher's Theorem (Theorem 10.8 in \citealp{villani2009}). The set $X^\star$ has full  measure under $\mu_0$ by Lemma~\ref{l:regular}. Thus, since $\mu_0$ has a density on $X$, the set $\tilde X=X^L\cap X^R\cap X^\star$ has full  measure under $\mu_0$.

Fix $\omega\in \tilde X$. Define ${\mathcal  X}^\star (\omega)=\{z\in S^\star:\omega\in\Gamma^\star_z\}$, which is non-empty, because $\omega\in X^\star$. We claim that $\mathcal X^\star(\omega)$ is a singleton. Suppose, by contradiction, that there exist distinct $x,\, y\in  {\mathcal X}(\omega )$, so that $\omega\in \Gamma^\star_x$ and $\omega\in \Gamma^\star_y$ with $x,y\in S^\star$, and thus $\omega\in \Gamma_x$ and $\omega\in \Gamma_y$ since $\Gamma^\star_x\subset \Gamma_x$ and $\Gamma^\star_y\subset \Gamma_y$. Note that $\omega\in \relint(\Gamma_x)$ and $\omega\in \relint (\Gamma_y)$ because $\omega\in X^L$. Thus, by part \textit{2(b)} of Lemma \ref{l:face}, we have $\Gamma_x=\Gamma_y$. Moreover, $\Gamma^\star_x=\Gamma_x$ (and  $\Gamma^\star_y=\Gamma_y)$, as otherwise $\Gamma^\star_x$ would be a face of $\Gamma_x$ different from $\Gamma_x$ itself, and thus it would be entirely contained in $\rbd (\Gamma_x)$, by Corollary 18.1.3 in \citet{rockafellar}, but then, by $\omega\in X^L$, we would have $\omega\notin \Gamma^\star_x$, yielding a contradiction. In sum, $\Gamma^\star_x=\Gamma_x=\Gamma_y=\Gamma^\star_y$.

Since $\omega \in \interior (X)\cap X^R$, $p^\star$ is differentiable at $\omega $, so part \textit{1} of Lemma \ref{l:face} yields
\[\nabla p^\star (\omega )=\nabla v(x)=\nabla v(y).\]
 Since, $x\in \relint (\Gamma_x^\star)$, $y\in \relint (\Gamma_y^\star)$, and $\Gamma^\star_x=\Gamma^\star_y$, it follows that there exists $\varepsilon>0$ such that 
\[\lambda x+(1-\lambda)y\in \relint (\Gamma_x^\star)\subset X,\quad \text{for all $\lambda\in [-\varepsilon,1+\varepsilon]$,}\]
and thus,
\begin{align*}
\lambda v(x)+(1-\lambda)v(y) &=\lambda p^\star(x)+(1-\lambda)p^\star(y)\\
&=p^\star(\lambda x+(1-\lambda )y)\\
&=p_{S^\star}(\lambda x+(1-\lambda )y)\\
&\geq v(\lambda x+(1-\lambda)y), \quad  \text{for all $\lambda\in [-\varepsilon,1+\varepsilon]$,}
\end{align*}
where the first equality is by $x\in \Gamma_x$ and $y\in \Gamma_y$, the second equality is by affinity of $p^\star$ on $\Gamma_x^\star=\Gamma_y^
\star$, the third equality is by $p^\star=p_{S^\star}$ on $\Omega=X$, and the inequality is by $p_{S^\star}\geq v$ on $X$. This contradicts the conditions of the theorem. Thus, $\mathcal X^\star (\omega)$ is a singleton $\{\chi (\omega)\}$ for each $\omega\in \tilde X$.

Finally, for any optimal $\pi\in \Pi(\mu_0)$, we have
\[
1=\pi(\Gamma^\star)=\pi \left(\cup_{\omega\in \tilde X} \{\chi(\omega)\}\times\{\omega\}\right),
\]
where the first equality is by Lemma \ref{l:regular}, and the second equality is by $\Gamma^\star=\cup _{\omega\in X^\star}\left(\mathcal X^\star(\omega)\times \{\omega\}\right)$, $\mathcal X^\star (\omega)=\{\chi(\omega)\}$ for $\omega\in \tilde X$, and $\mu_0(\tilde X)=1$.  Since $\chi(\omega)$ is determined by $p^\star$ for $\mu_0$-almost all $\omega\in X$, and $p^\star$ is independent of $\pi$, we conclude that $\pi$ is uniquely determined by
\[
\pi(A,\, B)=\int_B \1 \{\chi (\omega)\in A\}\df \mu_0(\omega),\quad \text{for all measurable $A\subset X$ and $B\subset X$}.
\]

Conversely, suppose that there exist distinct $x,\, y\in X$ and $\varepsilon>0$ such that the condition from Theorem \ref{t:Brenier1} holds. Let $\tilde \Omega=\{\omega\in \Omega:\, \omega=\lambda x+(1-\lambda)y,\, \lambda \in [-\varepsilon,1+\varepsilon]\}$. There exists $\mu_0$ with a strictly positive density on $\tilde \Omega$ such that  there exist $\ul \lambda_x\in (-\varepsilon,0)$, $\ol \lambda _x\in (0,1+\varepsilon)$ and $\ul \lambda_y\in (-\varepsilon,1)$, $\ol \lambda_y\in (1,1+\varepsilon)$ satisfying
{\allowdisplaybreaks
\begin{gather*}
\E_{\mu_0}[\omega\in \tilde \Omega|\, \omega=\lambda x+(1-\lambda)y,\, \lambda \in [\ul \lambda_x,\ol \lambda_x]]=x,\\
\E_{\mu_0}[\omega\in \tilde \Omega|\, \omega=\lambda x+(1-\lambda)y,\, \lambda \in [-\varepsilon,1+\varepsilon]\setminus[\ul \lambda_x,\ol \lambda_x]]=y,\\
\E_{\mu_0}[\omega\in \tilde \Omega|\, \omega=\lambda x+(1-\lambda)y,\, \lambda \in [\ul \lambda_y,\ol \lambda_y]]=y,\\
\E_{\mu_0}[\omega\in \tilde \Omega|\, \omega=\lambda x+(1-\lambda)y,\, \lambda \in (-\varepsilon,1+\varepsilon)\setminus[\ul \lambda_y,\ol \lambda_y]]=x.
\end{gather*}
}
Consider $\pi_x,\pi_y\in \Pi(\mu_0)$ determined by
\begin{gather*}
\pi_x(\cdot|\omega)=\delta_y+(\delta_x-\delta_y)\1\{\omega=\lambda x+(1-\lambda )y,\ \lambda\in [\ul \lambda_x,\ol \lambda_x]\},\\
\pi_y(\cdot|\omega)=\delta_x+(\delta_y-\delta_x)\1\{\omega=\lambda x+(1-\lambda )y,\ \lambda\in [\ul \lambda_y,\ol \lambda_y]\}.
\end{gather*}
By Theorem \ref{t:moment}, $\pi_x$ and $\pi_y$ are optimal. But, by construction, they are distinct and not convex-partitional.  
\end{proof}

\subsection{An example with infinite $S_x$ for Section \ref{sec_beyond}}\label{app_beyond} 

Suppose that $\mu_0$ is uniformly distributed on 
\[
\Omega=X= \{(\omega_1,\omega_2)\in [-2,2]\times[0,1]\}\cup \{(\omega_1,\omega_2)\in [-2,2]\times[-1,0]:\, (\omega_1/2)^2+\omega_2^2\leq 1\},
\]
and suppose that the objective function is
\[
v(x)=
\begin{cases}
-(x_1^2-1)^2, & x_2\geq 0,\\
-(x_1^2+x_2^2-1)^2, & x_2\leq 0. 
\end{cases}
\]
The optimal solution to \eqref{dual} is $p(x)=0$ for all $x\in X$. Moreover, 
\[
S^\star =\{x\in X:\, x_1^2=1,\, x_2\geq 0\}\cup \{x\in X:\, x_1^2+x_2^2=1,\, x_2\leq 0\},
\]
and $\Gamma_x=\Omega$, so that $S_x=\supp(\pi_X)$, for all $x\in S^\star$. Thus, $\pi\in \Pi(\mu_0)$ is an optimal solution to \eqref{primal_*} if and only if $S_x\subset S^\star$. Applying Jensen's inequality to a strictly concave function $\omega_2\mapsto \sqrt {1-\omega_2^2}$ and a strictly convex function  $\omega_2\mapsto -\sqrt {1-\omega_2^2}$, we conclude that $\pi\in \Pi(\mu_0)$ satisfies $S_x\subset S^\star$ only if $S_x$ contains the set $\{x\in X:\, x_1^2+x_2^2=1,\, x_2\leq 0\}$. That is, for $x_2<0$, each optimal signal must pool the states within a line segment $\{(t,x_2)\in \Omega:\, t> 0\}$ to induce a posterior mean $(\sqrt {1-x_2^2},\,x_2)$ and pool the states within a line segment $\{(t,x_2)\in \Omega:\, t<0\}$ to induce a posterior mean $(-\sqrt {1-x_2^2},\,x_2)$. This shows that there does not exist an optimal signal such that $S_x$ is a finite set. Since the set $\{x\in X:\, x_1^2=1,\, x_2\geq 0\}$ consists of line segments, there are multiple optimal signals differing in how states with $\omega_2\geq 0$ are pooled.  There exists an optimal signal with $S_x=\{x\in X:\, x_1^2+x_2^2=1,\, x_2\leq 0\}\cup \{(-1,1/2),(1,1/2)\}$, so that $S_x=\ext(S_x)$, in line with Theorem \ref{t:KMS}, but there also exists an optimal signal with $S_x=S^\star$, so that $S_x\neq \ext(S_x)$.

\subsection{An explicit formula for property 1 in Proposition \ref{t:RSsmooth}}\label{app_cond_exp}

Let $g$ denote the density of the prior distribution $\mu_0$ on $\Omega$. 

\begin{OAprop} 
Property 1 in Proposition \ref{t:RSsmooth} holds if and only if for almost all $t\in [\ul x_1,\ol x_1]$,
\begin{align*}
\int_{\ul l_t}^{\ol l_t} l(2f'(t)-f''(t)l)  g(t+l,f(t)-f'(t)l)\df l=0.
\end{align*}
\end{OAprop}
\begin{proof} Define $\widetilde \Omega=\bigcup_{t\in [\ul x_1,\ol x_1]} I_t$ and recall that $\mu_0(\widetilde \Omega)=1$. By footnote \ref{f:Ew1}, $\E[\omega|\, \omega\in I_t]=(t,f(t))$ is equivalent to $\E [\omega_1|\, \omega\in I_t]=t$. Let $G$ be the distribution function of the posterior mean of $\omega_1$ induced by $\pi^\star$, so that, for all $t\in [\ul x_1,\ol x_1]$, we have
\[
G(t)=\int_{\bigcup_{s\in [\ul x_1,t]}I_{s}}  g (\omega_1,\omega_2)\df \omega_1\df \omega_2.
\]
By the definition of the conditional expectation, property 1 in Proposition \ref{t:RSsmooth} holds if and only if, for all $t\in [\ul x_1,\ol x_1]$, we have
\[
\int_{\ul x_1}^t s\df G(s)= \int_{\bigcup_{s\in [\ul x_1,t]}I_{s}} \omega_1  g (\omega_1,\omega_2)\df \omega_1\df \omega_2.
\]
Consider a change of variables on $\widetilde \Omega$ given by the following transformation: $(\omega_1,\omega_2)=(t+l,f(t)-f'(t)l)$ where $t\in [\ul x_1,\ol x_1]$ and $l\in (\ul l_t,\ol l_t)$. This transformation is a diffeomorphism, as \textit{(1)} it is one-to-one and onto $\widetilde \Omega$, because $I_t\cap I_s=\emptyset$ for $t\neq s$, \textit{(2)} it is continuously differentiable, because $f$ is a twice continuously differentiable function, and \textit{(3) } the Jacobian determinant is negative on $\widetilde \Omega$,
\[
J(t,l)=\det 
\begin{pmatrix}
\frac{\partial \omega_1}{\partial t}	 & \frac{\partial \omega_2}{\partial t}	\\
\frac{\partial \omega_1}{\partial l}	 & \frac{\partial \omega_2}{\partial l}	
\end{pmatrix}
=\det
\begin{pmatrix}
1 & f'(t)-f''(t)l	\\
1 & -f'(t)
\end{pmatrix}
=-(2f'(t)-f''(t)l)<0,
\]
where the inequality follows from the second-order condition for the second property in Proposition \ref{t:RSsmooth} on the (relatively) open set $I_t$. Thus, by the Change of Variables Theorem (Theorem 13.49 in \citealp{aliprantis2006} and Remark 1.3 in \citealp{villani2009}), we have, for all $t\in [\ul x_1,\ol x_1]$,
{\allowdisplaybreaks
\begin{align*}
 G(t)
=&\int_{\ul x_1}^t \int_{\ul l_s}^{\ol l_s} |J(s,l)| g(s+l,f(s)-f'(s)l)\df l\df s\\
 = &\int_{\ul x_1}^t \int_{\ul l_s}^{\ol l_s} (2f'(s)-f''(s)l) g(s+l,f(s)-f'(s)l)\df l\df s,	
\end{align*}
}
and
\begin{align*}
\int _{\ul x_1}^ts\df G(s)
=&\int_{\ul x_1}^t \int_{\ul l_s}^{\ol l_s}(s+l)|J(s,l)|  g(s+l,f(s)-f'(s)l)\df l\df s\\
 = &\int_{\ul x_1}^t \int_{\ul l_s}^{\ol l_s} (s+l)(2f'(s)-f''(s)l) g(s+l,f(s)-f'(s)l)\df l\df s.	
\end{align*}
Substituting $G$ from the first equation to the last equation, we have, for all $t\in [\ul x_1,\ol x_1]$,
\[
\int_{\ul x_1}^t\int_{\ul l_s}^{\ol l_s} l(2f'(s)-f''(s)l) g(s+l,f(s)-f'(s)l)\df l\df s=0,
\]
which holds if and only if the inner integral is 0 for almost all $s\in[\ul x_1,\ol x_1]$.
\end{proof}

\end{document}